\documentclass[fleqn,usenatbib]{mnras}

\usepackage{newtxtext,newtxmath}
\usepackage{graphicx}    % Including figure files
\usepackage{amsmath}     % Advanced maths commands
\usepackage{hyperref}
\usepackage{xfrac} 
\usepackage{wrapfig}
\usepackage{savesym}\let\longtable*\relax
\savesymbol{tablenum}
\usepackage{siunitx}
\DeclareSIUnit\angstrom{\text{\AA}}
\restoresymbol{SIX}{tablenum}
\usepackage{needspace}
\usepackage{subcaption} 
\usepackage{lineno}
\usepackage{graphicx}
\usepackage{xspace}
\usepackage{tabularx}
\usepackage{booktabs}
\usepackage{makecell}
\usepackage{mathptmx}
\newcommand{\fion}[2]{\mbox{[#1\,\uppercase\expandafter{\romannumeral #2\relax}]}}
\newcommand{\sfion}[2]{\mbox{#1\,\uppercase\expandafter{\romannumeral #2\relax}]}}

\DeclareSIUnit\parsec{pc}

\usepackage[T1]{fontenc}

\DeclareRobustCommand{\VAN}[3]{#2}
\let\VANthebibliography\thebibliography
\def\thebibliography{\DeclareRobustCommand{\VAN}[3]{##3}\VANthebibliography}

\usepackage{graphicx}	% Including figure files
\usepackage{amsmath}	% Advanced maths commands

\title[Pop III Stars in JADES He II Emitters]{Sown at Cosmic Dawn: Searching for Pop III Signatures in JWST JADES He II Selected Line Emitters}

\author[R. Roberts et al.]{
R. Roberts,$^{1}$\thanks{E-mail: rosa.roberts@student.manchester.ac.uk}
G. Riley,$^{2}$
C. J. Conselice,$^{1}$
D. Austin,$^{1}$
T. Harvey,$^{3}$
L. Westcott$^{1}$
and J. C. J. D'Silva$^{1}$
\\
$^{1}$Jodrell Bank Centre for Astrophysics, University of Manchester, Oxford Road, Manchester M13 9PL, UK\\
$^{2}$Mathematical, Physical and Life Sciences Division, University of Oxford, Mansfield Road, Oxford OX1 3TA, UK\\
$^{3}$Max-Planck-Institut für Astronomie, Königstuhl 17, D-69117 Heidelberg, Germany\\
}

\date{Accepted XXX. Received YYY; in original form ZZZ}

\pubyear{\the\year{}}

\begin{document}
\label{firstpage}
\pagerange{\pageref{firstpage}--\pageref{lastpage}}
\maketitle

\begin{abstract}
Population~III (Pop~III) stars, the first generation of metal-free stars, remain undetected, partly because entirely pristine galaxies represent a short-lived phase. We instead target the longer-lived \textit{self-polluted} and \textit{hybrid} Pop~III--Pop~II phases, in which surviving metal-free Pop~III stars coexist with enriched gas and newly formed Pop~II stars, producing strong He~II alongside detectable UV metal lines. We search for this signature in 84 strong He~II-emitting galaxies at $z \simeq 3$--7 in the JWST/JADES GOODS-South and GOODS-North fields, combining UV emission-line diagnostics, gas-phase metallicities, AGN screening, and spatially resolved SED fitting. We find no evidence for any pristine Pop~III galaxies. Instead, \textit{The Blueberry} (ID~13176, $z=5.94$) is our strongest candidate for a \textit{hybrid} Pop~III--Pop~II system and is the only source with robust detections of all diagnostic UV lines. Its resolved stellar populations are extremely young, blue, and weakly attenuated, with high star formation rates, consistent with intense recent star formation. While these properties alone cannot distinguish it from a chemically enriched starburst, its robust UV metal-line ratios provide the strongest evidence in our sample for a surviving Pop~III contribution within an enriched host. \textit{The Carrot} (ID~13577, $z=5.57$) further demonstrates the power of resolved analysis, revealing a central region substantially bluer and younger than the galaxy as a whole. Together, these results suggest that Pop~III signatures may be spatially localised and diluted in integrated measurements, highlighting resolved SED fitting as a tool for identifying environments where the Universe's first stars may survive.
\end{abstract}


%%%%%%%%%%%%%%%%%%%%%%%%%%%%%%%%%%%%%%%%%%%%%%%%%%

%%%%%%%%%%%%%%%%% BODY OF PAPER %%%%%%%%%%%%%%%%%%

\section{Introduction}

Detecting Population~III (Pop~III) stars, the first stellar generation to form from the Universe's primordial gas, is a major goal of modern astronomy, offering direct insight into the earliest stages of star and galaxy formation, the onset of chemical enrichment, and the sources driving cosmic reionisation \citep{Bromm_1999,Bromm2004,Bromm_2013,McQuinn_2016_EoR}. Forming within dark matter minihaloes during \textit{cosmic dawn} \citep[$z \sim 15$--30;][]{Bovino_2014,Shimabukuro_2022_cosmicdawn,Zier_2025}, these stars are predicted to be metal-free and massive \citep[$\gtrsim 10,M_\odot$; e.g.][]{Bromm_1999,Bromm2004,Hirano2014}. Although primordial gas can cool through molecular hydrogen and atomic hydrogen transitions, these channels are inefficient compared with metal-enriched cooling \citep{Bromm_2002}. This limits fragmentation during collapse and favours a top-heavy initial mass function \citep[IMF; e.g.][]{Bromm_2002,venditti2025_sim}, producing stellar populations with extremely hard He$^+$-ionising spectra \citep[e.g.][]{Schaerer_2002,Schaerer2003_HeIIlowmet}. These hard radiation fields are expected to produce strong hydrogen and helium recombination lines, particularly He~II~$\lambda1640$, He~II~$\lambda4687$, and H$\alpha$ \citep[e.g.][]{Sibony_2022,Maschmann_2024_WR,trussler2024_popIII_properties}.

The prevalence of nebular He~II emission at high redshift, compared with the local Universe, suggests that hard ionising sources are considerably more common during the early stages of galaxy evolution \citep{Cassata2013_HeIIhighz,Kehrig2015_HeIIhighz,Kehrig2018_HeIIhighz,Mondal2025_HeIIhighz}. Because He~II line strength is expected to increase with decreasing metallicity, it is widely used as a key tracer of possible Pop~III stellar populations \citep{Schaerer2003_HeIIlowmet,Bromm_2013}. However, He~II emission is not unique to primordial stars and may also arise from Wolf--Rayet stars, binary evolution, radiative shocks, or active galactic nuclei (AGN) \citep[e.g.,][]{Maschmann_2024_WR,Dutta_2024,Gonz_lez_Tor__2025, Hovis_Afflerbach_2025}. Identifying genuine Pop~III systems therefore requires additional observational diagnostics capable of distinguishing metal-free stellar populations from chemically enriched galaxies hosting alternative hard ionising sources.

Formed from pristine, metal-free gas, Pop~III stars are expected to exhibit extremely low gas-phase metallicities ($Z\lesssim10^{-4}Z_\odot$; \citealt{Bromm_2002,Maio_2011,Bromm_2011_stellarmass,Pallottini_2014,Jaacks_2018}). Their pristine composition and extreme stellar temperatures are expected to leave distinct imprints on the spectra of Pop~III-dominated galaxies, including high ionisation parameters ($\log U\sim-1$ to $-2$) and very blue UV continua ($\beta\lesssim-3$), indicative of intense ionising radiation and minimal dust attenuation \citep[e.g.][]{Cullen2024,Topping2024_beta,Austin2025a,Morales_2025}. However, individual Pop~III star-forming clusters are expected to have low total stellar masses ($M_\star\sim10^5,M_\odot$; \citealt{Bromm_2011_stellarmass}), making them intrinsically faint in the UV ($M_{\rm UV}\gtrsim-14$ to $\sim-18$; \citealt{Zackrisson_2011,Schauer_2020,Ventura2024,trussler2024_popIII_properties}). Moreover, star formation in these shallow potential wells is expected to occur in short-lived, bursty episodes ($\sim$1--30~Myr), as strong radiative and supernova feedback rapidly regulates star formation and drives chemical enrichment, limiting the duration of the Pop~III phase \citep{Venditti_2023,Venditti_2024_sim,venditti2025_sim,Rusta_2025,Rusta_2026}. Together, their intrinsic faintness and short-lived nature make Pop~III-dominated galaxies highly challenging to identify observationally.

Simulations indicate that Pop~III star formation may persist into the \textit{Epoch of Reionisation} ($z\gtrsim6-7$), and in some cases even to \emph{cosmic noon} ($z \sim 3$) within chemically isolated regions \citep[e.g.][]{Tornatore_2007,Pallottini_2014,Liu2020_sim,Saccardi_2023,Venditti_2024_sim, Zier_2025}. 
The advent of the \textit{James Webb Space Telescope} \citep[JWST;][]{Gardner2006_JWST} has enabled the detection of several Pop~III candidate galaxies at redshifts $z \sim 3-11$ \citep[e.g.][]{Wang_2024_candidate,nakajima2025_candidate,Cullen2025,Cai_2025_candidate,Mondal2025_HeIIhighz, morishita2025_glimpsereject, Vanzella_2026, Reumert_2026, pollock2026, Rusta_2026, Maiolino_2026, Ubler_2026}, with very blue continua ($\beta \sim -3$) and low inferred metallicities ($\sim10^{-3}$–$10^{-2} Z_\odot$). However, these metallicities remain $\sim1$--$2$ dex above the metallicity threshold of $Z\sim10^{-4}Z_\odot$ expected for truly pristine Pop~III systems, and to date, no galaxy has been conclusively confirmed to host a genuinely metal-free stellar population. Several candidates have been disqualified following more precise chemical analysis. A notable example is GLIMPSE-16043 ($z \sim 6.5$), which was initially proposed as a strong Pop III candidate based on its extremely low photometric-inferred metallicity \citep{fujimoto2025_glimpse}. However, subsequent spectroscopic follow-up by \citet{morishita2025_glimpsereject} detected metal emission lines, including [O~III], and measured $12 + \log(\mathrm{O/H}) = 6.9$, ruling out a pristine Pop~III origin.

Given the difficulty of finding pristine systems, the search for Pop~III stars has shifted toward identifying galaxies that retain chemical signatures of early star formation despite some enrichment. To address this, \citet{Rusta_2025} developed a diagnostic framework using combinations of UV line ratios to distinguish between different evolutionary stages, enabling galaxies to be classified into pristine systems (containing only Pop~III stars and metal-free gas), self-polluted systems (Pop~III stars embedded in gas enriched by earlier Pop~III supernovae), and hybrid regimes where Pop~III and Pop~II stellar populations coexist. Within the hybrid category, further subdivisions capture the relative contribution of metal-free stars. This diagnostic is complicated by AGN, compact sources powered by accretion onto supermassive black holes whose emission can outshine that of their host galaxies \citep{Nakos_2008,Abraham_2012,Berghea_2017}. Photoionisation models show that AGN and low-metallicity stellar populations can produce overlapping spectral signatures \citep{Feltre_2016,Gutkin_2016}. In particular, \citet{Nakajima2022_PopIIIproperties} showed that Pop~III and metal-poor stellar populations can occupy the same regions of diagnostic diagrams as AGN and direct-collapse black holes with hard ionising spectra. This overlap highlights the need for additional observational constraints when assessing whether He~II emission contains a genuine Pop~III contribution.

AGN are particularly relevant in this context because their hard ionising continua extend well beyond the 54.4~eV threshold required to ionise He$^+$, allowing them to produce strong nebular He~II emission that can mimic the signatures expected from Pop~III stellar populations. AGN can in principle be identified through broad permitted emission lines ($\sim10^3$--$10^4$~km~s$^{-1}$) arising from the dense broad-line region (BLR) near the central black hole. Forbidden lines instead trace lower-density gas, such as the narrow-line region and AGN-driven outflows \citep{Du_2023}. However, this diagnostic is not always available: the BLR can be obscured by the surrounding torus, making broad-line visibility orientation-dependent \citep{Tadhunter_2008,Gaskell_2009,Czerny_2011}, and at the low spectral resolution of \textit{JWST}/NIRSpec PRISM, the instrumental line-spread function can smear out intrinsic broadening entirely. Consequently, the non-detection of broad permitted lines in PRISM spectra cannot by itself exclude an AGN contribution, motivating the use of complementary spectroscopic and photometric diagnostics when assessing the nature of He~II-emitting sources.

AGN often appear unresolved in imaging, with light profiles consistent with the telescope point spread function (PSF) rather than the extended morphology of star-forming galaxies \citep{Stark_2018, Zhuang_2024, Vietri_2024}. While Pop~III-dominated systems are expected to be more extended, compact high-redshift starbursts can mimic point-source morphologies, complicating this distinction \citep{Whalen_2022, Langeroodi_2023}. Recent \textit{JWST} studies address this using aperture-based concentration measurements to isolate unresolved AGN-like sources from extended star-forming systems \citep[e.g.][]{Kokorev_2024, Greene_2024, Ortiz_2024}. Broadband spectral energy distribution (SED) fitting provides a further independent diagnostic by simultaneously modelling AGN and stellar contributions: an AGN accretion disc can modify the UV--optical continuum, while dust reprocessing enhances longer-wavelength emission, producing SEDs distinct from those of stellar populations alone \citep{Fritz_2006,Feltre_2012,Thorne2022_AGNfraction}. Because AGN emission spans the entire electromagnetic spectrum, this approach captures multi-wavelength signatures that single-wavelength or colour criteria may miss \citep{Padovani_2017}. However, weak or obscured AGN emission can be degenerate with stellar populations, limiting the reliability of SED-based identification alone and making SED fitting most effective when combined with independent morphological and spectroscopic diagnostics \citep{Thorne2022_AGNfraction}.

Previous Pop~III searches have relied on galaxy-integrated measurements, implicitly assuming uniform conditions across a system. This assumption is problematic at high redshift, where galaxies are highly heterogeneous, with sharp spatial variations in star formation, metallicity, and ionisation state \citep{Conselice_2014,Schreiber_2020_cosmicnoon,Scholte2025_metallicities}. Theoretical models predict that any surviving Pop~III star formation persists within exactly this kind of substructure: small, chemically isolated pockets embedded within more evolved, enriched hosts \citep{Rusta_2025, Zier_2025, storck_2026}. Integrated measurements can dilute genuine Pop~III signatures, while the narrow \textit{JWST}/NIRSpec MSA slit may miss compact regions or blend them with surrounding enriched gas. The challenge is therefore not simply detecting Pop~III signatures, but spatially isolating them from their enriched surroundings \citep{Tornatore_2007,Pallottini_2014}. 

Resolved SED fitting offers a powerful way to address this challenge by leveraging the high spatial resolution of \textit{JWST} to recover sub-galactic structure that integrated fitting erases. \citet{Sorba_2018} showed that integrated fitting can underestimate stellar masses by a factor of five, as light from young, luminous populations "outshines" older or fainter stellar components. More recent JWST-era \texttt{piXedfit} modelling by \citet{Abdurro_2023} demonstrates that resolved fitting can break the age-dust-metallicity degeneracy on sub-kpc scales, revealing internal transitions, such as nuclear starbursts and inside-out quenching, that appear uniform in galaxy-integrated data. Applied to Pop~III searches, this approach can identify localised, ultra-blue regions that would otherwise be diluted in the integrated light \citep{iglesiasnavarro2026_sbi}. Crucially, unlike slit spectroscopy, resolved SED fitting is not dependent on the placement of a narrow slit: imaging samples the full spatial extent of the galaxy, allowing candidate regions to be identified wherever they occur. It therefore provides a systematic way to locate potential Pop~III regions before, or independently of, targeted spectroscopic follow-up.

Recent efforts have identified Pop~III candidates through spatially resolved spectroscopy. In the halo of GN-z11 \citep[$z = 10.6$;][]{Bunker2023}, a compact region offset by $\sim 3$~pkpc, dubbed "Hebe", exhibits strong He~II $\lambda1640$ and H$\gamma$ emission and no detectable metal lines \citep{Rusta_2026, Maiolino_2026, Ubler_2026}. Its detection relied on a chance alignment of the narrow slit and could easily have been missed or contaminated. Comparable evidence has been reported in a $z = 5.1$ galaxy from the JWST CAPERS survey \citep[PID: 6386,][]{CAPERS_2024}, where a blue companion located $\sim 3$~pkpc from the primary system is likewise dominated by hydrogen and helium emission \citep{Reumert_2026}. Together, these detections suggest that Pop~III-like conditions may be confined to small, spatially distinct regions in the outskirts of massive haloes, rather than tracing the bulk properties of their host galaxies \citep{Zier_2025}, consistent with simulations predicting that only $\sim0.06\%$ of Pop~III stars form within the virial radius of galaxies with $M_{\rm UV}<-17$ \citep{storck_2026}.

Despite these localised detections, a systematic, survey-wide search combining He~II and H$\alpha$ emission remains absent from the literature. While individual studies have used He~II to trace hard ionising radiation, a comprehensive census exploiting both recombination lines, and testing candidates against resolved stellar-population diagnostics, has yet to be undertaken, limiting constraints on the prevalence of candidate Pop~III systems. In this work, we develop a multi-stage framework that combines emission-line diagnostics, gas-phase metallicities, AGN screening, and spatially resolved SED fitting to identify and characterise candidate Pop~III galaxies within strong He~II-emitting galaxies. By combining integrated spectroscopy with resolved measurements of stellar populations, dust, and star formation, we test whether candidate systems contain localised regions consistent with Pop~III-like star formation, and distinguish these from AGN or chemically enriched stellar populations.

The remainder of this paper is organised as follows: Section~\ref{sec:Data} describes the spectroscopic and photometric datasets and the construction of the parent sample. Section~\ref{sec:Methods} details the emission-line measurements, He~II emitter selection, derived galaxy properties, Pop~III diagnostic framework, resolved SED fitting, and AGN identification. Section~\ref{sec:Results} presents the main results, including the identification and global properties of the Pop~III candidates, AGN screening, and spatially resolved properties. Section~\ref{sec:Discussion} evaluates the physical nature of the candidates, discusses systematic uncertainties, and explores implications for future searches. Finally, Section~\ref{sec:Summary} summarises our main conclusions.

Throughout this work, we adopt a flat $\Lambda$CDM cosmology consistent with the Planck 2018 results \citep{Plank18}, with $H_0=67.66$~km~s$^{-1}$~Mpc$^{-1}$, $\Omega_{\rm M}=0.3097$, and $\Omega_{\Lambda}=0.6903$.

\section{Data}\label{sec:Data}

\subsection{Spectroscopic and Photometric Data}

We utilise publicly available JWST observations from the GOODS-S \citep[Great Observatories Origins Deep Survey-South;][]{GOODS_2004} and GOODS-N survey fields, providing deep spectroscopic and photometric coverage over two of the best-studied extragalactic regions. Spectroscopic data is obtained from the DAWN JWST Archive (DJA)\footnote{\url{https://github.com/dawn-cph/dja.git}}, using observations from the JADES survey \citep[JWST Advanced Deep Extragalactic Survey;][]{JADES_DR1_2023}, including publicly available NIRSpec observations prepared for the forthcoming EPOCHS v2 release (Austin et al. in prep.). We use the v4 DJA data products, reduced using the standard \texttt{jwst} calibration pipeline\footnote{\url{https://github.com/spacetelescope/jwst}} and extracted with \texttt{msaexp}\footnote{\url{https://github.com/gbrammer/msaexp}}. Detailed descriptions of the reduction and extraction procedures are presented in \citet{Valentino_2025}, \citet{deGraaff_2025} and \citet{Heintz_2024}. The resulting catalogues provide calibrated 1D and 2D spectra together with spectroscopic redshifts, source coordinates, and quality information required for source identification and analysis. The extracted 1D spectra are corrected for slit losses in the DJA data products, such that the reported spectral flux densities account for flux falling outside the NIRSpec microshutter aperture.

Photometric measurements are obtained from the \texttt{EPOCHS-DR2} catalogue (Austin et al. 2026 in prep.), previously used in \citet{austin2025resolvingionizingphotonbudget}. Spectroscopic sources are matched to their photometric counterparts through nearest-neighbour cross-matching in sky coordinates using a matching radius of \SI{1}{\arcsecond}. Image cutouts are generated using the \texttt{galfind} pipeline\footnote{Available on github (\url{https://github.com/duncanaustin98/galfind}) and Zenodo (\doi{10.5281/zenodo.18613231})}, while NIRSpec slit geometries are reconstructed from the MSA configuration files and projected onto the imaging data using the World Coordinate System (WCS). This enables direct verification of slit placement relative to the observed galaxy morphology.

\subsection{Parent Sample}\label{sec:Parent Sample}

We begin with all unique NIRSpec sources available in the GOODS-N and GOODS-S fields through the DJA catalogue, yielding an initial sample of approximately 5,000 objects. One- and two-dimensional spectra, associated exposure information, and MSA slit geometries are extracted from the DJA summary FITS products using \texttt{msaexp} within the \texttt{galfind} pipeline. To ensure reliable redshift measurements, we exclude all sources with redshift inspection grades below 3. Spectra are shifted to the rest frame using the spectroscopic redshifts provided by the DJA catalogue, and only sources with spectral coverage across the rest-frame UV interval
$1250 < \lambda_{\mathrm{rest}}/\mathrm{\mathring{A}} < 3000$ are retained.
The primary spectroscopic dataset consists of observations obtained with the NIRSpec PRISM/CLEAR configuration, which provides continuous wavelength coverage over the full NIRSpec spectral range and enables robust measurements of the UV continuum. For each source, we calculate the average signal-to-noise ratio (SNR) across the rest-frame UV continuum and require signal-to-noise ratio in the UV to be $\geq5$. This criterion ensures reliable determinations of continuum properties and emission-line measurements. 

To enable accurate emission-line fitting and line-ratio diagnostics, we additionally require each source to possess at least one medium-resolution NIRSpec observation. These observations include combinations of the G140M, G235M, and G395M gratings with the F070LP, F170LP, and F290LP long-pass filters, providing significantly improved spectral resolution relative to the PRISM data and allowing blended emission features to be separated and measured robustly. After applying these quality and observational requirements, the final parent sample consists of 834 galaxies.

\section{Methods}\label{sec:Methods}

\subsection{Emission Line Measurements}\label{sec:Emission Line Measurements}

Emission lines in the NIRSpec 1D spectra are fitted using \texttt{specFitMSA}\footnote{Available on github (\url{https://github.com/vadimrusakov/specFitMSA.git}) and Zenodo (\doi{10.5281/zenodo.18486501})}, implemented using \texttt{scipy.optimize}. For each line, or group of blended lines within a common velocity window, the code fits a model consisting of a linear continuum plus a Gaussian profile per line, with line widths set by a shared kinematic FWHM combined with the instrumental dispersion at each wavelength. Lines that are spectrally unresolved are automatically merged into a single blended profile. The continuum and line parameters are optimised by minimising the $\chi^{2}$ statistic between model and data, using gradient-based minimisation (\texttt{scipy.optimize.minimize}) with analytic gradients from automatic differentiation using \texttt{JAX}\footnote{\url{http://github.com/jax-ml/jax}}. Parameter uncertainties are estimated from the covariance matrix at the best-fit solution. The code outputs best-fit parameters, integrated line fluxes, equivalent widths, and SNRs, with uncertainties propagated from the parameter covariance matrix.

\begin{figure*}
    \centering
    \begin{minipage}{0.48\linewidth}
        \centering
        \includegraphics[width=\linewidth]{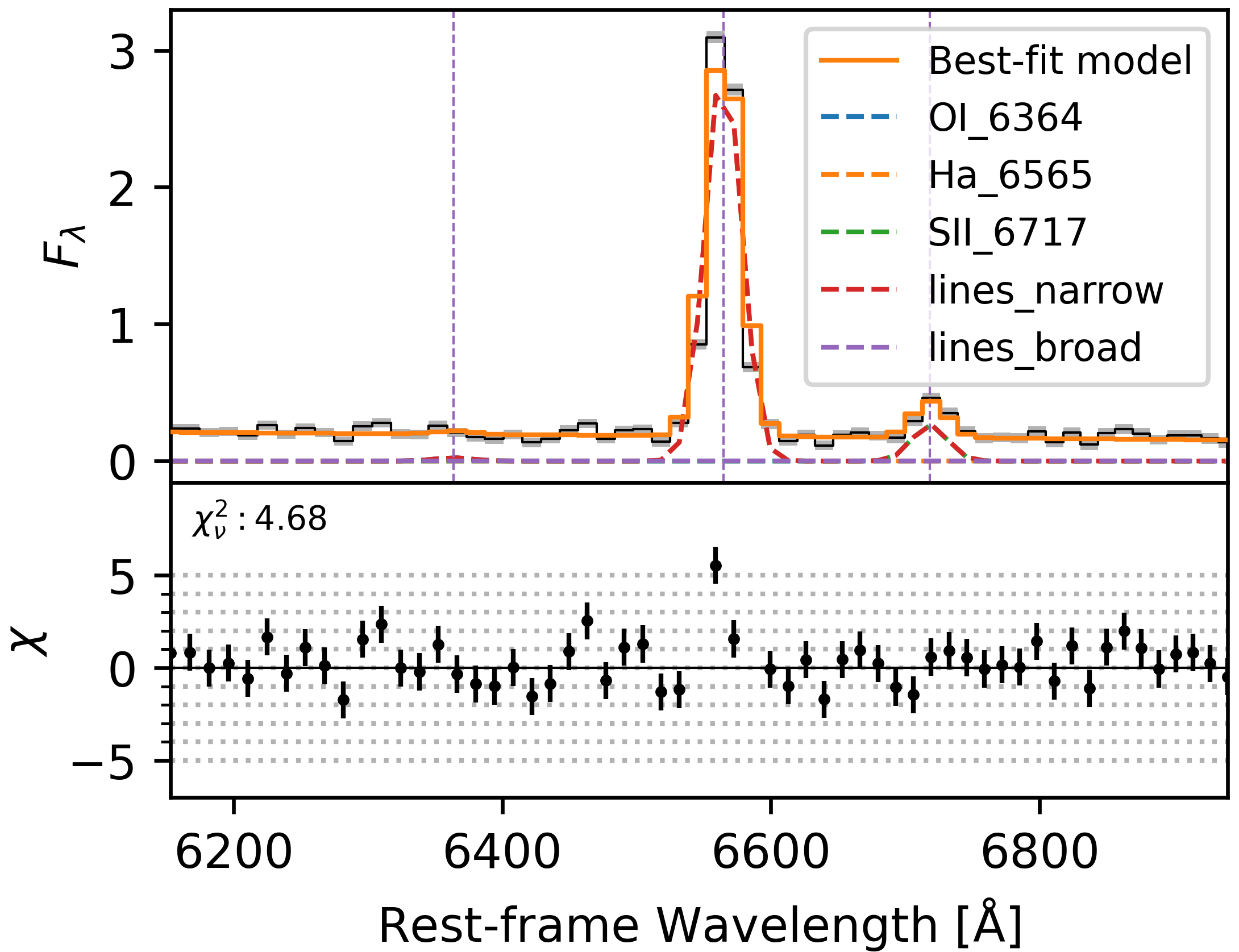}
    \end{minipage}
    \hfill
    \begin{minipage}{0.48\linewidth}
        \centering
        \includegraphics[width=\linewidth]{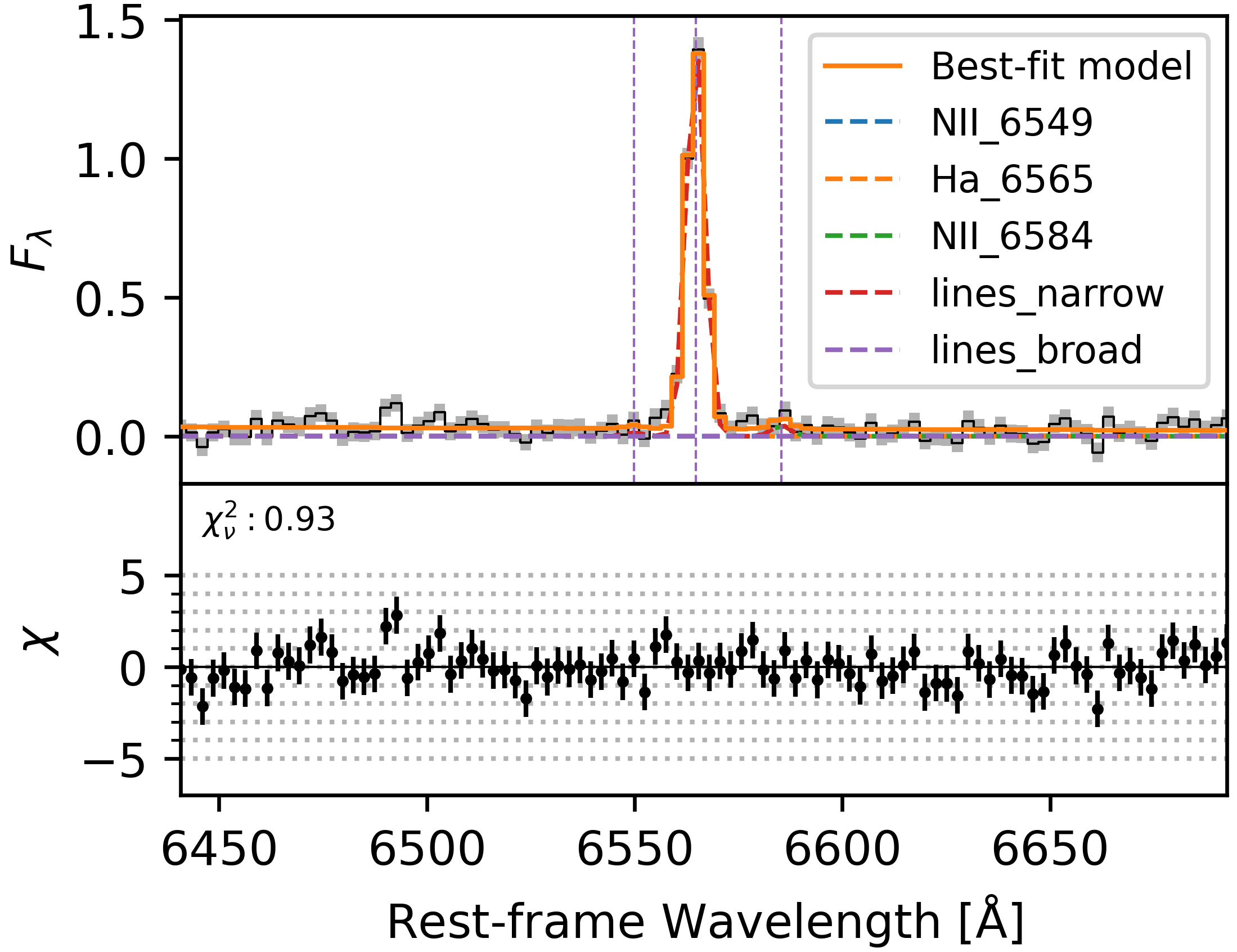}
    \end{minipage}
   \caption{Example \texttt{specFitMSA} fits to the H$\alpha$ $\lambda6565$ region of the NIRSpec 1D spectrum for \textit{The Carrot} (ID~13577) are shown for PRISM/CLEAR (left) and G140M/F070LP (right). The low-resolution PRISM spectrum exhibits blended emission features and a higher reduced $\chi^2$ of 4.68 compared to the higher-resolution G140M data. Observed fluxes are shown in black with $1\sigma$ uncertainties in grey, with the median posterior model overlaid in orange and its 68\% highest-density interval. Individual Gaussian components are indicated by coloured dashed curves, and vertical lines mark rest-frame transitions. Narrow-line components are indicated by red dashed curves and broad-line components by purple dashed curves; however, no broad components are fitted in this example. Residuals, normalised by the uncertainties, are shown in the lower panels.
}
\label{fig:specFitMSA}
\end{figure*}

\subsection{Selection of He II Emitters}\label{sec:Selection of He II Emitters}

We fit to the He~II~$\lambda1640$, He~II~$\lambda4687$, and H$\alpha$~$\lambda6565$ emission lines across all available grating spectra. Both He~II lines trace the same ionisation potential, but the $\lambda4687$ line is intrinsically weaker than the UV He~II~$\lambda1640$ line. However, it lies in the optical range, where the continuum is flatter and there are fewer nearby bright emission lines, facilitating more reliable fits in some cases. Low-resolution PRISM spectra are susceptible to line blending, which can bias individual line-flux measurements. Where available, medium-resolution grating observations provide better separation of neighbouring emission features and are therefore preferred for reliable line measurements. To ensure robust detections, we require a line-fit SNR $>2$ in H$\alpha$ and in at least one He~II line, using measurements from any available grating. Applying these criteria yields a final sample of 84 galaxies with significant He~II and H$\alpha$ emission suitable for further analysis. Figure~\ref{fig:specFitMSA} compares the fit quality between PRISM/CLEAR and G140M/F070LP for an example 1D spectrum, with the overlaid \texttt{specFitMSA} best-fit model.

\subsection{Derived UV Properties}\label{sec:Derived UV Properties}

\subsubsection{UV Continuum Slope}\label{sec:UV Continuum Slope}

The UV spectral slope, $\beta$, defined by $F_\lambda \propto \lambda^{\beta}$ \citep{Calzetti_1994}, serves as a proxy for the continuum colour and dust attenuation. To mitigate the impact of localised noise spikes and emission-line contamination, we sample the spectra through ten standard rest-frame windows spanning $1268$--$2580$\,\AA\ \citep{Calzetti_1994} avoiding strong rest-UV emission lines and the $2175$~\AA\ dust bump. The median flux is computed within each window to provide a robust estimate of the underlying continuum level. We perform a weighted least-squares fit in logarithmic space, fitting $\log_{10} F_\lambda$ as a function of $\log_{10} \lambda$ using \texttt{scipy.optimize.curve\_fit}. The fit is weighted by the inverse variance $1/\sigma^2$ of the median flux in each window. The best-fitting value of $\beta$ and its associated uncertainty are obtained from the covariance matrix of the fit. This weighting scheme ensures that higher SNR measurements contribute more strongly to the fit, yielding a statistically robust estimate of the continuum slope. 

\subsubsection{Absolute UV Magnitude}\label{sec:Absolute UV Magnitude}

The absolute UV magnitude, $M_{\rm UV}$, is calculated by integrating the rest-frame spectrum through a top-hat filter centred at $1500$~\AA\ with $100$~\AA\ width. The resulting flux density is converted to an AB magnitude and corrected for distance using the adopted cosmology and for bandwidth compression using the appropriate $K$-correction \citep{2002_K}. Uncertainties are propagated from the DJA spectral errors through the integration, capturing the contribution from spectral noise.

\subsection{Gas-Phase Metallicity from Strong-Line Ratios}\label{sec:Gas-Phase Metallicity Method}

Gas-phase metallicities are inferred for the 84 He~II-emitting galaxies using the Bayesian \texttt{gasp} pipeline (Westcott et al. in prep.), which simultaneously fits multiple strong-line diagnostics to the \texttt{specFitMSA} emission-line fluxes. Only emission lines with SNR $>3$ are treated as detections. We adopt the following diagnostics:
\begin{align}
R2 &= \log \left[ \frac{[\mathrm{O~II}]~3727,3729}{\mathrm{H}\beta} \right], \\
R3 &= \log \left[ \frac{[\mathrm{O~III}]~5007}{\mathrm{H}\beta} \right], \\
O32 &= \log \left[ \frac{[\mathrm{O~III}]~4959,5007}{[\mathrm{O~II}]~3727,3729} \right], \\
R23 &= \log \left[ \frac{[\mathrm{O~III}]~4959,5007 + [\mathrm{O~II}]~3727,3729}{\mathrm{H}\beta} \right], \\
\hat{R} &= 0.47\,R2 + 0.88\,R3,
\end{align}
which are calibrated against direct ($T_e$-based) abundance measurements \citep{Scholte2025_metallicities, Laseter_2024}. Simultaneously fitting all available diagnostics mitigates the degeneracies associated with individual strong-line indicators, including the double-valued nature of $R23$ and the ionisation dependence of single line ratios. While individual diagnostics such as $R2$ and $R3$ remain sensitive to the ionisation parameter, composite indicators including $R23$ and the $\hat{R}$ estimator \citep{Laseter_2024} reduce this dependence by combining multiple emission-line ratios, yielding more robust metallicity estimates, particularly at high redshift. As the diagnostic emission lines are closely spaced in wavelength, differential dust attenuation is negligible.

The inferred posterior distributions incorporate both measurement and calibration uncertainties. Gas-phase oxygen abundances are reported as the posterior median, with uncertainties given by the 16th--84th percentile credible interval. The fitting is subject to a lower metallicity bound of $12+\log(\mathrm{O/H})_\odot = 7.0$. Metallicities are normalised to the solar oxygen abundance of $12+\log(\mathrm{O/H})_\odot = 8.69$ \citep{Asplund_2009} according to
\begin{equation}
[O/H]_{\mathrm{rel}} = \left(12 + \log(\mathrm{O/H})\right) - 8.69,
\end{equation}
and linear metallicity ratios computed as $Z / Z_{\odot} = 10^{[\mathrm{O/H}]_{\mathrm{rel}}}$, where uncertainties are propagated directly from the posterior distributions.

\subsection{Pop III Candidate Selection}

\subsubsection{Pop III Diagnostic Framework}\label{sec:Pop III Diagnostic Framework}

To identify Pop~III candidates at different evolutionary stages, we adopt the diagnostic framework of \citet{Rusta_2025}. This method exploits a brief phase in the chemical evolution of a Pop~III system in which the surrounding gas becomes enriched before the stellar population transitions to metal-enriched Pop~II star formation. A newly formed Pop~III system is initially pristine, with both its stars and gas effectively metal-free. This phase is expected to last $\sim3$--$12$~Myr, depending on the stochastic sampling of the Pop~III IMF. Once the first, most massive Pop~III stars end their lives as supernovae, they pollute the surrounding interstellar medium (ISM) with metals. However, the stellar population can remain entirely metal-free during this period because star formation has not yet transitioned to Pop~II. \citet{Rusta_2025} refer to this as the "self-polluted" phase, which lasts $\sim1$~Myr in their fiducial model, until the gas metallicity exceeds the critical threshold for the onset of Pop~II star formation~($Z_{\rm gas}~>~10^{-4.5}\,Z_{\odot}$).

Once Pop~II star formation begins, the system enters a "hybrid" phase in which Pop~III and Pop~II stars coexist. This phase is subdivided according to the fraction of the total formed stellar mass remaining in Pop~III stars, $M_{\ast, \mathrm{PopIII}}/M_{\ast, \mathrm{tot}}$. Systems are classified as Pop~III-rich ($>50\%$), Pop~III-mid ($25$--$50\%$), or Pop~III-poor ($<25\%$). Since Pop~III star formation is largely complete by the onset of the hybrid phase, $M_{\ast,\mathrm{PopIII}}$ remains approximately fixed while $M_{\ast,\mathrm{tot}}$ increases as Pop~II star formation proceeds. In the models of \citet{Rusta_2025}, the Pop~III-rich and Pop~III-mid stages together persist for approximately $2$--$5$~Myr before the Pop~III mass fraction falls below $25\%$.

During the self-polluted phase and into the Pop~III-rich and Pop~III-mid stages, the ionising spectrum required to produce strong He~II emission is still dominated by the remaining metal-free Pop~III stars, while UV metal lines such as C~III]~$\lambda1908$, C~IV~$\lambda1551$, Si~III]~$\lambda1883$, and O~III]~$\lambda1663$ arise from the metal-enriched gas produced by earlier Pop~III supernovae. \citet{Rusta_2025} show that this combination of strong He~II emission and detectable metal lines can persist for up to $\sim20$~Myr, provided a substantial Pop~III stellar mass fraction remains.

This decoupling between the metal-free stellar population and the already enriched surrounding gas provides the physical basis for the diagnostic framework. Pristine Pop~III systems are expected to show strong He~II emission but no metal lines, while Pop~II star-forming galaxies have softer ionising spectra and consequently weaker He~II emission. In contrast, self-polluted and hybrid Pop~III systems can exhibit both strong He~II and metal-line emission. The relative strengths of these lines therefore provide an evolutionary diagnostic of the remaining Pop~III stellar mass fraction and allow candidate systems to be distinguished from both pristine Pop~III and chemically enriched Pop~II populations.

The C~III]~$\lambda1908$/He~II~$\lambda1640$ ratio is expected to be negligible in pristine Pop~III systems and low in the self-polluted phase, typically $<0.3$, when He~II emission remains dominated by the hard ionising spectrum of the surviving Pop~III stars while the surrounding gas has only recently been enriched by supernovae. The ratio increases through the hybrid phase as continued chemical enrichment strengthens the metal-line emission, while the declining Pop~III mass fraction reduces the contribution of the hard ionising spectrum and increases the relative contribution of the softer Pop~II population. Thus, C~III]/He~II provides a diagnostic of the evolutionary state and remaining Pop~III contribution.

The ratio C~IV~$\lambda1551$/He~II~$\lambda1640$, and correspondingly C~IV~$\lambda1551$/C~III]~$\lambda1908$, is expected to be high in Pop~III-dominated systems with high ionisation parameters ($\log U > -2$), because C~IV is a higher-ionisation species than C~III] and is preferentially excited by the harder radiation field of massive, metal-free stars; its strength therefore tracks both the ionisation parameter and the presence of a genuinely hard (Pop~III-like) spectrum, helping break the degeneracy between low-$U$ Pop~III models and Pop~III-poor hybrids that would otherwise overlap in C~III]/He~II alone. For the remaining diagnostics, Si~III]~$\lambda1883$/He~II~$\lambda1640$ and O~III]~$\lambda1663$/He~II~$\lambda1640$, \citet{Rusta_2025} derive empirical bounding relations that, used in combination, isolate a region of line-ratio space populated only by models with $>25\%$ of stellar mass in Pop~III stars, with no contamination from AGN or normal Pop~II/I star-forming galaxies. In this way, the method turns the fact that Pop~III galaxies pollute their own surroundings into an additional, longer-lived diagnostic channel, extending the detectable window for Pop~III signatures from the $\sim$few-Myr pristine phase to up to $\approx20$~Myr into the hybrid phase.

\subsubsection{Emission-line Ratio Measurements}

Emission-line fluxes are measured using \texttt{specFitMSA}, as described in Section~\ref{sec:Emission Line Measurements}. Flux ratios are computed from the highest signal-to-noise measurements and compared with the \citet{Rusta_2025} diagnostic grids. Uncertainties are propagated using the asymmetric $1\sigma$ upper and lower flux uncertainties (\texttt{sigup} and \texttt{siglo}) returned by \texttt{specFitMSA}, yielding asymmetric uncertainties in logarithmic line-ratio space. Emission lines with $\mathrm{SNR}\geq2$ are treated as detections, while lines with $\mathrm{SNR}<2$ are assigned $3\sigma$ upper limits calculated as $3\times\texttt{sigup}$, which are propagated into the corresponding diagnostic ratios.

Diagnostic density plots are constructed following the procedure of \citet{Rusta_2025}, using the logarithmic emission-line ratios defined in Section~\ref{sec:Pop III Diagnostic Framework}. The measured ratios are compared directly with the Pop~III model grids and bounding relations of \citet{Rusta_2025}, which define regions of line-ratio space corresponding to different Pop~III evolutionary stages and, in particular, identify systems with $>25\%$ of their stellar mass in Pop~III stars. We additionally overlay the AGN and star-forming galaxy model grids of \citet{Feltre_2016} and \citet{Gutkin_2016}, respectively, to assess whether the observed line ratios are also consistent with alternative ionising sources. This provides a direct comparison between the observed emission-line properties and the predicted Pop~III, AGN, and metal-enriched star-forming populations. The analysis is repeated using He~II~$\lambda4687$ in place of He~II~$\lambda1640$, which traces the same ionisation potential and provides an independent consistency check on the Pop~III candidate selection.

\subsubsection{Nebular Dust Attenuation}
\label{sec:nebular_dust_attenuation}

For galaxies with robust detections of both H$\alpha$~$\lambda6565$ and H$\beta$~$\lambda4863$ ($\mathrm{SNR}>2$; Section~\ref{sec:Emission Line Measurements}), we estimate the nebular dust attenuation from the observed Balmer decrement. Assuming Case~B recombination with $T_e=10^4$~K and $n_e=10^2\,\mathrm{cm}^{-3}$, we adopt an intrinsic ratio $(\mathrm{H}\alpha/\mathrm{H}\beta)_{\rm int}=2.86$ \citep{Osterbrock_1989,Osterbrock_2006, Groves_2012}. We adopt the \citet{Calzetti_2000} attenuation curve, with $k(\mathrm{H}\alpha)=3.33$, $k(\mathrm{H}\beta)=4.60$, and $R_V=4.05$. The nebular colour excess is

\begin{equation}
E(B-V)_{\rm gas} = \frac{2.5}{k(\mathrm{H}\beta)-k(\mathrm{H}\alpha)}
\log_{10}
\left[
\frac{(\mathrm{H}\alpha/\mathrm{H}\beta)_{\rm obs}}{2.86}
\right],
\label{eq:ebv_gas}
\end{equation}

\noindent with the corresponding $V$-band attenuation given by:
\begin{equation}
A_V = R_V E(B-V)_{\rm gas}.
\label{eq:av}
\end{equation}

\noindent Uncertainties are propagated from the asymmetric uncertainties on the measured Balmer decrement. Measurements with $(\mathrm{H}\alpha/\mathrm{H}\beta)~<~2.86$ yield formally negative $A_V$; these values are floored to $A_V=0$ and flagged with an asterisk, as such low decrements may indicate a departure from standard Case~B recombination rather than genuinely zero attenuation \citep{McClymont_2025}. The resulting nebular dust attenuations are presented in Section~\ref{sec:Metallicity Results}.

\subsubsection{Dust Vectors}
\label{sec:dust_vectors}

Because the \citet{Rusta_2025} diagnostic models predict intrinsic emission-line ratios and assume no dust, we show illustrative attenuation vectors in the diagnostic diagrams. We adopt a fixed $A_V=0.7$ to represent moderate attenuation; this is used only to illustrate the shift in line-ratio space and is not applied as a correction to the measured fluxes. For a line ratio $R=F_1/F_2$, the logarithmic shift is
\begin{equation}
\Delta \log_{10} R = -0.4\left[k(\lambda_1)-k(\lambda_2)\right]E(B-V)_{\rm gas},
\label{eq:dust_vector}
\end{equation}

where $E(B-V)_{\rm gas}$ is obtained from Equation~\eqref{eq:av} using~$A_V~=~0.7$~. The vectors therefore illustrate the expected effect of moderate attenuation on the diagnostic ratios.

\subsection{Resolved SED Fitting}\label{sec:Resolved SED Fitting}

The integrated properties considered in Sections~\ref{sec:Derived UV Properties}-\ref{sec:Gas-Phase Metallicity Method} may dilute signatures from spatially localised young, metal-poor stellar populations: a small, genuinely Pop~III-like region can be washed out when averaged with the light of an otherwise chemically evolved galaxy. Moreover, it is clear that individual Pop~III galaxies are, at least within the current archive of JWST spectra, either absent or extremely challenging to identify. We therefore take the next step and search instead for Pop~III-like regions within our candidate galaxies, rather than requiring the signature to dominate the integrated light of the system as a whole.

To this end, we perform resolved SED fitting to determine whether individual regions within the candidate galaxies exhibit properties consistent with Pop~III-like star formation. Specifically, we search for spatially coincident regions with low mass-weighted ages, negligible dust attenuation, steep UV slopes, and elevated star formation rate (SFR) surface densities, which together provide a strong signature of recent, intense star formation. We use the \texttt{EXPANSE} pipeline \citep{EXPANSE}, which extends \texttt{Bagpipes} \citep{Bagpipes_I, Bagpipes_II} SED fitting to 2D imaging and enables internal variations in stellar populations and star-formation properties to be characterised on sub-kpc scales.

\subsubsection{Imaging Preparation}\label{sec:Imaging Preparation}

\begin{figure}
    \centering
    \includegraphics[width=\linewidth]{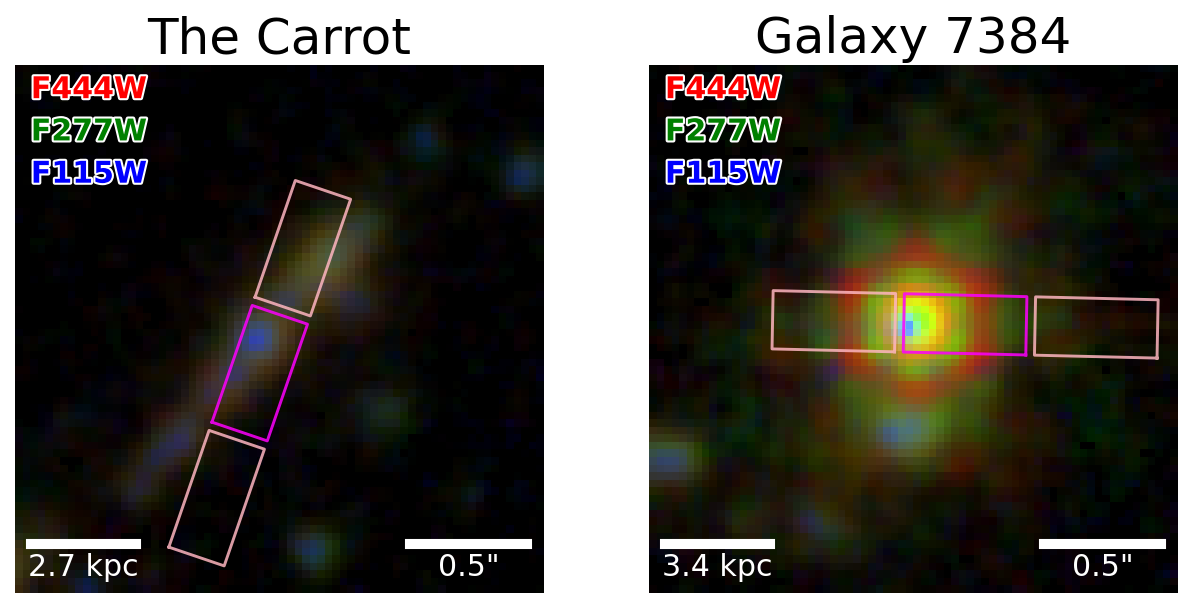}
    \caption{Example RGB composites (F115W, F277W, F444W) illustrating the quality of the images we use for morphological classification with the MSA slitlet configurations overlaid in pink and magenta. Left: an extended, clumpy galaxy, ID~13577 (\textit{The Carrot}), suitable for resolved SED fitting. Right: a compact, point-like source, ID~7384, with diffraction features, flagged as a potential AGN and unsuitable for resolved analysis. Scale bars are shown at the bottom of each cutout.}
    \label{fig:rgb}
\end{figure}

For each galaxy, \texttt{EXPANSE} \citep{EXPANSE} uses the galaxy ID, survey name, sky position, and redshift to automatically cross-match the spectroscopic catalogue with imaging from the \textit{Hubble Space Telescope} Advanced Camera for Surveys (ACS) \citep[\textit{HST/ACS};][]{Hubble_1998} and \textit{JWST}/NIRCam. For each galaxy, we extract $70 \times 70$ pixel cutouts centred on the target coordinates, incorporating imaging from 5 \textit{HST}/ACS bands (F435W, F606W, F775W, F814W, and F850LP) and 9 \textit{JWST}/NIRCam bands (F090W, F115W, F150W, F200W, F227W, F335W, F356W, F410M and F444W), spanning a total wavelength range of $\sim0.4$–$4.4~\mu$m. For each band, the data products include science images, associated error maps, \texttt{SExtractor} segmentation maps \citep{SExtractor_1, SExtractor_2}, and \texttt{EPOCHS-DR2} empirical PSFs \citep{EXPANSE}.

To enable consistent pixel-by-pixel analysis, all images are PSF-matched to a common resolution using precomputed convolution kernels. Each source is then initialised as a \texttt{ResolvedGalaxy} object, which organises the multi-band imaging, PSF information, and associated metadata for subsequent binning and SED fitting. To guide the analysis, we construct three-colour (RGB) composite images from the PSF-matched cutouts, adopting F115W, F277W, and F444W for the blue, green, and red channels, respectively.

We use \textit{JWST}/NIRCam RGB composites to assess the visual morphology of our galaxy sample, enabling identification of systems suitable for resolved SED fitting and flagging of compact or AGN-dominated sources. Sources are classified as either (i) point-like, with compact emission and possible diffraction spikes indicative of unresolved or AGN-dominated systems, or (ii) extended and/or clumpy, where spatial structure is clearly resolved. Only the latter are retained, as resolved structure is required for reliable spatial binning. This inspection also identifies nearby companions that may contaminate the spectroscopic slit and bias background subtraction. An example comparison between an extended galaxy and a point-like source is shown in Figure~\ref{fig:rgb}.

\subsubsection{Spatial Binning}

Spatial bins are constructed using the adaptive binning algorithm implemented in \texttt{piXedfit} \citep{Abdurro_pixedfit_2021, Abdurro_2023}. A segmentation mask generated from the F277W image defines the galaxy footprint, while pixels with $\mathrm{SNR}>3$ are grouped into contiguous bins that maximise signal-to-noise while preserving the underlying morphology. Fluxes and associated uncertainties are then measured within each bin across all available photometric bands, with uncertainties propagated directly from the input error maps.

The resulting binning maps are visually inspected to ensure that the full spatial extent of each galaxy is captured. Where necessary, the segmentation is interactively refined using the \texttt{EXPANSE} graphical interface. Rapid SED fitting with \texttt{EAZY} \citep{EAZY} is used to estimate photometric redshifts for neighbouring structures, allowing physically associated clumps to be distinguished from unrelated foreground or background sources. Nearby regions with photometric redshifts consistent with the target galaxy are incorporated into the final segmentation, ensuring that the spatial bins are both statistically robust and representative of the intrinsic galaxy morphology. The final binned photometry serves as the input for the resolved SED fitting.

\subsubsection{Bayesian SED Fitting}\label{sec:Bayesian SED Fitting}

Resolved SED fitting is performed independently for each spatial bin using the SED fitting software \texttt{Bagpipes}, implemented within \texttt{EXPANSE}. We adopt BPASS v2.2 stellar population synthesis models \citep{Eldridge_2017, Stanway_2018, Byrne_2022}, which incorporate binary stellar evolution and its effects on mass loss and UV emission, and simultaneously fit all available HST and JWST photometric bands. While BPASS does not include a dedicated Pop~III component, we adopt it as the available model that accounts for binary evolution and provides a more realistic description of the stellar populations and their spectra. Spectroscopic redshifts are fixed to the NIRSpec measurements, and the star formation history is described by the six-bin non-parametric continuity--bursty prior of \citet{Tacchella_2020}. The same fitting configuration is applied across the sample, with redshift-dependent parameter ranges adopted where required.

Posterior probability distributions are inferred for the physical properties of each spatial bin. Median posterior values are used to construct 2D maps of $\beta$, stellar mass surface density ($M_\star\,\mathrm{kpc}^{-2}$), star-formation-rate (SFR) surface density ($M_\star\,\mathrm{yr}^{-1}\mathrm{kpc}^{-2}$), $A_V$, and mass-weighted age. The resolved fits are summarised using the standard \texttt{EXPANSE} overview figures, which combine RGB imaging, segmentation and binning maps, resolved and integrated SED fits, posterior distributions, star formation histories, and the corresponding spatial distributions of the derived physical properties.

\subsubsection{Identification of Pop~III-like Regions}

The resolved property maps are used to identify spatially localised regions with properties consistent with theoretical expectations for Pop~III star formation. One key diagnostic is the mass-weighted stellar age, which traces the characteristic timescale over which the stellar mass was assembled. Pop~III-like regions are expected to have low mass-weighted ages ($\lesssim 10$--$20$ Myr), reflecting their predominantly young stellar populations and burst-dominated star-formation histories. This is expected to coincide with high SFR surface densities, indicative of intense and spatially concentrated star formation. Together, these properties can produce compact, high-density star-forming regions embedded within otherwise low-mass galaxies. 

These conditions may also produce very blue UV slopes ($\beta \lesssim -3$), as hot, metal-poor stars emit strongly in the UV and low dust attenuation minimises reddening. These properties are therefore expected to be correlated: low mass-weighted age and high SFR surface density indicate a recent, intense burst, while the low metallicity and limited dust content associated with an early evolutionary stage can produce a steep UV slope. As stellar mass builds up and successive generations of massive stars enrich the ISM, dust production may increase and lead to greater attenuation and redder UV slopes. However, this relationship is not necessarily monotonic, as feedback-driven outflows may expel or redistribute dust and metals on scales larger than the stellar component, allowing even relatively massive, chemically enriched galaxies to retain low attenuation and blue UV slopes \citep{Ferrara_2025}.

We therefore search for spatially coincident regions exhibiting low mass weighted age, negligible dust, steep UV slopes, and elevated SFR density as the clearest signatures of Pop~III-like star formation.  We carry this search out as it is clear that individual galaxies, at least in the current JWST archive of spectra, do not exist or are extremely challenging to find. Thus, we are taking the next step to search for Pop~III regions, as opposed to entire galaxies. 

\subsection{AGN Identification}

\subsubsection{Morphological AGN Screening}\label{sec:Morphological AGN Screening}

We assess potential AGN contamination through quantitative morphological measurements using \texttt{Morfometryka} \citep{Ferrari_2015, Westcott_2025}. For each source, $0.96''$ cutouts in the F444W filter are generated, and corresponding FITS files are produced for structural fitting. Morphological parameters are then obtained by fitting 2D Sérsic profiles, adopting a fixed PSF for F444W with $\mathrm{FWHM} \sim 4.55$ pixels \citep{conselice2024epochsidiscoverystar}. We extract the following model parameters: the half-light radius $R_n$, defined as the radius enclosing half of the total light from the best-fitting Sérsic model; the Sérsic index $n$; and the axis ratio $q$, given by the ratio of the minor to major axis and describing the source ellipticity. We therefore define compact sources as those satisfying $R_n < 4.55$ pixels, $n > 4$, and $q > 0.9$ \citep{Sersic_1963}. The requirement $R_n < 4.55$ pixels selects objects whose measured extent is smaller than the characteristic PSF scale and are therefore effectively unresolved, while $n > 4$ identifies highly centrally concentrated light profiles and $q > 0.9$ selects sources that are approximately circular in projection. Together, these criteria identify objects whose morphologies are consistent with compact, point-source-dominated emission.

As an independent measure of compactness, we additionally calculate aperture concentration following \citet{Greene_2024} and \citet{Kokorev_2024}. Fluxes are extracted within circular apertures of diameter $0.2''$ and $0.4''$. Source positions were re-centred prior to measurement using \texttt{centroid\_2dg} 
from \texttt{Photutils}\footnote{Available on github (\url{https://github.com/astropy/photutils}) and Zenodo (\doi{https://doi.org/10.5281/zenodo.19636730})} to ensure apertures were accurately placed on the flux centroid of each source. The concentration, 
$C = F_{0.4''}/F_{0.2''}$, uses fluxes computed with \texttt{Photutils}, via the \texttt{ApertureStats} class, which sums pixel values within each aperture. We classify sources with $C < 1.5$--$1.7$ as compact \citep{Kokorev_2024, Greene_2024}.

\subsubsection{SED-based AGN Identification}\label{sec:SED-based AGN Identification}

As a complementary test for AGN activity, we model the broadband SEDs of the sample using the \texttt{PROSPECT} \citep{Robotham_2020_Prospect} fitting framework, following the procedure described by \citet{Thorne2022_AGNfraction}. We use \texttt{PROSPECT} because a comparable AGN component is not yet fully integrated into \texttt{Bagpipes}, allowing us to assess AGN contributions independently of our primary \texttt{Bagpipes} analysis. We adopt the same stellar population modelling assumptions as described in Section~\ref{sec:Bayesian SED Fitting} for \texttt{Bagpipes} fitting, including BPASS v2.2 stellar population synthesis models \citep{Eldridge_2017} and the six-bin non-parametric continuity--bursty star formation history prior of \citet{Tacchella_2020}. This ensures consistency with the stellar population properties derived from our \texttt{Bagpipes} fits while allowing the \texttt{PROSPECT} fits to additionally test for an AGN contribution. The 84 He~II-emitting galaxies are cross-matched to the EPOCHS v2 photometric catalogue (Austin et al. in prep.) using their celestial coordinates with a matching radius of $0.3''$. We use the broadband fluxes from the matched catalogue as the input photometry for the SED fitting.

Two independent SED models are fitted to each galaxy. The first includes only stellar emission using the stellar population synthesis models of \citet{Bruzual_2003}. The second incorporates an AGN component using the dusty torus models of \citet{Fritz_2006}. The \citet{Fritz_2006} model simultaneously fits stellar emission and AGN emission arising from the accretion disc and its reprocessing by the surrounding dusty torus, allowing the relative contributions of the host galaxy and AGN to be constrained within a Bayesian framework. The intrinsic accretion-disc emission is modelled as a combination of power laws, following the functional forms given in \citet{Feltre_2012}. This power-law prescription is commonly adopted for modelling high-redshift AGN continua \citep[e.g.][]{Stalevski_2016_AGN,Maiolino_AGN_2024, Juod_balis_2023,Juod_balis_2026}.

Model selection between the stellar-only and AGN-inclusive fits is performed using the Deviance Information Criterion (DIC), for which a lower value indicates a preferred model. We define $\Delta\mathrm{DIC}~=~\mathrm{DIC}_{\mathrm{stellar}} - \mathrm{DIC}_{\mathrm{stellar+AGN}}$, such that positive values favour inclusion of an AGN component. We adopt $\Delta\mathrm{DIC} > 5$ as a conservative threshold for strong statistical evidence in favour of the AGN-inclusive model \citep{KassRaftery1995, DSilva_2025_DIC}.  We also compute the fractional contribution of the AGN component to the rest-frame 1500~\AA\ luminosity. Galaxies are identified as AGN candidates only if the AGN component contributes at least 10\% of the rest-frame UV luminosity and the AGN-inclusive model satisfies $\Delta\mathrm{DIC} > 5$. Requiring both criteria ensures that any inferred AGN component represents both a statistically significant improvement to the SED fit and a non-negligible contribution to the observed UV emission, rather than an unnecessary increase in model complexity.

\section{Results}\label{sec:Results}

\begin{figure*}
\centering
\includegraphics[width=\textwidth]{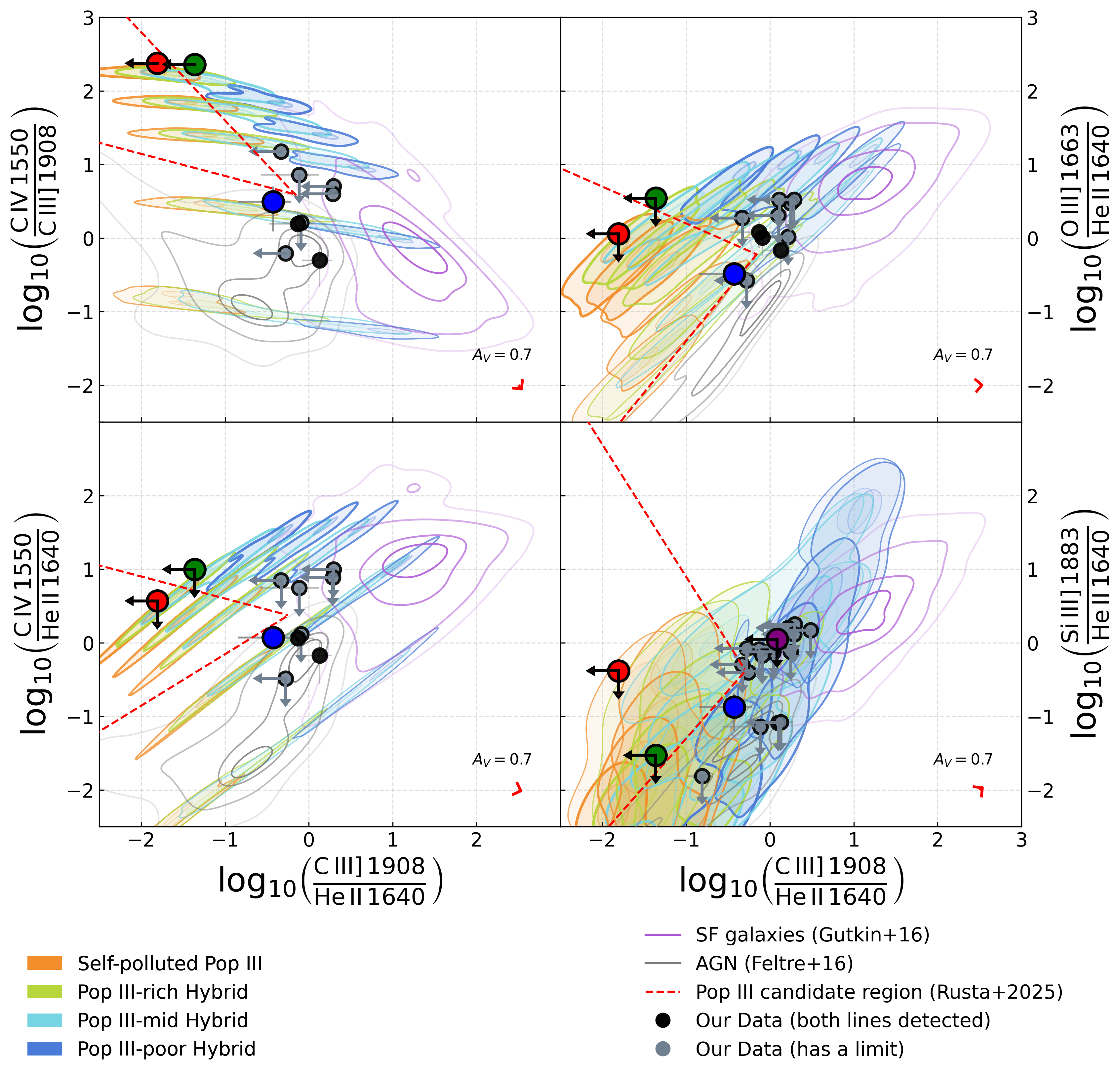}
\caption{UV emission-line diagnostic diagrams using He~II $\lambda1640$, comparing our sample with the Pop~III model predictions of \citet{Rusta_2025}. Coloured contours show the model density distributions, with increasing contour thickness corresponding to increasing ionisation parameter ($\log U = -3, -2, -1, -0.5, 0$). Empty contours show the AGN (grey) and star-forming galaxy (purple) model grids from \citet{Feltre_2016} and \citet{Gutkin_2016}, respectively. Red dashed lines delineate the region in which Pop~III stars are predicted to contribute more than 25\% of the total stellar mass. Black points show the observed line ratios, with grey error bars indicating $1\sigma$ uncertainties. For non-detected lines, ratios are instead shown in grey using the corresponding $3\sigma$ upper limits, with arrows indicating the direction in which the true ratio may lie. Objects identified as potential Pop~III candidates from their nominal line ratios are highlighted with larger red, green, blue and purple  markers; these markers retain the same limit treatment where one of the relevant UV metal lines is undetected. The red vectors in the lower-left corner of each panel show the effect of dust attenuation for $A_V=0.7$. Error bars are omitted where they are smaller than the symbol size.}
\label{fig:LimitContourPlot1}
\end{figure*}

\begin{figure*}
\centering
\includegraphics[width=\textwidth]{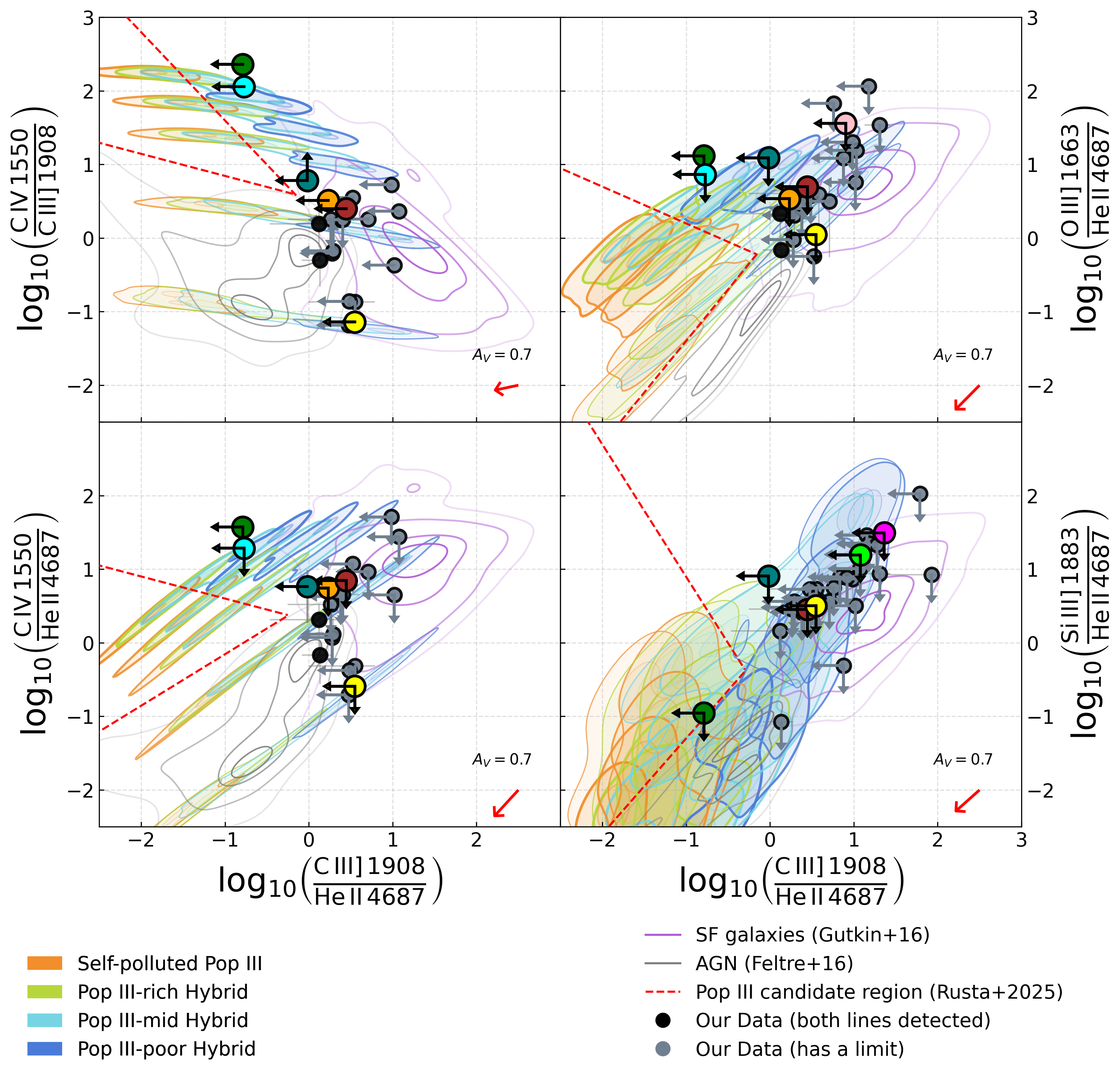}
\caption{UV emission-line diagnostic diagrams using He~II $\lambda4687$, comparing our sample with the Pop~III model predictions of \citet{Rusta_2025}. Coloured contours show the model density distributions, with increasing contour thickness corresponding to increasing ionisation parameter ($\log U = -3, -2, -1, -0.5, 0$). Empty contours show the AGN (grey) and star-forming galaxy (purple) model grids from \citet{Feltre_2016} and \citet{Gutkin_2016}, respectively. Red dashed lines delineate the region in which Pop~III stars are predicted to contribute more than 25\% of the total stellar mass. Black points show the observed line ratios, with grey error bars indicating $1\sigma$ uncertainties. For non-detected lines, ratios are instead shown in grey using the corresponding $3\sigma$ upper limits, with arrows indicating the direction in which the true ratio may lie. Objects identified as potential Pop~III candidates from their nominal line ratios are highlighted with larger coloured markers; these markers retain the same limit treatment where one of the relevant UV metal lines is undetected. The red vectors in the lower-left corner of each panel show the effect of dust attenuation for $A_V=0.7$. Error bars are omitted where they are smaller than the symbol size.}
\label{fig:LimitContourPlot2}
\end{figure*}

\subsection{Candidate Selection Workflow}

Our candidate selection proceeds through a sequence of diagnostics, with galaxies discarded as possible Pop~III hosts at each stage. We first identify He~II emitters from the parent sample and classify them using the Pop~III diagnostic framework, before deriving their global properties ($\beta$, $M_{\rm UV}$, and gas-phase metallicity) to explore the nature of our candidates. We then perform morphological screening to identify compact, AGN-like systems and assess suitability for resolved analysis, before finally applying spatially resolved SED fitting to our extended sources. This section details this workflow and presents our main results.

\subsection{Pop III Candidates}\label{sec:Pop III Candidates}

To assess the potential contribution of Pop~III stars using the UV emission-line diagnostics of \citet{Rusta_2025}, we construct diagnostic diagrams based on the He~II~$\lambda1640$ and He~II~$\lambda4687$ emission lines. Figures~\ref{fig:LimitContourPlot1} and \ref{fig:LimitContourPlot2} show these diagnostics, respectively, including the effects of upper limits on low-SNR emission-line measurements. Following the \citet{Rusta_2025} criterion, whereby galaxies with Pop~III stellar mass fractions exceeding 25\% are classified as candidates, we identify 12 galaxies for further study: \texttt{EPOCHS-DR2} ID~76015 (red), ID~2868 (green), ID~9564 (purple), ID~1130 (cyan), ID~212506 (teal), ID~28746 (brown), ID~61712 (yellow), ID~210262 (magenta), ID~126865 (lime), ID~9598 (pink), ID~13577 (\textit{The Carrot}; orange), named for its elongated morphology (Section~\ref{sec:The Carrot}), and ID~13176 (\textit{The Blueberry}; blue), named for its point-like appearance and very blue UV emission (Section~\ref{sec:The Blueberry}).

Several of these galaxies occupy regions associated with significant Pop~III contributions when considering the measured line ratios alone, as shown in the corresponding diagnostic diagrams in Appendix~\ref{sec:Pop III Candidate Selection and Emission-Line Diagnostics}. The rationale for retaining these sources for further study, despite the observational limitations affecting their line-ratio constraints, is also discussed there. However, the majority have at least one upper limit on a low-SNR UV metal line, which substantially affects the constraints on their locations in the diagnostic planes. In particular, incorporating these limits places many of the sources outside the regions corresponding to Pop~III stellar mass fractions exceeding 25\%. Thus, although the measured line ratios can place several galaxies within the Pop~III candidate regions, these classifications are generally not robust once the observational constraints from the spectroscopic data are taken into account (see Section~\ref{sec:caveats and systematic uncertainties}).

With the exception of ID~2868, the candidates have only one of He~II~$\lambda1640$ or He~II~$\lambda4687$ available for the corresponding diagnostics, further limiting the constraints on their positions. The stronger constraints provided by the $\lambda1640$ diagnostics are important because He~II~$\lambda4687$ is weaker than He~II~$\lambda1640$, while the larger wavelength separation between $\lambda4687$ and the UV metal lines makes the corresponding ratios more sensitive to attenuation. This is illustrated by the $A_V=0.7$ dust vectors: in Figure~\ref{fig:LimitContourPlot1}, the short UV wavelength baseline makes dust effects negligible, leaving candidate positions essentially unchanged. In Figure~\ref{fig:LimitContourPlot2}, the larger baseline between optical He~II~$\lambda4687$ and UV metal lines produces a substantially longer vector, indicating that dust attenuation can significantly shift candidate positions and should therefore be considered when interpreting their locations relative to the Pop~III regions.

ID~76015 and ID~2868 illustrate the effect of incorporating upper limits: ID~76015 becomes consistent with a candidate region in Figure~\ref{fig:LimitContourPlot1} when limits are included, while ID~2868 is consistent with candidate regions in both the Figure~\ref{fig:LimitContourPlot1} and Figure~\ref{fig:LimitContourPlot2} when the limits are taken into account.

Notably, \textit{The Blueberry} is the only candidate with robust measurements of all UV emission lines used in these diagnostics, providing the most complete set of line-ratio constraints without requiring limits. In Figure~\ref{fig:LimitContourPlot1}, it lies within the red-dashed candidate region in the upper-right panel and close to the candidate boundaries in the remaining panels, while consistently occupying regions of the model grids associated with Pop~III-rich hybrid populations. The corresponding emission-line fits are presented in Appendix~\ref{sec:Emission-Line Fits for The Blueberry}. It therefore represents our strongest candidate for a potential Pop~III contribution based on the UV diagnostics.

\subsection{Global Properties of the Candidate Sample}\label{sec:Global Properties of the Candidate Sample}

Of the 84 strong He~II-emitting galaxies in our parent sample, 75 have sufficient rest-frame UV spectral coverage to enable reliable measurements of the UV continuum slope, $\beta$. The remaining nine galaxies were selected based on the detection of He~II~$\lambda4687$ rather than the UV He~II lines and do not have sufficient spectral coverage below $\sim2000$~\AA\ to measure $\beta$. However, these galaxies are retained in our broader analysis, as they still have spectral coverage over the upper end of the UV continuum. We therefore restrict the UV continuum diagnostics presented here to the 75 galaxies with adequate wavelength coverage for reliable measurements of $\beta$. We examine the most notable sources individually using their spectroscopic and imaging properties. A visual overview of their \textit{JWST}/NIRSpec spectra, NIRCam imaging, and NIRSpec slitlet placement is provided in Appendix~\ref{sec:Spectroscopic and Imaging Properties of Selected Galaxies}, with coordinates, redshifts, and derived properties summarised in Table~\ref{tab:selected_galaxies} and corresponding emission-line fluxes and dust attenuation in Table~\ref{tab:line_fluxes_snrs}.

\subsubsection{UV Continuum Slopes}

\begin{figure*}
\centering
\begin{minipage}{0.48\linewidth}
    \centering
    \includegraphics[width=\linewidth]{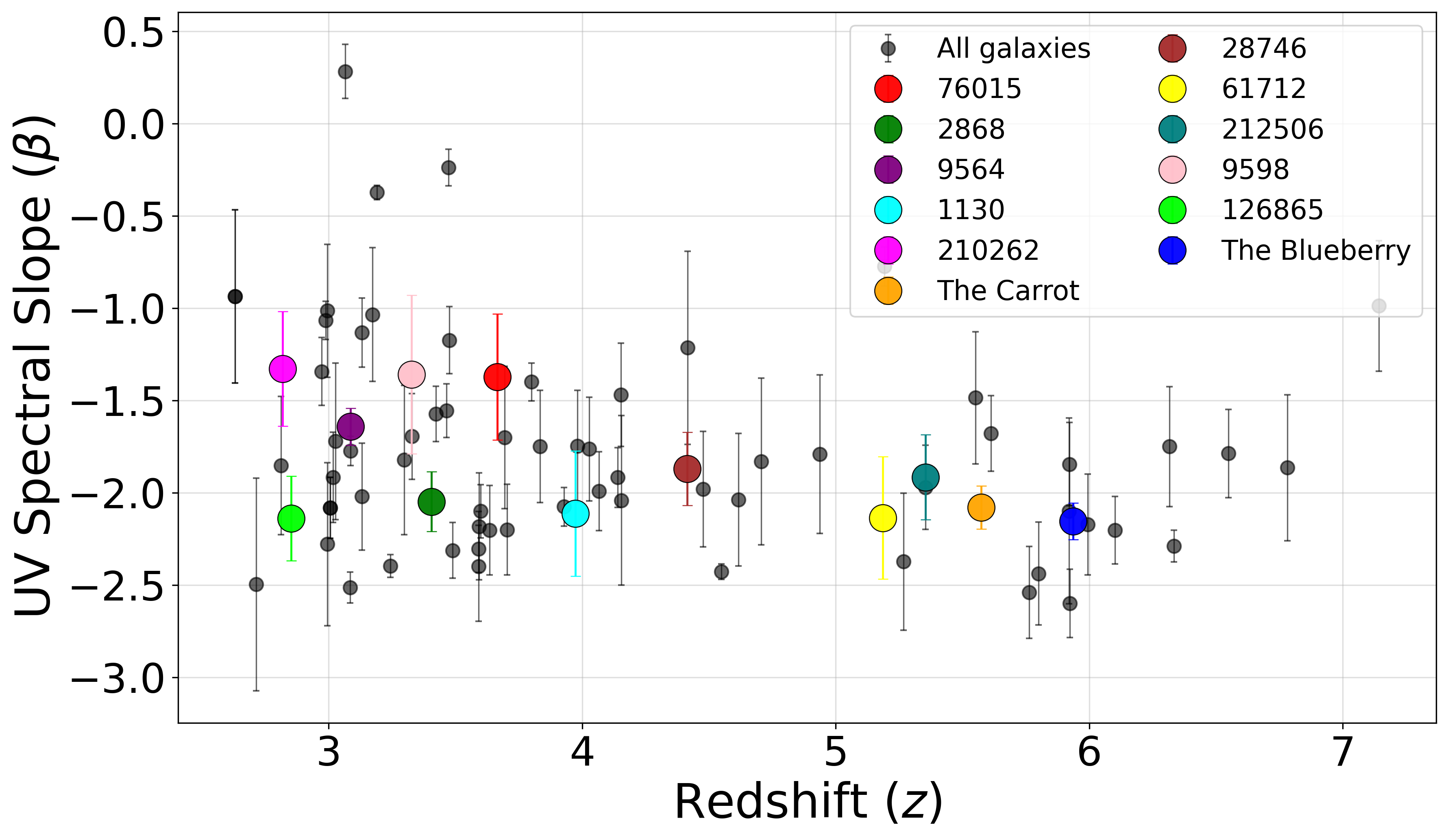}
\end{minipage}
\hfill
\begin{minipage}{0.48\linewidth}
    \centering
    \includegraphics[width=\linewidth]{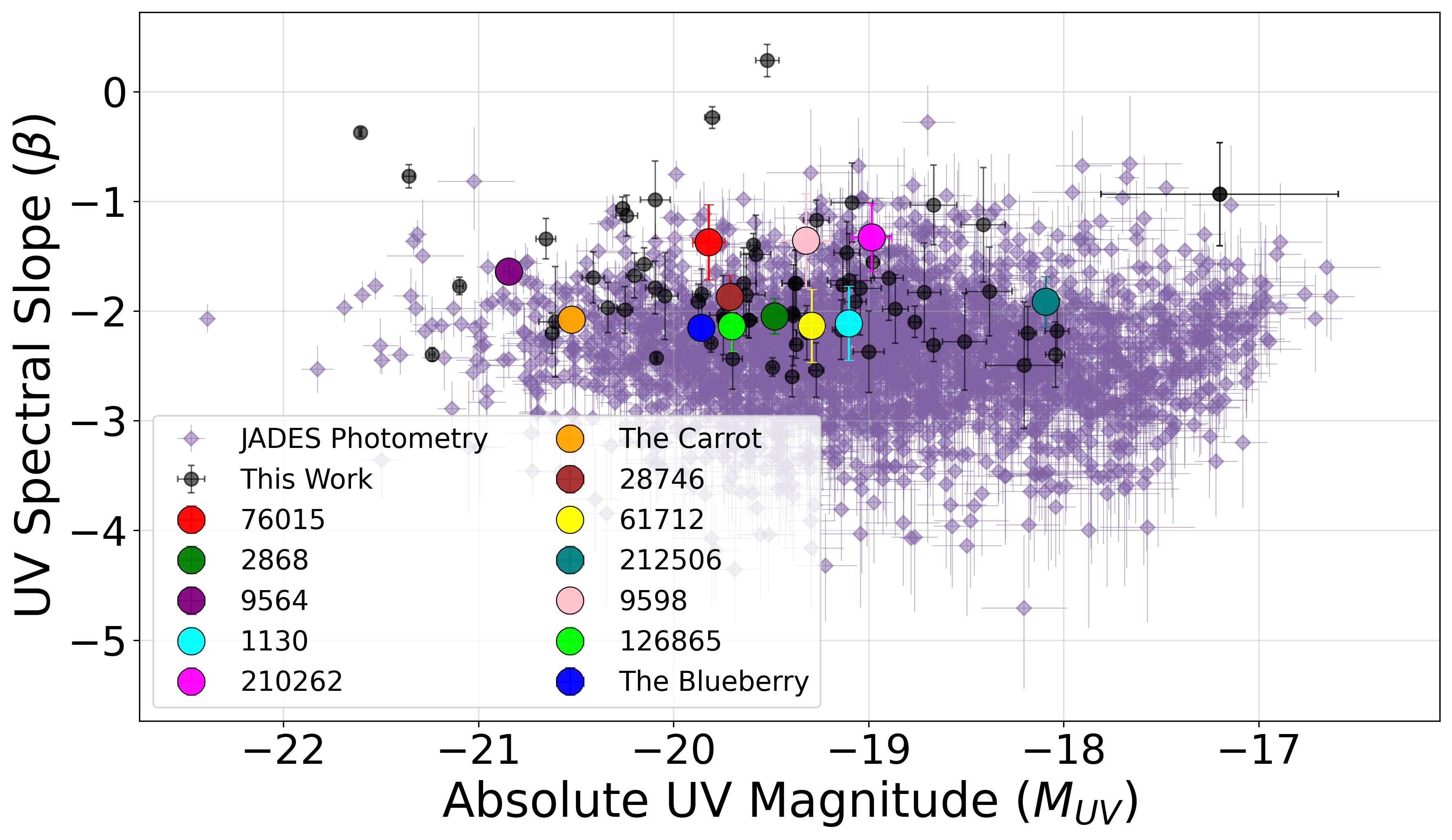}
\end{minipage}
\caption{UV continuum slope diagnostics for the 75 strong He~II-emitting galaxies with sufficient rest-frame UV spectral coverage.
\textit{Left:} UV continuum slope $\beta$ as a function of redshift ($z$). Points show the best-fitting power-law slopes measured from rest-frame UV spectra, with $1\sigma$ uncertainties from the weighted least-squares fits. Redshift uncertainties are omitted as they are not provided by the DJA. The 12 Pop~III candidates are highlighted with enlarged coloured markers and occupy a predominantly blue range of $\beta \approx -2.1$ to $-1.3$, although they do not reach the extremely blue slopes expected for pristine, metal-free stellar populations \citep{Austin2025a}.
\textit{Right:} $\beta$ as a function of absolute UV magnitude, $M_{\rm UV}$, for the same sample. The Pop~III candidates span a range of UV luminosities but preferentially exhibit blue UV continua. Photometrically derived estimates of $\beta$ and $M_{\rm UV}$ for JADES galaxies from \texttt{EPOCHS-DR2} (Austin et al. in prep.) are shown in lilac for comparison, demonstrating broad consistency with the spectroscopic measurements.}
\label{fig:beta_diagnostics}
\end{figure*}

Figure~\ref{fig:beta_diagnostics} (\textit{left}) shows the UV continuum slope, $\beta$, as a function of redshift for these 75 strong He~II-emitting galaxies, with the 12 Pop~III candidates highlighted. Across the sample, $\beta$ spans $-2.60$ to $0.28$ over the redshift range $2.63 \leq z \leq 7.14$, while the candidate galaxies occupy a narrower range of $-2.14 \leq \beta \leq -1.33$ between $2.82~\leq~z~\leq~5.94$. Although the candidates are systematically characterised by blue UV continua, none exhibit the extremely blue slopes ($\beta \lesssim -3$) predicted for galaxies dominated by pristine, metal-free stellar populations in the absence of nebular reprocessing. We note, however, that strong nebular continuum emission, expected for young, highly ionising populations, can itself redden the observed slope well above this intrinsic value regardless of the underlying stellar population \citep{Katz_2025}; we return to this point in Section~\ref{sec:Discussion}. No clear correlation is observed between $\beta$ and redshift for either the full sample or the Pop~III candidates, indicating that the UV continuum colour alone is not a reliable discriminator of candidate systems. Instead, the candidates are distributed across the full redshift range of the sample while clustering around $\beta \approx -2$, consistent with young, actively star-forming galaxies.

The bluest galaxy in the full sample is candidate ID~13620, with $\beta = -2.60 \pm 0.18$ at $z = 5.92$. Despite its exceptionally blue UV continuum, it is not the highest-redshift Pop~III candidate; this distinction belongs to \textit{The Blueberry} at $z = 5.94$, which has $\beta = -2.15 \pm 0.10$. \textit{The Carrot}, at $z = 5.57$, also exhibits a similarly blue UV slope of $\beta = -2.08 \pm 0.12$. Among the candidate population, ID~126865 is the bluest ($\beta = -2.14 \pm 0.23$) at $z = 2.85$, while IDs~9598, 76015, and 210262 form a redder subset with $\beta \approx -1.5$. These objects demonstrate that candidate Pop~III systems are not exclusively associated with the bluest UV continua, suggesting that UV colour alone is insufficient to distinguish galaxies hosting significant Pop~III stellar populations.

\subsubsection{UV Luminosities}

To place our spectroscopic measurements in context, we overlay photometrically derived estimates of $\beta$ and $M_{\rm UV}$ from \texttt{EPOCHS-DR2} (Austin et al. in prep.). Figure~\ref{fig:beta_diagnostics} (\textit{right}) shows the relationship between the UV continuum slope and absolute UV magnitude for the 75 strong He~II-emitting galaxies with sufficient rest-frame UV spectral coverage below $\sim2000$~\AA\, with the Pop~III candidates highlighted. The photometric measurements are plotted alongside our spectroscopically derived values, enabling a direct comparison between the two approaches. Overall, the two methods show good agreement, although the spectroscopically derived $\beta$ values are systematically slightly redder than the photometric estimates. The extremely blue photometric estimates, extending to $\beta < -4$, are driven by photometric uncertainties and scatter in the measured broadband fluxes rather than representing genuinely ultra-blue stellar populations. This highlights the advantage of spectroscopic measurements, which provide a more direct constraint on the UV continuum shape and are less susceptible to extreme deviations caused by photometric noise.

The full sample spans a wide luminosity range,~$-21.6~\lesssim~M_{\rm UV}~\lesssim~-17.2$, while the candidate population occupies~$-20.8~\lesssim~M_{\rm UV}~\lesssim~-18.1$. The broad luminosity distribution of the candidates demonstrates that potential Pop~III-hosting galaxies are not confined to either the brightest or faintest systems in the sample.
No significant correlation is evident between $\beta$ and $M_{\rm UV}$ for either the full sample or the candidate population. Instead, galaxies with similarly blue UV continua are found across a wide range of luminosities, suggesting that UV colour is largely independent of intrinsic UV brightness within this He~II-selected sample.

\textit{The Carrot} is among the most UV-luminous candidates, with $M_{\rm UV} = -20.52 \pm 0.02$, while the bluest galaxy in the full sample, ID~13620 ($\beta = -2.60 \pm 0.18$), is also relatively luminous, with $M_{\rm UV} = -19.39 \pm 0.03$. \textit{The Blueberry} is similarly luminous, with $M_{\rm UV} = -19.85 \pm 0.01$, placing it between these two sources in UV luminosity. At the opposite extreme, the faintest galaxy in the sample, ID~88540 ($M_{\rm UV} \approx -17.20$, $z = 2.63$), has a measured UV slope of $\beta = -0.94 \pm 0.47$.

\subsubsection{Metallicity and Nebular Dust Attenuation}\label{sec:Metallicity Results}
 
\begin{figure}
    \centering
    \includegraphics[width=\linewidth]{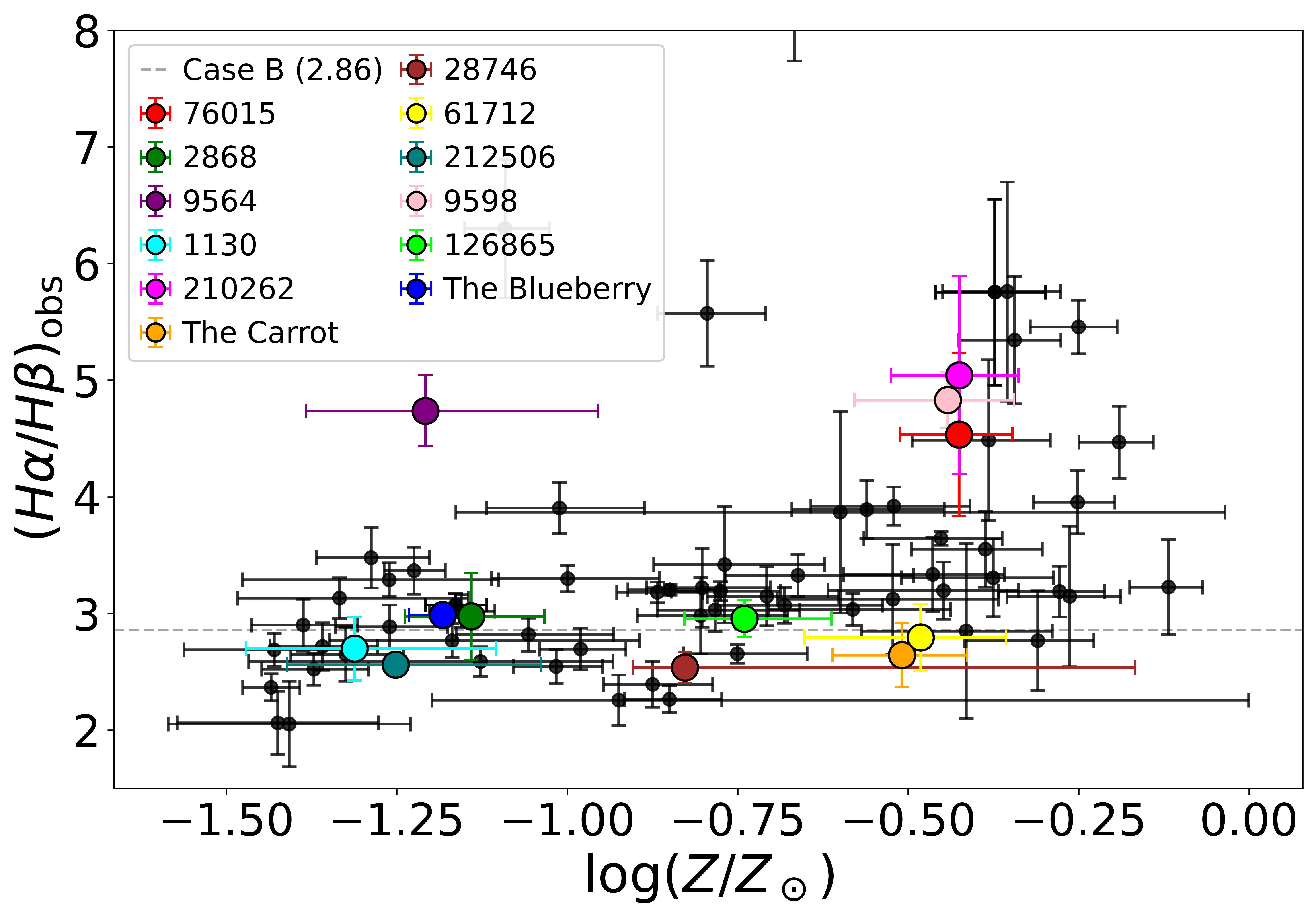}
    \caption{Observed Balmer decrement, $(\mathrm{H}\alpha/\mathrm{H}\beta)_{\rm obs}$, as a function of gas-phase metallicity $\log(Z/Z_{\odot})$ for the 75 strong He~II-emitting galaxies with robust metallicity measurements, spanning $\log(Z/Z_{\odot}) \approx -1.4$ to $-0.1$. Error bars denote propagated uncertainties (omitted where smaller than the symbol size). The horizontal dashed line marks the Case~B recombination value $(\mathrm{H}\alpha/\mathrm{H}\beta)_{\rm int}=2.86$, corresponding to zero nebular attenuation ($A_V=0$); points below this line yield formally negative $A_V$. The 12 Pop~III candidates are highlighted with enlarged, coloured markers. The sample median, $(\mathrm{H}\alpha/\mathrm{H}\beta)=3.13$ ($A_V=0.31$~mag), indicates generally modest nebular attenuation across the population, with no clear trend between the Balmer decrement and metallicity.}
    \label{fig:balmer_dec_metallicity}
\end{figure}

Figure~\ref{fig:balmer_dec_metallicity} shows the Balmer decrement $(\mathrm{H}\alpha/\mathrm{H}\beta)_{\rm obs}$ as a function of $\log(Z/Z{\odot})$ for 75 strong He~II-emitting galaxies with robust metallicity estimates, spanning $\log(Z/Z_{\odot}) \approx -1.4$ to $-0.1$, with Pop~III candidates highlighted. Seven galaxies lack the emission-line detections required for reliable strong-line metallicity estimates. However, all but one show detectable emission [O~III] $\lambda4959$ and $\lambda5007$ (see Appendix~\ref{sec:Oxygen Spectra}), indicating prior chemical enrichment and ruling out genuinely primordial metal-free systems. However, their available line measurements are insufficient to quantitatively constrain the metallicity. The exception, ID~171147, lacks detections of both [O~III] and [O~II].

Two sources (IDs~721 and 208134) are excluded from the metallicity comparison because their inferred metallicities reach the lower boundary of the \texttt{gasp} model grid. In these cases, the posterior distributions are truncated by the imposed metallicity limits, preventing reliable constraints below the calibrated range. For ID~208134 ($z \approx 1.85$), the weak [O~III] $\lambda5007$/H$\beta$ ratio is suggestive of low metallicity; however, the presence of strong H$\alpha$ and [N~II] emission provides an alternative interpretation in which the galaxy is instead relatively metal-enriched, with reduced [O~III] emission resulting from efficient gas cooling. Notably, ID-721 has been identified as a candidate AGN by \citet{Maiolino_AGN_2025}, based on its detection as an X-ray source with a flat, Compton-reflection-like spectrum consistent with a heavily obscured (Compton-thick) nucleus.

The most metal-poor galaxy in the full sample is ID~63685 at $z=3.59$, with $\log(Z/Z_{\odot})=-1.43\pm0.04$. The Pop~III candidates span $\log(Z/Z_{\odot})\approx-1.3$ to $-0.4$, with IDs~1130, 9564, 212506, and 2868 forming the metal-poor tail at $\log(Z/Z_{\odot})\lesssim-1.1$. ID~1130 has the lowest inferred metallicity, $\log(Z/Z_{\odot})=-1.31^{+0.21}_{-0.16}$, while \textit{The Carrot} is among the most enriched at $\log(Z/Z{\odot})=-0.51\pm0.10$. \textit{The Blueberry} has $\log(Z/Z_{\odot})=-1.18\pm0.05$, placing it towards the metal-poor end of the candidate distribution. Overall, most candidates have moderate metallicities of $-0.8\lesssim\log(Z/Z_{\odot})\lesssim-0.4$, demonstrating that their strong ionising signatures do not arise from galaxies that are globally chemically pristine. 

Figure~\ref{fig:balmer_dec_metallicity} shows no clear overall correlation between gas-phase metallicity and the Balmer decrement, with galaxies spanning the full metallicity range at broadly similar levels of nebular attenuation. Nevertheless, the largest Balmer decrements are preferentially found among the more metal-rich galaxies. The median Balmer decrement is $(\mathrm{H}\alpha/\mathrm{H}\beta)=3.13$, corresponding to $A_V=0.31$~mag, indicating generally modest attenuation across the He~II-emitting population. Eight of the twelve Pop~III candidates are consistent with negligible nebular attenuation within the uncertainties (Table~\ref{tab:line_fluxes_snrs}). Five of these (IDs~1130, 212506, 28746, 61712, and \textit{The Carrot}) have Balmer decrements at or below the intrinsic Case~B ratio of 2.86, yielding formally negative $A_V$ values that we floor to $A_V=0$ (marked with an asterisk in Table~\ref{tab:line_fluxes_snrs}); the remaining three (\textit{The Blueberry}, IDs~2868 and 126865) have small positive $A_V$ values that are nonetheless statistically consistent with zero attenuation. For example, \textit{The Blueberry} has $A_V=0.15\pm0.08$, indicating essentially unobscured star formation and consistent with the low attenuation independently recovered from its resolved SED fitting (Section~\ref{sec:The Blueberry}).

The remaining candidates, IDs~76015, 9598, and 210262, show greater diversity. They combine relatively high metallicities, $\log(Z/Z_{\odot})\approx-0.4$, with elevated attenuations of $A_V\approx1.6$--$2.0$~mag. This is qualitatively consistent with the expectation that dust becomes more prominent in chemically enriched systems, although the lack of a population-wide correlation indicates that metallicity alone does not determine the nebular attenuation. This is particularly evident for ID~9564, which has one of the highest attenuations in the candidate sample ($A_V=1.75\pm0.22$) despite its relatively low metallicity, $\log(Z/Z_{\odot})=-1.21^{+0.25}_{-0.18}$. Its high attenuation may instead reflect localised or clumpy dust that is not captured by the integrated gas-phase metallicity. Overall, the majority of the candidate sample shows Balmer decrements consistent with little-to-no nebular attenuation.

\subsection{Morphological AGN Candidates}\label{sec:Morphological AGN Candidates}

Visual inspection of the \textit{JWST}/NIRCam RGB composites identifies four galaxies within the 84 strong He~II-emitting sample that exhibit prominent diffraction spikes consistent with point-like morphologies. We therefore flag these sources as potential AGN candidates, while noting that diffraction spikes alone do not provide definitive evidence for an AGN contribution. These sources are IDs~7384, 171147, 197911, and 158273. Notably, ID~171147 is the only galaxy in the sample lacking reliable [O~III] and [O~II] detections, preventing robust strong-line metallicity constraints.

Quantitative morphological measurements from \texttt{Morfometryka} provide additional constraints on source compactness. ID~7384 is identified as a particularly compact system, exhibiting a small effective radius ($R_n \sim 1.7$--$2.2$ pixels), high Sérsic index ($n \gtrsim 10$), and near-unity axis ratio ($q \gtrsim 0.93$). These properties are consistent with a centrally concentrated, PSF-dominated morphology, supporting the interpretation of an unresolved nuclear component.

An independent aperture concentration analysis identifies two additional compact sources, IDs~1130 and 3608, in addition to ID~7384. Of these, ID~1130 is of particular interest as it is also classified as a Pop~III candidate and exhibits a relatively low metallicity ($\log(Z/Z_{\odot})~=~-1.31^{+0.21}_{-0.16}$). However, its concentration index ($C\sim1.5$) places it close to the adopted compactness threshold. Together with visual inspection, these measurements indicate that IDs~1130, 3608, and 7384 possess compact, circular morphologies consistent with unresolved emission, warranting caution when interpreting their high-ionisation line ratios as purely stellar in origin. 

We note that ID~3608 has independently been proposed
as a candidate broad-line AGN by \citet{Maiolino_AGN_2024}, based on a tentative broad component of H$\alpha$. Consistent with this AGN interpretation, \citet{Maiolino_AGN_2025} report a marginal ($\sim1.5$--$2\sigma$) X-ray signal for this source, one of only two objects in their sample with above-zero net counts.

\subsection{SED-based AGN Candidates}\label{sec:SED-based AGN Candidates}

Applying the SED-fitting procedure described in Section~\ref{sec:SED-based AGN Identification} to our 84 galaxies with complete photometric coverage identifies two SED-selected AGN candidates: IDs~212506 and 7384. Both galaxies satisfy our joint selection criteria, with an AGN component contributing more than 10\% of the rest-frame 1500~\AA\ luminosity and an AGN-inclusive SED model that is statistically preferred over the stellar-only fit according to the DIC ($\Delta\mathrm{DIC} > 5$; see Section~\ref{sec:SED-based AGN Identification}). The remaining 82 galaxies are adequately described by purely stellar SED models, indicating that any AGN contribution to their broadband photometry is either negligible or cannot be robustly distinguished from stellar emission given the available data. Together with the morphological analysis presented in Section~\ref{sec:Morphological AGN Candidates}, these results suggest that significant AGN contamination is rare within the He~II-selected sample as a whole.

\begin{figure}
    \centering
    \includegraphics[width=\linewidth]{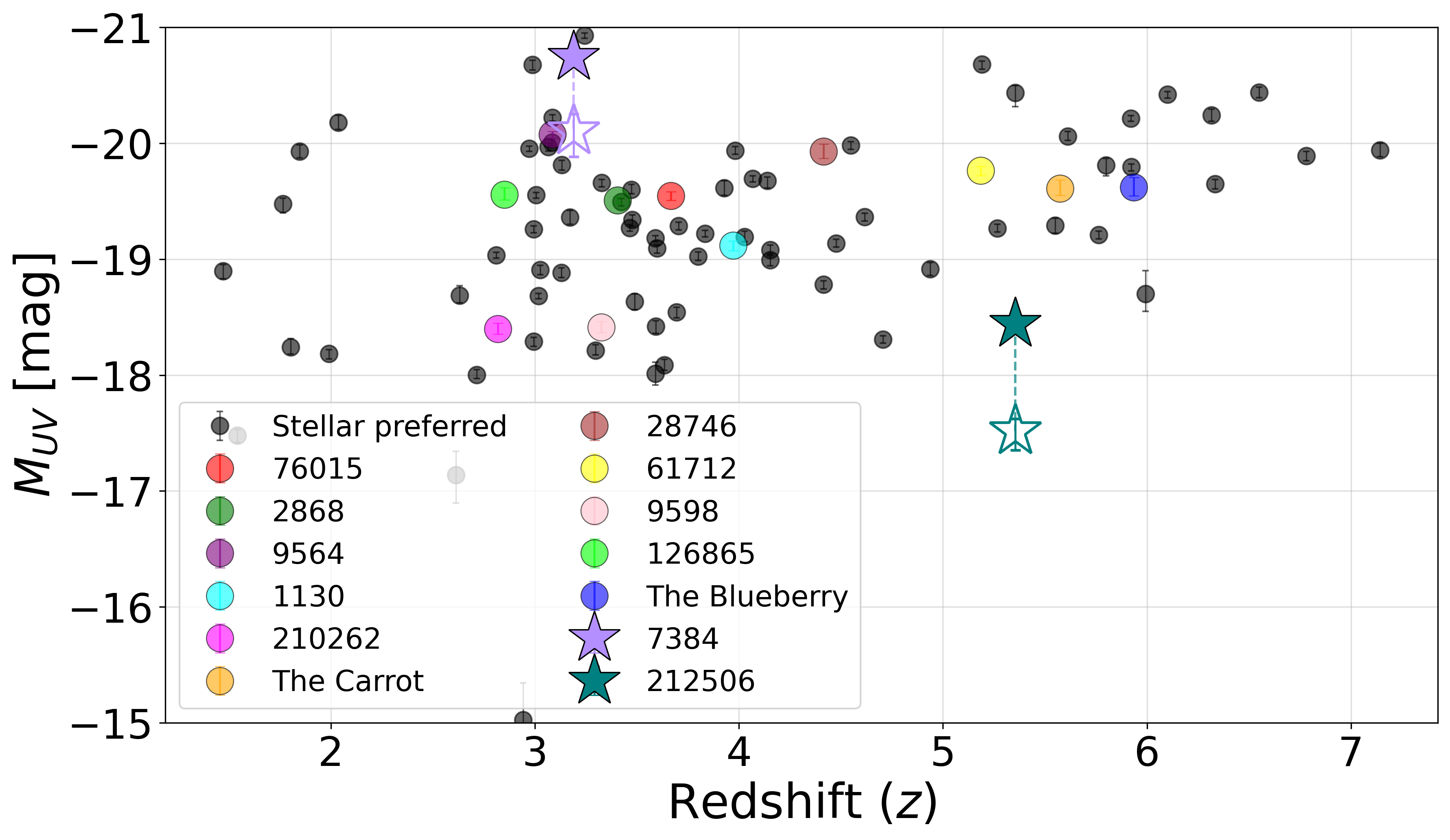}
    \caption{Rest-frame UV absolute magnitude ($M_{\mathrm{UV}}$) as a function of redshift for the 84 galaxies analysed with \texttt{PROSPECT}. UV magnitudes are computed using a top-hat filter centred at 1500~\AA\ with a width of 100~\AA. Grey circles show galaxies whose SEDs are best described by purely stellar emission; enlarged coloured circles denote the Pop~III candidates identified via the \citet{Rusta_2025} diagnostics in this work. The filled lilac star marks ID~7384, independently flagged as an AGN candidate by all morphological diagnostics in Section~\ref{sec:Morphological AGN Candidates} and its SED is statistically preferred to include an AGN component. The filled teal star marks ID~212506, one of our Pop~III candidates, whose SED is also statistically preferred to include an AGN component. For both sources, the filled star is connected by a dashed line to an unfilled star of the same colour, showing the rest-frame UV magnitude of the fitted AGN component alone, as distinct from the total (stellar~+~AGN) magnitude shown by the filled star. Uncertainties are plotted but in some cases are too small to be visible.}
    \label{fig:AGN fraction results}
\end{figure}

Figure~\ref{fig:AGN fraction results} shows the distribution of the full sample in the $M_{\mathrm{UV}}$--redshift plane, with the two SED-selected AGN candidates marked as filled stars and the rest-frame UV magnitude of their fitted AGN components shown as unfilled stars connected by dashed lines. Of the two, ID~212506 is also one of the 12 Pop~III candidates identified via the \citet{Rusta_2025} UV emission-line diagnostics in Section~\ref{sec:Pop III Candidates}, and we retain the same teal colour coding used in Figure~\ref{fig:LimitContourPlot2} to identify it here. For the remaining ten Pop~III candidates, their observed properties therefore appear to be predominantly driven by stellar populations rather than luminous AGN activity, as judged by this SED-based test.

Notably, ID~7384, highlighted in lilac in Figure~\ref{fig:AGN fraction results}, is independently identified as an AGN candidate by all of the morphology-based diagnostics described in Section~\ref{sec:Morphological AGN Candidates}, as well as by this SED-based method. Its RGB image (Figure~\ref{fig:rgb}) provides a complementary visual view of its morphology. This source therefore represents our most robust AGN candidate, with its classification corroborated by four independent diagnostics spanning morphology and broadband SED shape. Its spectrum and imaging are included alongside the Pop~III candidates in Appendix~\ref{sec:Spectroscopic and Imaging Properties of Selected Galaxies} for direct comparison.

\subsection{Spatially Resolved Properties of Candidate Galaxies}\label{sec:Spatially Resolved Properties of Candidate Galaxies}

Having established from the integrated broadband SED fits with \texttt{PROSPECT} that the vast majority of galaxies in our sample do not require an AGN component, we next perform spatially resolved SED fitting with \texttt{Bagpipes}, as implemented in \texttt{EXPANSE} \citep{EXPANSE}. Following the morphological selection cuts, 62 galaxies are suitable for resolved SED fitting. We discuss the properties of their resolved structures in the following subsections.

\subsubsection{Population trends from Resolved SED Fitting}\label{sec:Population trends from Resolved SED Fitting}

\begin{figure*}
\centering
\includegraphics[width=\textwidth]{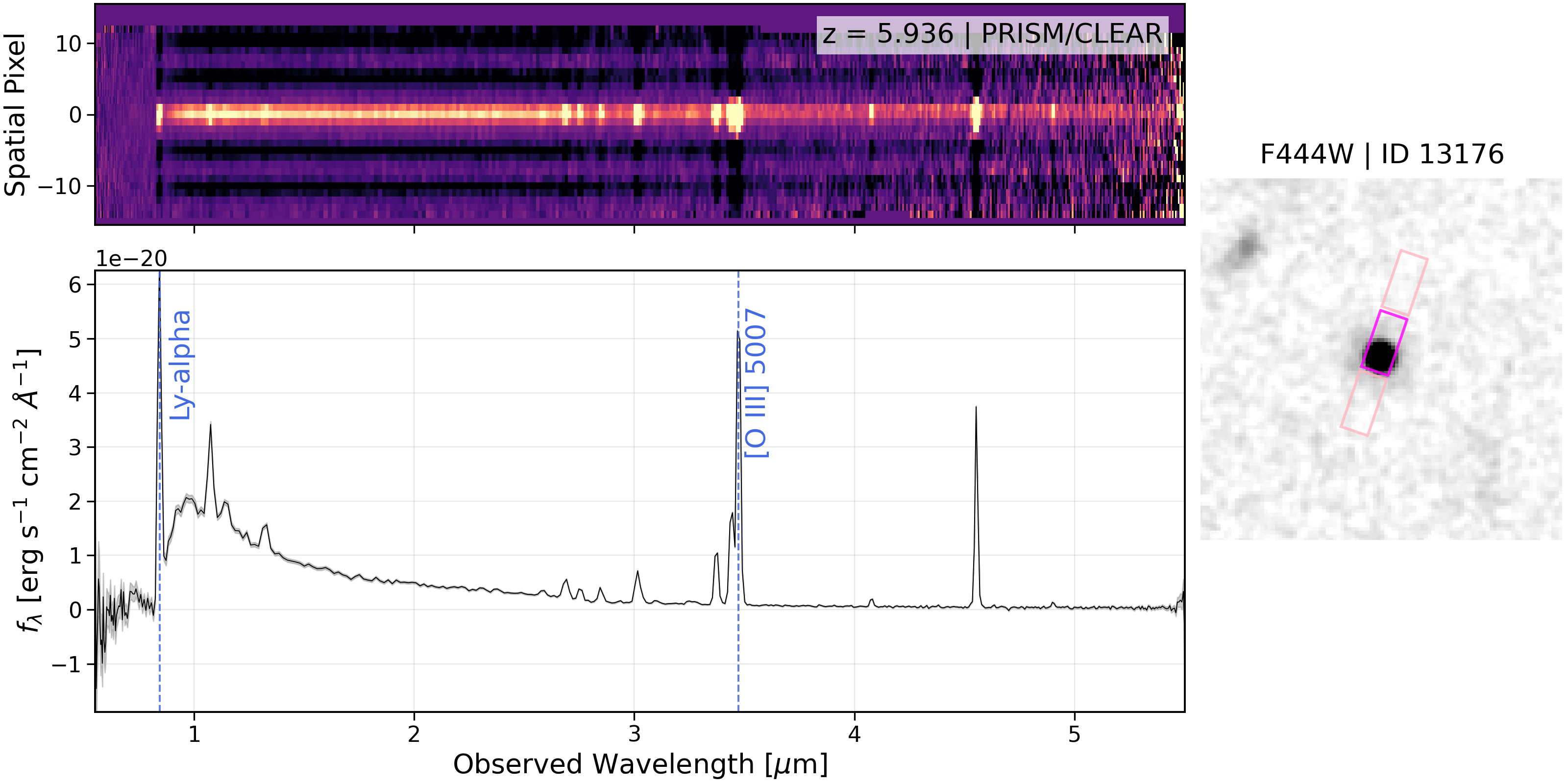}
\caption{Combined spectroscopic and imaging diagnostics for \textit{The Blueberry} (ID~13176, $z=5.94$), observed with \textit{JWST}/NIRSpec in \texttt{PRISM/CLEAR} mode.
\textit{Top left:} 2D spectrum showing prominent emission features, including [O~III]~$\lambda5007$ and Ly$\alpha$.
\textit{Bottom left:} extracted 1D spectrum in $\mathrm{ergs^{-1}cm^{-2}\mathring{A}^{-1}}$, showing the principal emission lines at the systemic redshift.
\textit{Right:} \textit{JWST}/NIRCam F444W cutout with the NIRSpec MSA slitlet configuration overlaid in magenta. The slitlet extends across the observed projected extent of the galaxy, providing spectroscopic coverage of the full visible source.}
\label{fig:slitlet_config_13176}
\end{figure*}

\begin{figure*}
\centering
\includegraphics[width=\textwidth]{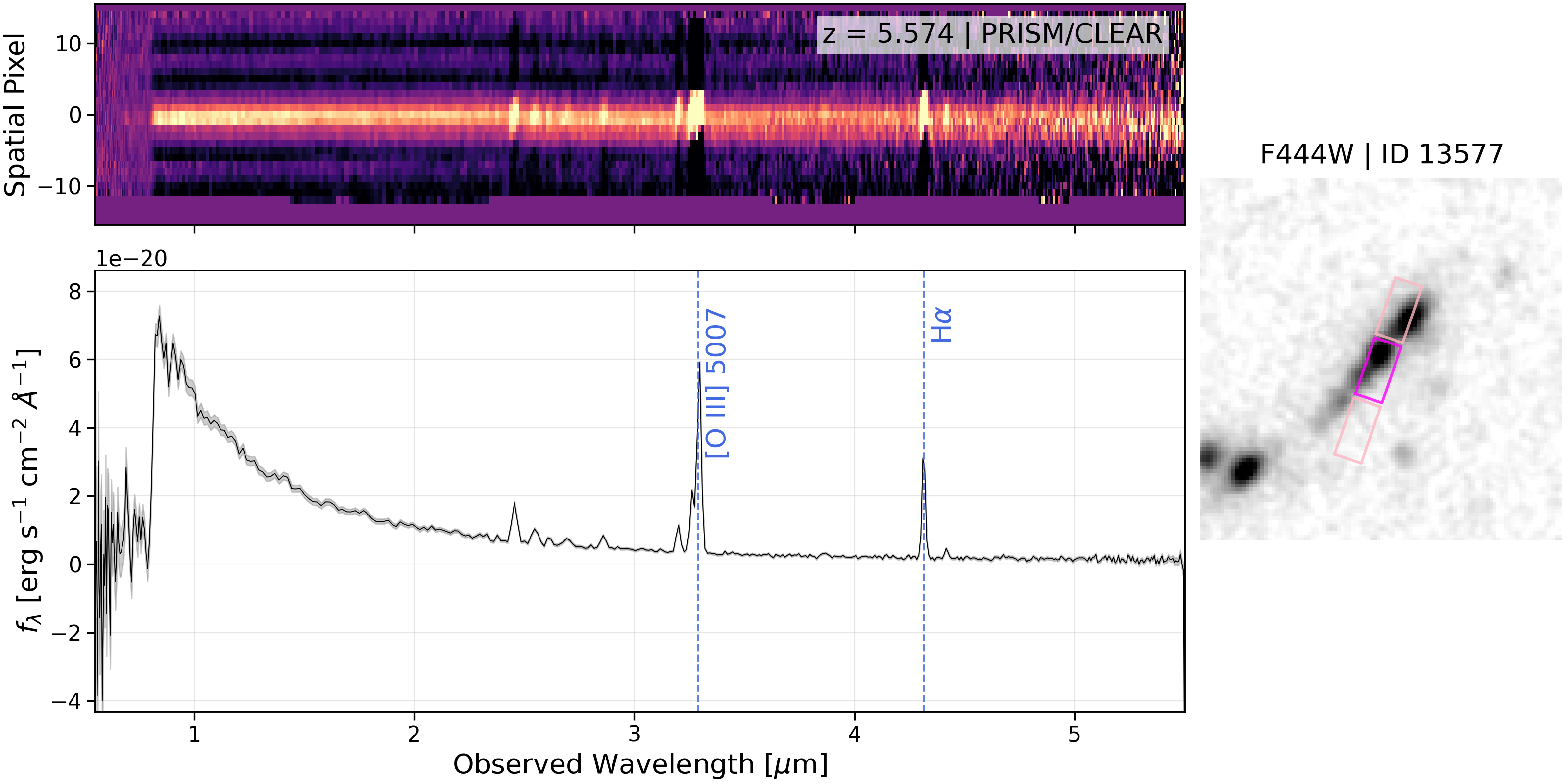}
\caption{Combined spectroscopic and imaging diagnostics for \textit{The Carrot} (ID~13577, $z = 5.57$), observed with \textit{JWST}/NIRSpec in \textit{PRISM/CLEAR} mode. 
\textit{Top left:} 2D spectrum showing prominent emission features, including [O~III]~$\lambda5007$ and H$\alpha$.
\textit{Bottom left:} 1D spectrum ($\mathrm{erg\,s^{-1}\,cm^{-2}\,\mathring{A}^{-1}}$), showing key emission lines redshifted to $z = 5.57$. 
\textit{Right:} \textit{JWST}/NIRCam F444W cutout with the MSA slitlet configuration overlaid in magenta, showing that the spectroscopic extraction centres on the brightest region of a spatially resolved, extended system.}
\label{fig:slitlet_config_13577}
\end{figure*}

Two galaxies in the Pop~III candidate sample (IDs 1130 and 212506) exhibit point-like morphologies in the imaging data. For these sources, spatially resolved SED fitting does not yield reliable internal structure, and they are excluded from the resolved analysis.
The ten remaining resolved candidates show significant spatial variation in their derived stellar population properties. Stellar mass surface density and SFR surface density are typically centrally concentrated, while outer regions exhibit lower mass surface densities and systematically bluer UV continuum slopes. The UV slope $\beta$ varies across individual galaxies, typically spanning $\beta \sim -1$ to $-2.5$, with no uniform radial trend across the sample. The dust attenuation $A_V$ also shows strong spatial variation and is not consistently centrally peaked. No clear monotonic relationship is observed between $\beta$ and $A_V$, with regions of relatively blue continuum slopes persisting even at moderate attenuation.

The sample divides into two broad populations based on their spectral and star-formation properties. Galaxies exhibiting a Balmer break (IDs 2868, 76015, 9564, and 210262) show older mass-weighted ages ($\gtrsim 50$--$60$ Myr) and star formation histories that decline after $\sim 25$--$30$ Myr. The Balmer break arises from a substantial population of intermediate-age stars, particularly A-type stars, whose strong Balmer absorption produces a pronounced discontinuity at 3645~\AA ~\citep{Pawlik_2018}. This spectral signature is commonly associated with galaxies transitioning from active star formation towards quiescence and is characteristic of post-starburst populations \citep[e.g.][]{Pawlik_2018, Wild_2020, Looser_2024,Onoue_2025}. The declining star formation histories of these sources are therefore consistent with an early post-starburst phase.

In contrast, galaxies without a pronounced Balmer break (\textit{The Carrot}, \textit{The Blueberry}, 61712, 28746, 126865, and 9598) are dominated by younger stellar populations ($\lesssim 20$--$30$ Myr) and display bursty star formation histories, often characterised by recent peaks at $\sim 10$ Myr. Across the sample, lower mass-weighted ages are generally associated with higher SFR surface densities.

Although morphological analysis identifies several sources with potential AGN-like properties, some candidates exhibit mixed or ambiguous diagnostics. In particular, ID~9598 ($\log(Z/Z_{\odot}) = -0.44^{+0.10}_{-0.14}$ at $z \approx 3.33$) shows a combination of very young mass-weighted stellar age and high SFR surface density, alongside a red central UV slope ($\beta \simeq -0.5$) and elevated dust attenuation ($A_{\mathrm{V}} \simeq 1.5$). Its integrated properties are $\beta = -1.36 \pm 0.43$ and $M_{\rm UV} = -19.34 \pm 0.06$. 

\subsubsection{The Blueberry}\label{sec:The Blueberry}

We first focus on \textit{The Blueberry} (ID~13176, $z=5.94$), which provides the strongest combination of spectroscopic and spatially resolved evidence for a potential Pop~III contribution in our sample. Unlike most other candidates, \textit{The Blueberry} satisfies the \citet{Rusta_2025} Pop~III criterion using robust measurements of all UV emission lines entering the diagnostic framework, without relying on upper limits (see Section~\ref{sec:Pop III Candidates}).

The NIRSpec slit geometry for \textit{The Blueberry} is shown in Figure~\ref{fig:slitlet_config_13176}. The slitlets are positioned across the full projected extent of the galaxy, minimising the possibility that an extended component is excluded from the spectroscopic extraction. This provides greater confidence that the integrated emission-line measurements are representative of the observed galaxy as a whole rather than being strongly biased by incomplete spatial coverage.

\begin{figure*}
\centering
\includegraphics[
width=\textwidth,
height=0.85\textheight,
keepaspectratio
]{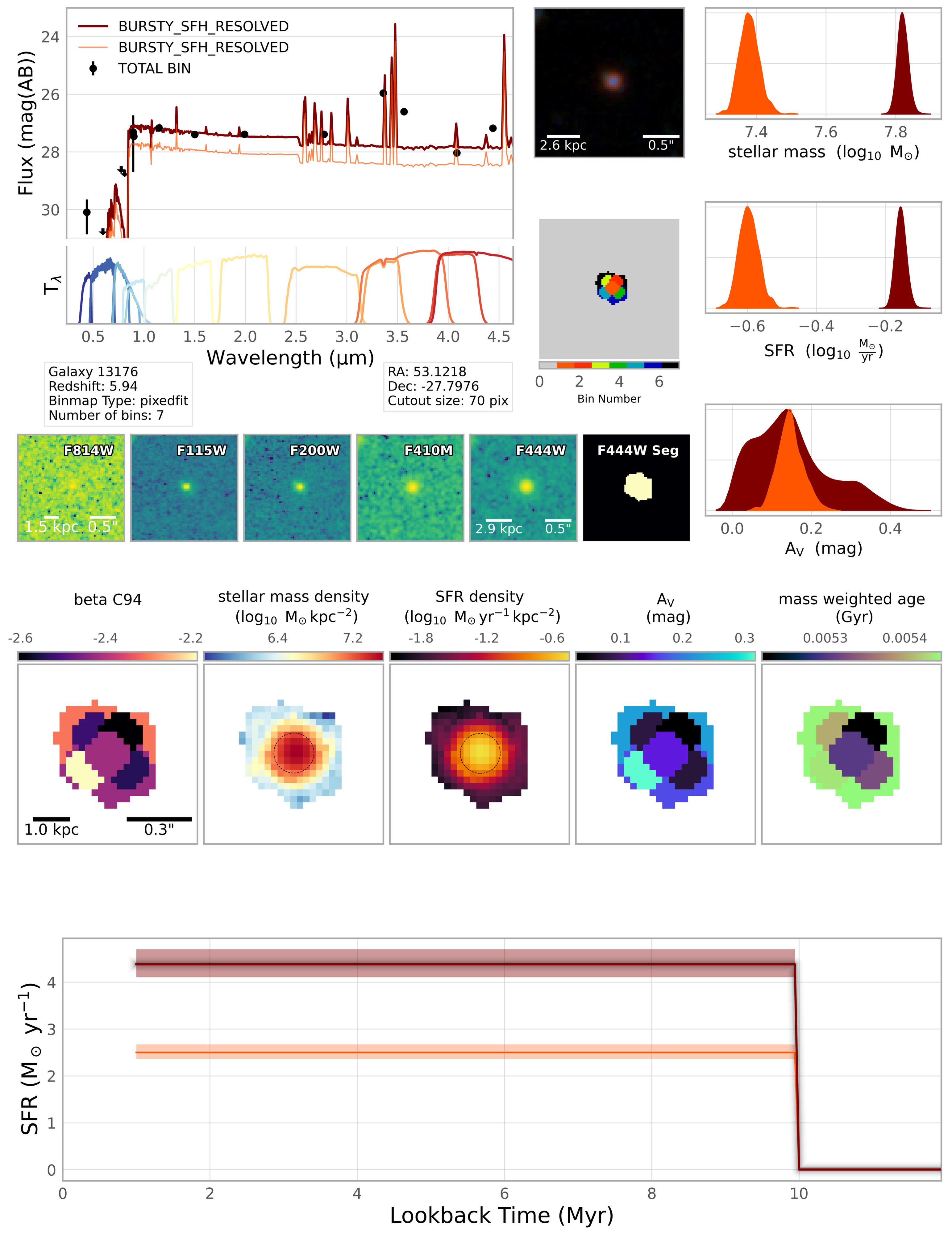}
\caption{Spatially resolved \texttt{Bagpipes} SED-fitting overview for \textit{The Blueberry} at $z=5.94$. \textbf{Top-left:} Integrated photometry (black) compared with the \texttt{Bagpipes} SED fit constructed from the binned photometry (dark red) and the fit to spatial bin 1 (orange). \textbf{Top-right:} Posterior distributions of the inferred stellar mass, SFR, and dust attenuation, $A_V$, for the corresponding fits. \textbf{Middle:} Multi-band \textit{JWST}/NIRCam imaging, the RGB composite, and the spatial binning map generated using \texttt{piXedfit}. \textbf{Bottom-middle:} Spatially resolved maps of the UV continuum slope $\beta$, stellar mass surface density, SFR surface density, $A_V$, and mass-weighted stellar age. \textbf{Bottom:} Inferred star-formation histories as a function of lookback time for the integrated \texttt{Bagpipes} fit (dark red) and spatial bin 1 (orange). The resolved maps show consistently young, blue and weakly attenuated stellar populations across the spatially resolved regions.}
\label{fig:bagpipes_overview_blueberry}
\end{figure*}

The spatially resolved SED fitting provides a complementary, resolved view of the stellar populations across \textit{The Blueberry}. Figure~\ref{fig:bagpipes_overview_blueberry} shows that the resolved regions are consistently characterised by blue UV continua, with $\beta \lesssim -2.2$, low dust attenuation ($A_V < 0.3$), elevated SFR surface densities of $\log_{10}(\Sigma_{\rm SFR}/M_{\odot}\mathrm{yr}^{-1}\mathrm{kpc}^{-2}) \approx -2$ to $-0.5$, and very young mass-weighted stellar ages of $\lesssim5.4$~Myr. The combination of young stellar populations and elevated SFR surface density indicates an intense and recent episode of star formation.

\textit{The Blueberry} has $\beta=-2.15\pm0.10$, $M_{\rm UV}=-19.85\pm0.01$, and $\log(Z/Z_{\odot})=-1.18\pm0.05$. Its blue UV continuum and moderate UV luminosity are consistent with a young, actively star-forming system, while its integrated metallicity demonstrates that it is chemically enriched on global scales. This disfavours an interpretation in which the galaxy as a whole is chemically pristine, but does not preclude the presence of a more metal-poor stellar component within the enriched system. The resolved attenuation is also consistent with the integrated value, $A_V=0.15\pm0.08$ (Table~\ref{tab:line_fluxes_snrs}), supporting the interpretation that the low dust content is representative of the galaxy rather than arising solely from spatial averaging. We note, however, that this combination of young age, blue UV slope, low attenuation, and high SFR surface density is not by itself unique to a Pop~III-like population: chemically enriched starbursts can produce very similar resolved signatures, particularly during bursty phases of star formation (Section~\ref{sec:Discussion}).

\textit{The Blueberry} has also been independently studied in previous JWST/NIRSpec analyses of the JADES GOODS-S field. \citet{Simmonds_2024} and \citet{Saxena_2024} find a high ionising photon production efficiency, $\log\xi_{\rm ion}\approx25.5$--$25.7 ~\mathrm{Hz~erg^{-1}}$, with \citet{Saxena_2024} additionally measuring an extreme ionisation parameter ($\mathrm{O32}=70.6\pm4.1$). These measurements highlight the extreme ionising output of the source. \citet{Witstok_2024} further identify it as a strong Ly$\alpha$ emitter, with $\mathrm{EW}_{\mathrm{Ly}\alpha}\approx124$~\AA\ and $f_{\rm esc~,~Ly\alpha}~\approx~0.26$, and place it within a significant overdensity at $z\sim5.9$, with an inferred required ionised region of order $R_{\rm ion}\sim0.5$~pMpc to reproduce its observed Ly$\alpha$ transmission. Together, these studies provide additional evidence that \textit{The Blueberry} is a young and highly ionising galaxy in an interesting reionisation-era environment.

Taken together, \textit{The Blueberry} represents our strongest candidate for a Pop~III contribution. Its UV emission-line ratios satisfy the \citet{Rusta_2025} criterion for a Pop~III stellar mass fraction exceeding $25\%$ and are consistent with a Pop~III-rich hybrid. Its resolved properties further reveal very young stellar populations, low attenuation, blue UV continua, and elevated SFR surface densities. The previous measurements of its ionising efficiency, ionisation state and Ly$\alpha$ emission provide complementary evidence for the extreme physical conditions inferred from our diagnostics. Nevertheless, because the resolved SED modelling does not explicitly include Pop~III templates and the integrated gas-phase metallicity demonstrates that the galaxy is not pristine, we interpret it as a \emph{candidate} for a hybrid Pop~III population rather than as a definitive detection of Pop~III stars.

\begin{figure*}
\centering
\includegraphics[
    width=\textwidth,
    height=0.85\textheight,
    keepaspectratio
]{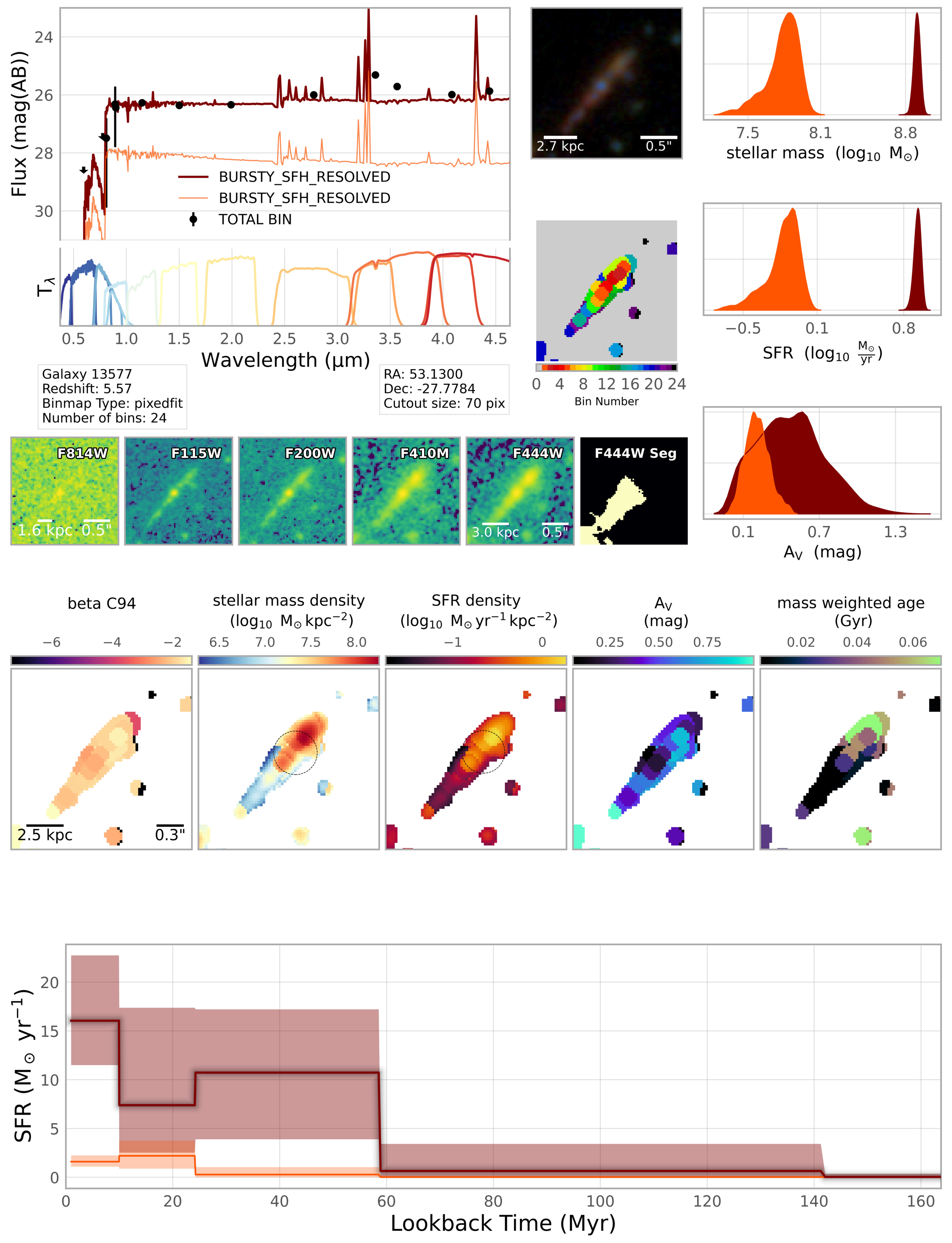}
\caption{Spatially resolved \texttt{Bagpipes} SED-fitting overview for \textit{The Carrot} at $z=5.57$. \textbf{Top-Left:} Integrated photometry (black) compared with the \texttt{Bagpipes} SED fit constructed from the binned photometry (dark red) and the \texttt{Bagpipes} fit to spatial bin 1 (orange). \textbf{Top-Right:} Posterior distributions of the inferred stellar mass, SFR, and dust attenuation, $A_V$, for the corresponding fits. \textbf{Middle:} Multi-band \textit{JWST}/NIRCam imaging, the RGB composite, and the spatial binning map generated using \texttt{piXedfit}. \textbf{Bottom-Middle:} Spatially resolved maps of the UV continuum slope $\beta$, stellar mass surface density, SFR surface density, $A_V$, and mass-weighted stellar age. \textbf{Bottom:} Inferred star formation histories as a function of lookback time for the \texttt{Bagpipes} SED fit (dark red) and spatial bin 1 (orange).}
\label{fig:bagpipes_overview_carrot}
\end{figure*}

\subsubsection{The Carrot}\label{sec:The Carrot}

We also examine \textit{The Carrot} (ID~13577, $z=5.57$) as a second case study. Unlike \textit{The Blueberry}, it exhibits a spatially localised combination of properties expected for a Pop~III-like star-forming region within a more evolved host galaxy, providing a complementary environment in which to investigate the potential contribution of Pop~III stars.

Figure~\ref{fig:bagpipes_overview_carrot} presents the full resolved SED-fitting overview for \textit{The Carrot}. The galaxy contains a central region characterised by a very blue UV continuum ($\beta \approx -2.5$), low dust attenuation ($A_V \sim 0.2$), high SFR surface density, and young stellar populations with mass-weighted ages $\lesssim 30$ Myr. In contrast, the surrounding regions exhibit markedly different physical conditions: the upper-right portion of the galaxy is characterised by high dust attenuation and older mass-weighted stellar ages, whereas the lower-left region shows younger stellar populations but enhanced dust content. This spatial variation may indicate a more complex assembly history, potentially consistent with secondary star-formation episodes occurring in chemically enriched gas, possibly through feedback-driven outflows from the central starburst. However, such variations may also arise from the age--dust degeneracy, with different regions favouring different solutions for similar observed colours.

We note that the NIRCam segmentation map for \textit{The Carrot} was adjusted to include the upper extension identified following the \texttt{EAZY} SED-fitting redshift estimates, ensuring that this extended component was incorporated into the subsequent resolved SED fitting. The NIRSpec slit geometry shown in Figure~\ref{fig:slitlet_config_13577} places the slit predominantly across the galaxy core, while the extended upper and lower components are only partially sampled. Consequently, the integrated spectrum contains the central region together with incomplete contributions from the surrounding extensions and consequently provides a luminosity-weighted view of the system.

The integrated measurements for \textit{The Carrot}, presented in Section~\ref{sec:Global Properties of the Candidate Sample}, are $\beta = -2.08 \pm 0.12$ and $M_{\rm UV} = -20.52 \pm 0.02$, with an integrated metallicity of $\log (Z/Z_{\odot}) = -0.51^{+0.10}_{-0.10}$. The central region inferred from the resolved SED fitting is therefore substantially bluer, less attenuated, and younger than suggested by the integrated properties. The integrated Balmer decrement yields a formally negative $A_V$, floored to $A_V=0$ (Table~\ref{tab:line_fluxes_snrs}), despite substantial dust attenuation in the resolved outskirts, likely because the NIRSpec slit only partially covers the more attenuated extended components. This discrepancy highlights the importance of spatially resolved measurements when interpreting candidate low-metallicity or Pop~III-like systems, as compact extreme regions can be hidden within globally enriched galaxies. Extremely blue values ($\beta < -4$) are confined to low signal-to-noise peripheral bins and are therefore not considered statistically significant, as each spatial bin is fitted independently and such low-SNR measurements are particularly susceptible to noise fluctuations.

\subsection{Resolved SED Properties of the Full He~II-emitter Population}\label{sec:Resolved Population Properties}

\begin{figure*}
\centering
\includegraphics[
    width=\textwidth,
    height=\textheight,
    keepaspectratio
]{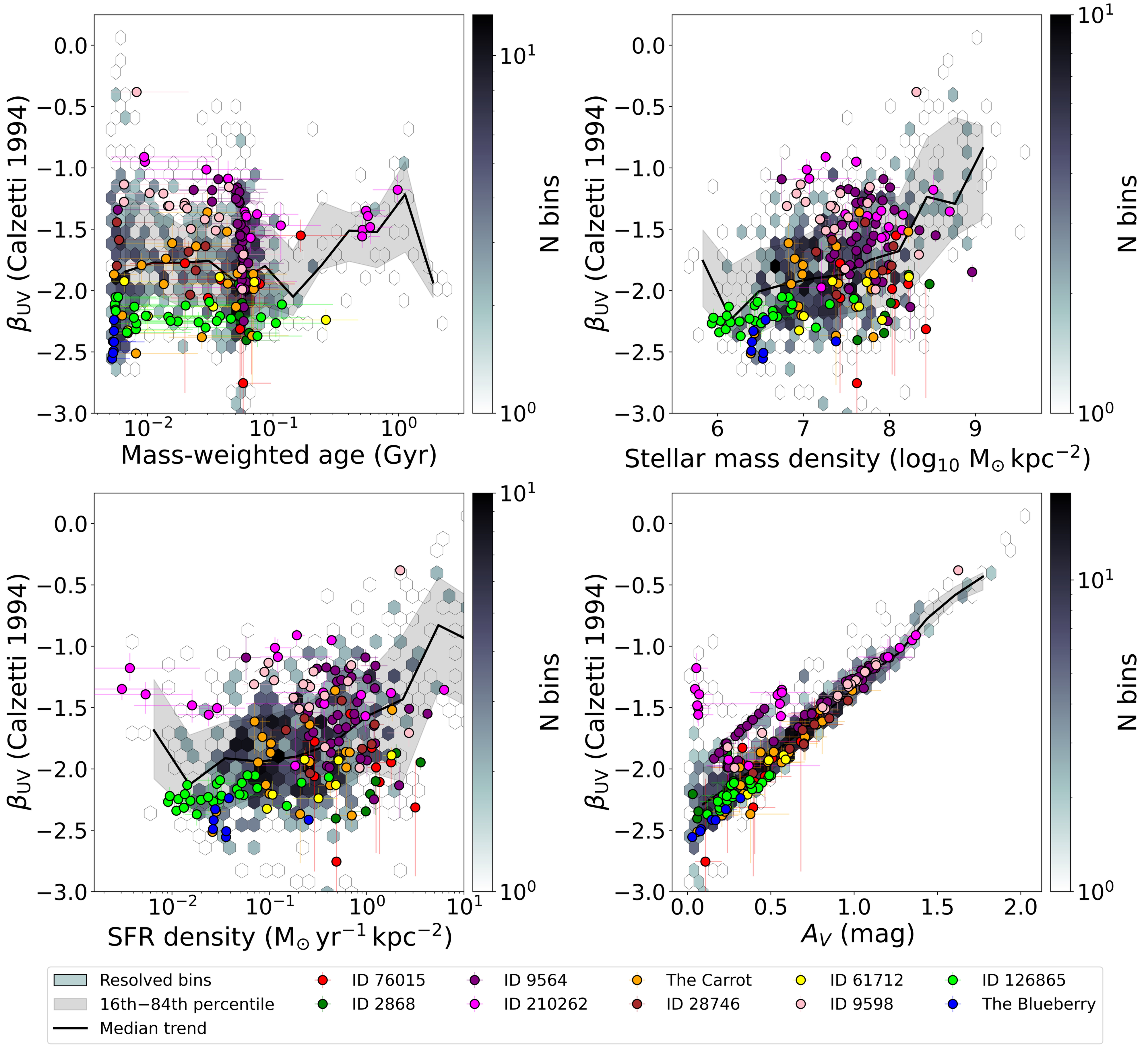}
\caption{
Resolved UV continuum slope, $\beta$, as a function of four physical properties derived from the resolved SED fitting for every spatial bin in the resolved He~II-emitter sample. The four panels show $\beta$ as a function of (top left) mass-weighted age, (top right) stellar mass surface density, (bottom left) star-formation-rate surface density, and (bottom right) dust attenuation, $A_V$. Grey hexagonal bins show the distribution of all resolved spatial bins, with darker shading indicating regions containing a larger number of bins. The solid black curve shows the median relation of the full sample, while the shaded region denotes the 16th--84th percentile range. Spatial bins belonging to 10 Pop~III candidate galaxies are plotted using coloured markers corresponding to each candidate, with horizontal and vertical error bars indicating the uncertainties from the resolved SED fitting.
}
\label{fig:beta_resolved_population}
\end{figure*}

\begin{figure*}
\centering
\includegraphics[
    width=\textwidth,
    height=\textheight,
    keepaspectratio
]{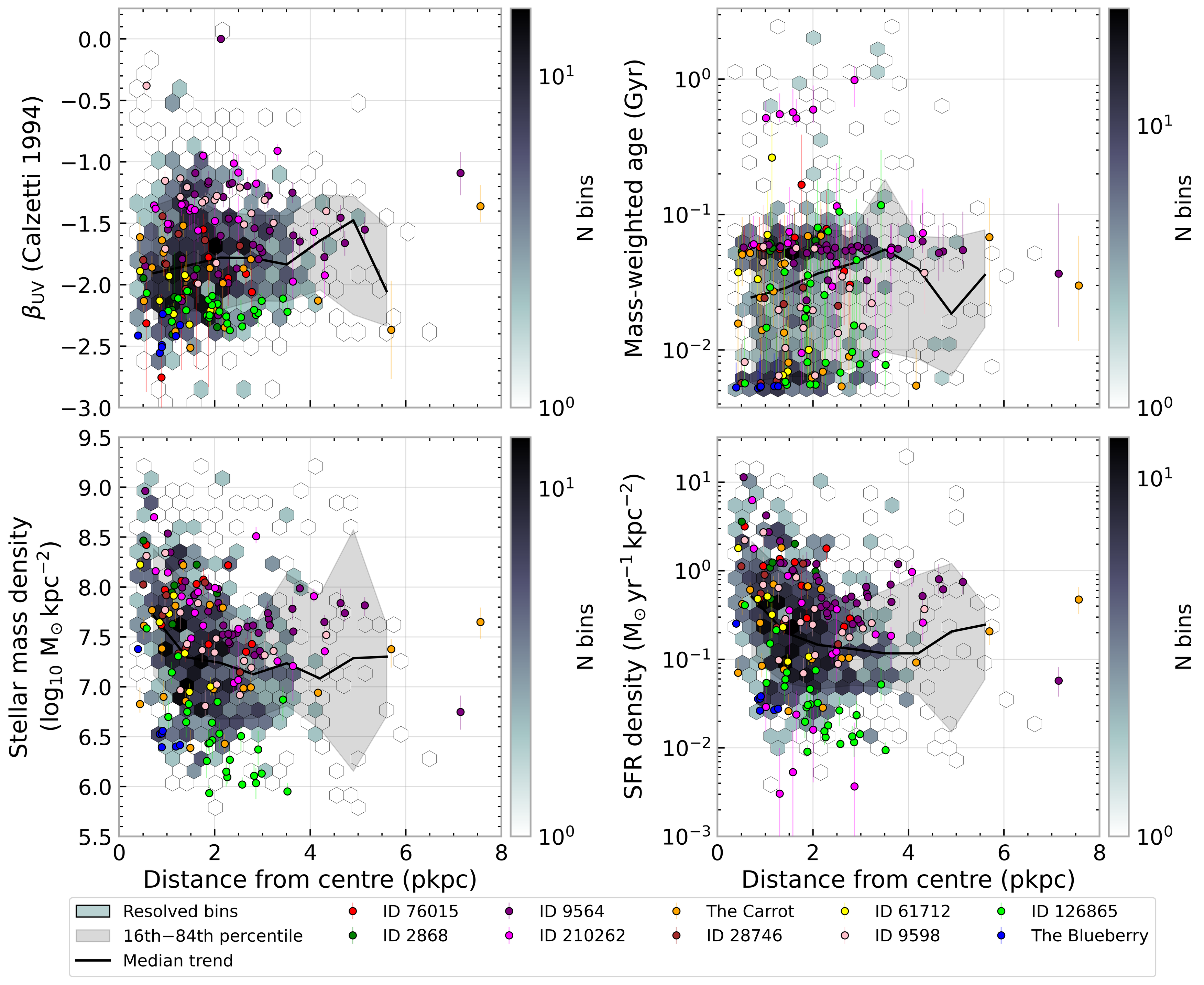}
\caption{
Radial variation of resolved physical properties for every spatial bin in the resolved He~II-emitter sample as a function of projected distance from the galaxy centre. Distances were calculated using the \texttt{EXPANSE} \texttt{radial\_scatter\_of\_property\_map} routine and are given in proper kiloparsecs (pkpc). The panels show (top left) UV continuum slope $\beta$, (top right) mass-weighted age, (bottom left) stellar mass surface density, and (bottom right) star-formation-rate surface density. Grey hexagonal bins show the distribution of all resolved bins, with darker colours corresponding to higher numbers of spatial bins. The black line shows the median radial trend, while the grey shaded region indicates the 16th--84th percentile range. Spatial bins belonging to 10 Pop~III candidate galaxies are highlighted using coloured markers, allowing their internal resolved structures to be compared directly with the full He~II-emitter population.
}
\label{fig:radial_population}
\end{figure*}

Having examined the resolved properties of \textit{The Blueberry} and \textit{The Carrot} individually, we now place the full resolved candidate sample in the context of the broader population of 62 resolved He~II emitters, testing whether the extreme resolved properties identified above are genuinely distinctive or instead fall within the normal scatter of the resolved population. Figures~\ref{fig:beta_resolved_population} and \ref{fig:radial_population} provide two complementary population-level views: the first shows how the resolved UV slope, $\beta$, correlates with mass-weighted age, stellar mass surface density, SFR surface density, and dust attenuation across every spatial bin in the sample (Section~\ref{sec:resolved_beta_properties}); the second examines how these same properties vary with projected galactocentric distance, testing whether extreme, potentially Pop~III-like regions are preferentially centrally located or can occur throughout a galaxy (Section~\ref{sec:resolved_radial_properties}). In both figures, all spatial bins are combined into population-level distributions, with the 10 resolved candidate galaxies highlighted individually; grey hexagonal bins show the distribution of the full resolved sample, with darker shading indicating regions containing a larger number of bins, and the black curves show the median relation with shaded 16th--84th percentile ranges. Peripheral, low-S/N bins producing extreme inferred properties, such as those visible in the resolved maps of \textit{The Carrot} in Figure~\ref{fig:bagpipes_overview_carrot}, are excluded from these population-level comparisons as they are not considered physically significant.

\subsubsection{Resolved UV Slopes and Physical Properties}\label{sec:resolved_beta_properties}

Figure~\ref{fig:beta_resolved_population} compares the resolved UV continuum slope, $\beta$, with mass-weighted age, stellar mass surface density, SFR surface density, and dust attenuation for the resolved He~II-emitter population. The majority of the candidates occupy parameter space shared with the broader population rather than forming a distinct resolved locus, with their distinguishing properties arising from the combination and spatial distribution of individual regions rather than uniformly extreme values across all spatial bins. \textit{The Blueberry} is the clear exception to this, with its resolved bins consistently clustering towards young ages, blue UV slopes, and low dust attenuation.

The $\beta$--mass-weighted age relation shows the expected tendency for older regions to have redder UV continua, with the candidate galaxies broadly following the same relation as the parent population. Most resolved bins are concentrated around mass-weighted ages of $\sim10^{-2}$--$10^{-1}$~Gyr; this concentration should be interpreted with some caution, as the \texttt{Bagpipes}/\texttt{EXPANSE} fits adopt a six-bin bursty star formation history parameterisation, which can produce clustering of solutions around the discrete temporal bins rather than reflecting a continuous age distribution. Several candidates nevertheless contain individual bins extending towards younger and bluer populations. ID~210262 is particularly notable, with its resolved bins separating into two groups with substantially different inferred mass-weighted ages, possibly indicating genuine spatial variation in stellar populations and dust content, although the age--dust degeneracy inherent to broadband SED fitting prevents these possibilities from being cleanly distinguished.

The relations with stellar mass and SFR surface density show substantial scatter, with candidate bins spanning much of the parameter space occupied by the full resolved sample. There is a general tendency for higher-density regions to have redder UV slopes, but the candidates do not occupy a unique locus: several combine relatively high SFR or stellar mass surface densities with comparatively blue UV slopes, such as a resolved region in ID~76015 with $\beta \sim -2.25$ despite high SFR surface density and elevated dust attenuation. This illustrates that high star-formation intensity does not necessarily correspond to a red UV continuum, and that the observed $\beta$ values instead reflect the combined effects of recent star formation, stellar population age, and attenuation.

The strongest and most continuous trend is observed between $\beta$ and dust attenuation, with increasing $A_V$ associated with systematically redder UV slopes. The candidates largely follow the same sequence as the wider He~II-emitter population, indicating that dust is an important contributor to the variation in resolved UV colour. At the same time, some candidate regions remain relatively blue at low-to-moderate $A_V$, suggesting that their colours cannot be explained by attenuation alone.

\subsubsection{Radial trends in resolved physical properties}\label{sec:resolved_radial_properties}

Figure~\ref{fig:radial_population} provides a complementary view by examining how the resolved properties vary with distance from the galaxy centre. The galaxy centre was defined as the stellar-mass-weighted centre of mass, calculated from the resolved Bagpipes stellar-mass density map using the \texttt{EXPANSE} \texttt{\detokenize{radial_scatter_of_property_map}} routine. For each spatial bin, the galactocentric distance was calculated from the centre using the mean radial distance of the pixels assigned to that bin, and converted to proper kiloparsecs (pkpc).

The radial $\beta$ and mass-weighted age profiles show no strong universal gradients, indicating that UV colour and stellar population age are not systematically dependent on galactocentric distance across the He~II-emitter population. Instead, individual galaxies exhibit substantial internal variation, consistent with spatially heterogeneous and bursty star formation. Stellar mass surface density generally declines with increasing galactocentric distance, while SFR surface density shows a weaker decline with substantial scatter. Elevated star-formation activity and blue UV slopes are not confined to galaxy centres: \textit{The Carrot} and ID~76015 contain particularly blue central regions, while several candidates also contain blue bins at larger galactocentric distances. ID~126865, for example, contains consistently blue resolved regions, with several bins combining $\beta \lesssim -2$, low dust attenuation, and young mass-weighted ages of $\sim10^{-2}$~Gyr, demonstrating that individual extreme or near-extreme regions can occur within otherwise non-uniform systems.

\textit{The Blueberry} is the clearest example of a galaxy with uniformly extreme resolved properties. Its bins occupy a relatively distinct locus, with the central bin reaching $\log(\Sigma_{M_\star}/M_\odot\mathrm{kpc}^{-2})\sim7.5$, $\beta\sim-2.4$, and $\Sigma_{\mathrm{SFR}}\sim10^{-0.5}~M_\odot\mathrm{yr}^{-1}\mathrm{kpc}^{-2}$, while mass-weighted ages remain below $10^{-2}$~Gyr and dust attenuation is low, together suggesting a compact region undergoing an intense, recent burst of star formation.

Overall, the candidate galaxies are broadly representative of the wider He~II-emitter population in their resolved structure and radial trends, while containing individual regions with more extreme physical conditions. Young, blue, and relatively unobscured regions can occur at a range of galactocentric distances rather than being universally concentrated in galaxy centres, supporting a picture in which the properties relevant to Pop~III-like star formation are spatially localised rather than characteristic of an entire galaxy.

\section{Discussion} \label{sec:Discussion}

\subsection{Interpreting the Pop III Candidates}

Our analysis identifies a hierarchy of evidence among the 12 galaxies selected by the \citet{Rusta_2025} emission-line framework. Most candidates are subject to substantial uncertainty because their positions in the diagnostic diagrams depend on low-SNR measurements or upper limits on UV metal lines. \textit{The Blueberry} (ID~13176) is distinct: it occupies the $>25\%$ Pop~III region and Pop~III-rich hybrid contours using robust detections of all of the UV emission lines entering the diagnostic framework, without relying on upper limits. It therefore provides the strongest spectroscopic evidence in our sample for a substantial Pop~III contribution.

The resolved SED fitting provides independent spatial support for this interpretation. Across \textit{The Blueberry}, the resolved regions are characterised by very young ages, blue UV continua, low attenuation, and elevated SFR surface densities, indicating a compact and intense episode of recent star formation. These conditions are consistent with the short-lived, burst-dominated phases predicted for Pop~III systems \citep{Venditti_2023,Venditti_2024_sim,venditti2025_sim,Rusta_2025}. However, they are not unique to Pop~III: young, chemically enriched starbursts can produce similarly blue continua, high SFR surface densities and strong nebular emission \citep{Simmonds_2024}. Moreover, our BPASS-based SED models do not include an explicit Pop~III component, so the resolved fits cannot independently demonstrate the presence of metal-free stars. Independent observations further indicate unusually hard ionising conditions, with $\log\xi_{\rm ion}\approx25.5$--$25.7~\mathrm{Hz~erg^{-1}}$ and $\mathrm{O32}=70.6\pm4.1$ \citep{Simmonds_2024,Saxena_2024}. While these properties can also arise in young, highly ionised, chemically enriched systems, their combination with the UV diagnostics strengthens the Pop~III interpretation.

This picture is consistent with the evolutionary framework of \citet{Rusta_2025}, in which Pop~III supernovae enrich the surrounding gas while metal-free stars survive until the onset of Pop~II star formation. During the subsequent hybrid phase, Pop~III and Pop~II stars coexist, allowing enriched gas to produce UV metal lines while surviving Pop~III stars maintain a hard ionising spectrum. The simultaneous strong He~II and UV metal lines in \textit{The Blueberry}, together with its extreme resolved stellar populations and ionising conditions, make \textit{The Blueberry} the most compelling \textit{candidate} in our sample for a surviving Pop~III population embedded within an increasingly enriched host. 

Its globally inferred metallicity, $\log(Z/Z_\odot)=-1.18\pm0.05$, together with its very blue UV slope ($\beta=-2.15\pm0.10$) and high UV luminosity ($M_{\rm UV}=-19.85\pm0.01$), indicates a chemically enriched rather than pristine system, but does not exclude a surviving metal-free stellar component embedded within an enriched host. This interpretation is consistent with recent JWST studies, which identify Pop~III candidates across a broad parameter space ($z\sim3$--11, $\beta\sim-3$ to $-2$, $M_{\rm UV}\sim-22$ to $-10$, $Z\sim10^{-3}$–$10^{-2} Z_\odot$), rather than exclusively in the faint, ultra-blue, high-redshift, very low metallicity regime expected for pristine systems \citep{Venditti2026_POPIIIcandidates}.

The remaining candidates provide substantially weaker evidence: most lack secure UV line measurements and/or resolved stellar-population constraints, several overlap with conventional star-forming or AGN models, and their metallicities indicate chemically enriched systems. The observed He~II emission in these sources therefore likely requires alternative hard ionising sources, such as Wolf--Rayet stars, which can produce hard spectra in metal-poor but non-pristine galaxies \citep{Maschmann_2024_WR, Gonz_lez_Tor__2025}, or binary stellar evolution, where mass transfer and envelope stripping can sustain hard radiation fields in young populations \citep{Dutta_2024, Hovis_Afflerbach_2025}. We therefore interpret these candidates as chemically enriched galaxies hosting spatially localised high-ionisation regions, rather than as robust evidence for surviving Pop~III populations. 

\textit{The Carrot}, for example, has an enriched integrated metallicity ($\log(Z/Z_{\odot})=-0.51^{+0.10}_{-0.10}$) but a substantially more extreme central region, illustrating how a localised young population can be diluted by a more evolved host in integrated measurements. This is particularly relevant given the growing number of JWST Pop~III candidates associated with compact, off-centre, or kpc-scale regions \citep[e.g.,][]{Maiolino_2026, Rusta_2026, Ubler_2026, Reumert_2026}. Such regions may dominate the ionising emission while leaving only a weak signature in integrated stellar populations, highlighting the importance of spatially resolved measurements for identifying Pop~III-like star formation.

The absence of extremely blue UV slopes ($\beta \lesssim -3$) among our candidates is not unexpected for the polluted/hybrid Pop~III systems targeted here. \citet{Katz_2025} show that nebular continuum emission from free--free, free--bound, and two-photon processes can substantially redden the UV slopes of young, highly ionising populations, with intrinsically blue stellar continua reaching $\beta\sim-2.5$ or redder when nebular emission is included. Thus, the absence of $\beta\lesssim-3$ slopes does not rule out a polluted/hybrid Pop~III scenario. \citet{Katz_2025} further identify a Balmer jump at rest-frame 3645~\AA, high ionising photon production efficiencies, and a UV downturn from two-photon emission as signatures of strong nebular continuum emission. Our PRISM/CLEAR spectra cover this wavelength range, and we discuss Balmer jumps in our candidate sample in Section~\ref{sec:Population trends from Resolved SED Fitting}; a quantitative assessment of the nebular continuum contribution is beyond the scope of this work.

Additionally, extremely blue UV continua are not unique to primordial stellar populations. Chemically enriched galaxies can exhibit similarly blue slopes through mechanisms such as feedback-driven gas clearing; for example, \citet{marqueschaves2026prismsu37126blueismnaked} identify a potential "ISM-naked" starburst at $z=10.26$ with $\beta\approx-2.9$ and $f_{\rm esc}\approx1$. Early galaxies are also increasingly characterised as highly bursty \citep[][]{Dome_2023,Tacchella_2023,Dressler_2024}: following a star-formation burst, ionising photon production declines faster than the UV continuum, reducing nebular emission and allowing the blue stellar continuum to dominate, driving $\beta$ toward $\approx-2.8$ even in chemically enriched systems. These effects further motivate combining emission-line diagnostics with spatially resolved stellar-population constraints, particularly when assessing whether extreme ionising conditions arise from localised Pop~III-like populations or from conventional, chemically enriched star formation \citep{Bouwens_2010, Topping2024_beta}.

\subsection{Assessing AGN Contamination}

Morphological analysis using visual inspection, quantitative Sérsic-profile fitting, and aperture concentration provides complementary tests for compact or unresolved nuclear emission. Visual inspection identifies four galaxies with AGN-like morphologies based on the presence of a visually unresolved, point-like central component and/or prominent diffraction-spike features characteristic of a bright nuclear source. In contrast, the quantitative morphological diagnostics test the degree of central concentration through Sérsic-profile fitting and aperture concentration, recovering only one of the four visually identified sources (ID~7384) as quantitatively compact and independently flagging two additional compact galaxies not identified by eye (ID~3608 and the Pop~III candidate ID~1130). The limited overlap between the visual and quantitative classifications reflects the fact that  compact morphology alone cannot uniquely establish an AGN interpretation: an AGN may contribute insufficient light relative to its extended host for the galaxy to appear morphologically compact, while compact nuclear star formation can produce similarly concentrated, unresolved emission. Morphological measurements are therefore most appropriately used to identify sources warranting further AGN investigation rather than as a standalone AGN classification.

To provide an independent assessment of AGN contamination, we additionally model the broadband SEDs of the sample using \texttt{PROSPECT}. Only two galaxies (IDs~212506 and 7384) are statistically preferred to include an AGN component, satisfying both the DIC and AGN-fraction selection criteria (Section~\ref{sec:SED-based AGN Identification}). Source~212506 is one of the 12 Pop~III candidates selected using the \citet{Rusta_2025} UV emission-line diagnostics; however, none of the remaining 11 Pop~III candidates require an AGN component to reproduce their broadband SEDs. The SED fitting result for ID~212506 is particularly notable because its broadband photometry favours an AGN component despite its position outside the AGN contour regions in Figure~\ref{fig:LimitContourPlot2}. This highlights the difficulty of distinguishing AGN and stellar contributions with the available data, such that the absence of an AGN classification from line-ratio diagnostics does not rule out an AGN contribution to the broadband continuum. Additionally, ID~212506 appears morphologically compact (see Section~\ref{sec:Population trends from Resolved SED Fitting}), potentially indicating an AGN.

A particularly compelling result is that ID~7384 is independently flagged as an AGN candidate by every diagnostic applied in this work. The convergence of four independent diagnostics on a single source, despite the otherwise limited overlap between methods, makes ID~7384 the most consistently supported AGN candidate in the sample. This agreement provides substantially stronger evidence for an AGN interpretation than any individual diagnostic alone.

Taken together, these results indicate that a substantial AGN contribution to the broadband UV emission is not required for the majority of the sample. However, lower-luminosity or heavily obscured AGN may contribute little to the observed UV continuum or produce weak morphological signatures while still influencing the ionising radiation field and emission-line ratios. Such components cannot be robustly constrained with the available photometry, which extends only to F444W, particularly given degeneracies between AGN and stellar emission \citep{Thorne2022_AGNfraction}. Longer-wavelength observations, such as JWST/MIRI or ALMA, could provide complementary constraints through their sensitivity to dust emission associated with obscured AGN \citep[e.g.][]{Yang__2023,Lyu__2024,Li_2024,Leung_2024,leung2026miriearlyobscuredagnwide,Hewitt_2026,Juod_balis_2026}.

Future higher-quality spectroscopy will also be important for distinguishing AGN-powered from stellar ionising sources. Where present, permitted broad emission-line components provide a more direct AGN diagnostic by tracing high-velocity gas in the BLR that cannot be produced by stellar processes \citep{Czerny_2011,Fu_2023}. However, the BLR can remain obscured by intervening torus material, so the absence of broad permitted lines likewise does not exclude an AGN. Higher-resolution spectroscopy capable of resolving line profiles and improving weak UV metal-line measurements would therefore provide an important independent constraint on the nature of the ionising source.

\subsection{Caveats and Systematic Uncertainties}\label{sec:caveats and systematic uncertainties}

The principal limitation of this work arises from the quality of the available UV spectroscopy for the majority of the sample. The low spectral resolution of the \textit{JWST}/NIRSpec PRISM observations results in significant blending of neighbouring UV emission lines, allowing \texttt{specFitMSA} to assign similar fluxes to multiple lines. This produces an artificial clustering in Figure~\ref{fig:LimitContourPlot1} and Figure~\ref{fig:LimitContourPlot2}, near $(x,y)\approx(0,0)$ in diagnostic space, reflecting ratios close to unity that are numerical artefacts rather than physical. Combined with the low SNRs of the UV lines, with most UV metal-line measurements at $\mathrm{SNR}<2$, this results in diagnostic diagrams dominated by upper limits rather than secure detections, making candidate selection sensitive to measurement uncertainties. Importantly, these limitations do not apply to \textit{The Blueberry}, whose position in the UV diagnostic diagrams is determined entirely by robustly measured emission lines. For the remaining candidates, however, the available data introduce a fundamental ambiguity: without higher-resolution or higher-SNR spectroscopy, it is difficult to establish whether weak UV metal-line measurements are genuine or affected by spectroscopic noise. Their classifications should therefore be treated with caution and confirmed with higher-quality spectroscopy.

Additional uncertainty arises from the \citet{Rusta_2025} diagnostic framework itself. The theoretical grids rely on assumptions regarding Pop~III supernova yields and instantaneous, homogeneous metal mixing that are unlikely to capture the complexity of early galaxy evolution. Realistic turbulence, delayed enrichment, and stochastic star formation broaden the expected parameter space, increasing the overlap between Pop~III, AGN, and conventional star-forming galaxies. Furthermore, the short-lived nature of the polluted and hybrid phases~(~$\sim$1~--~5~Myr) makes capturing these transitions statistically improbable. Consequently, diagnostic regions should be considered to identify promising candidates rather than provide definitive classifications.

The resolved SED analysis is subject to a selection effect, as it is only applicable to galaxies with sufficient angular extent and spatial structure. Of the 84 He~II-emitting galaxies in our sample, 62 could be reliably analysed; the remainder were too compact, unresolved, or affected by diffraction features. Compact sources provide insufficient independent spatial information to recover reliable internal gradients in stellar age, dust attenuation, or star-formation activity, limiting our ability to distinguish localised populations from their host galaxies. This limitation is particularly relevant for extreme sources identified in the integrated analysis that could not be spatially resolved, including the compact Pop~III candidate ID~1130, the SED-selected AGN candidate ID~212506, and the galaxy with the bluest UV continuum slope, ID~13620. The resolved analysis should consequently be interpreted as representative of the spatially extended subset of the sample and not necessarily of the most compact or extreme systems.

A further limitation is that our resolved SED fitting does not explicitly include Pop~III stellar population templates. Instead, we use the BPASS stellar population models within \texttt{Bagpipes}, meaning that the resolved fits provide constraints on the age, dust attenuation, and star-formation properties of each region but do not directly measure its Pop~III contribution. This is particularly important because the broadband properties recovered by the fitting are not uniquely diagnostic of Pop~III populations, as discussed above. We therefore interpret the resolved SED results as identifying regions with properties consistent with young, extreme star formation rather than as independent evidence for Pop~III stars. Incorporating Pop~III-specific stellar population templates into the resolved SED fitting would provide a more direct test of the Pop~III contribution and is an important avenue for future work.

The inferred stellar population properties are also dependent on the assumed initial mass function (IMF), which is built into the stellar population synthesis models. Our BPASS-based fits adopt a broken-power-law IMF. This assumption may not be appropriate for a genuinely Pop~III-dominated region, for which a more top-heavy IMF is expected \citep{Bromm_2002}. A top-heavy IMF produces a larger proportion of massive stars per unit stellar mass and therefore more ionising and UV emission per unit mass. If such a population is instead fitted with a conventional IMF, the inferred stellar masses and star-formation rates may be systematically biased. The effect on broad-band colours is comparatively degenerate, meaning that the IMF assumption cannot readily be identified from the photometry alone. The stellar masses and star-formation rates derived from the resolved SED fitting should therefore be interpreted with this systematic uncertainty in mind, particularly for regions that could plausibly host a substantial Pop~III contribution.

\subsection{Implications for Future Pop~III Searches}

Our results highlight the importance of combining integrated emission-line diagnostics with spatially resolved analyses. The resolved SED fitting demonstrates that galaxy-integrated measurements can significantly dilute or entirely mask compact regions exhibiting much bluer UV continua, younger stellar populations, and lower dust attenuation than inferred from global photometry. Extending such resolved analyses to larger samples will require faster inference techniques, such as simulation-based inference (SBI), which could enable efficient pixel-level inference across large JWST surveys while allowing simultaneous comparison of Pop~III and enriched stellar-population models \citep{iglesiasnavarro2026_sbi}.

Future searches for Pop~III stars should therefore prioritise spatially resolved spectroscopy capable of isolating individual star-forming clumps, rather than relying solely on integrated galaxy spectra. Integral field spectroscopy with the \textit{JWST}/NIRSpec IFU represents a critical next step, as it enables simultaneous mapping of nebular emission, ionisation structure, and chemical enrichment across entire galaxies. This approach mitigates slit-loss effects and reduces the dilution of compact Pop~III-like star-forming regions by the integrated light of more evolved stellar populations, although limitations from spatial resolution and surface-brightness sensitivity will remain \citep{Maiolino_2026}.

Given that $\mathrm{He~II}$ is a non-unique tracer, identifying "polluted" Pop~III candidates requires high-resolution spectroscopy to detect UV metal lines, alongside expanded UV coverage and constraints on the ionising photon production efficiency ($\xi_{\mathrm{ion}}$). More generally, progress in this field will require extending searches to $z>10$, where Pop~III star formation is more likely to persist in minimally enriched environments. Gravitational lensing offers a complementary avenue for accessing intrinsically faint systems, with cluster lensing surveys such as GLIMPSE and VENUS providing magnification of compact star-forming regions and their potential supernovae to observable flux levels \citep{Windhorst_2019, atek_2025, allingham_2026}. Together with spatially resolved spectroscopy, these approaches offer the most promising pathway toward detecting surviving Pop~III star formation and distinguishing it from chemically enriched but spectrally extreme high-redshift galaxies.

Improved metallicity constraints will also be essential. Current strong-line calibrations remain uncertain in the low-metallicity, high-ionisation regime occupied by many high-redshift galaxies \citep{Scholte2025_metallicities, Sanders_2024}. Although recent frameworks incorporate $U$-dependence \citep{Nakajima_2022_metallicities} or employ $U$-insensitive diagnostics such as $\hat{R}$ \citep{Laseter_2024}, these rely on specific line ratios. In the absence of $R_2$ or $R_3$, $\hat{R}$ cannot be robustly constrained, rendering inferred metallicities implicitly dependent on $U$ and prone to systematic error. Addressing these limitations will require deeper spectroscopy from future JWST cycles and next-generation 30–40 m class ground-based facilities. The Extremely Large Telescope (ELT) will be particularly valuable for spatially resolved studies of high-redshift star-forming regions and star clusters, extending the spatial sensitivity of such analyses beyond what is currently achievable with JWST \citep{MICADO,Padovani_2023}. Targeting faint auroral lines like $[\mathrm{O\,III}]\,\lambda4363$ will enable direct electron temperature measurements, providing robust constraints on $U$ and confirming metallicities.

Ultimately, confirming the presence of Pop~III stars may require complementary observational strategies beyond He~II diagnostics, such as detecting pair-instability supernovae (PISNe) rather than relying on degenerate spectral signatures. Non-rotating Pop~III stars with masses $140$--$260~M_{\odot}$ are predicted to explode as highly energetic thermonuclear events, producing large metal yields and driving early chemical enrichment \citep{Chen_2014, Moriya_2019, Ventura2024}. Their extreme luminosities and extended light curves make them, in principle, detectable at $z>10$ with next-generation near-infrared facilities such as JWST, Roman, and Euclid. PISNe are distinguished by slow-evolving, multi-year observer-frame light curves and relatively red late-time colours. However, their rates remain highly uncertain and likely very low, depending on the Pop~III IMF and early enrichment efficiency \citep{Moriya_2019, Venditti_2024}. While a few candidates have been proposed, complementary evidence from stellar archaeology, such as PISNe-like abundance patterns in a $z=7.54$ quasar \citep{Yoshii2022}, offers an independent route to confirming Pop~III stars through their explosive endpoints.

\section{Summary} \label{sec:Summary}

We have presented a multi-diagnostic search for Population~III (Pop~III) stellar populations in 84 strong He~II-emitting galaxies at $z\sim3$--7 in the JWST JADES GOODS-S and GOODS-N fields. By combining UV emission-line diagnostics, gas-phase metallicities, AGN screening, and spatially resolved SED fitting, we assess the extent to which these systems exhibit signatures consistent with having a least some Pop~III star formation. Our main conclusions are:

\begin{itemize}

\item \textbf{We find no evidence for purely Pop~III He~II emitters.} None of the strong He~II emitters shows the combination of properties expected for a globally pristine Pop~III system. Instead, the systems identified as candidates have enriched gas and are therefore more consistent with the self-polluted or hybrid phases targeted by this study.

\item \textbf{Strong He~II emission alone is not a unique tracer of Pop~III stars or low metallicity.} The He~II-emitting sample spans a broad range of UV slopes, luminosities, and metallicities, demonstrating that strong He~II can arise in chemically diverse galaxies and may be powered by multiple hard-ionising sources.

\item \textbf{Twelve galaxies satisfy the adopted Pop~III candidate criterion, but most classifications remain uncertain.} Their positions in the \citet{Rusta_2025} diagnostic diagrams frequently depend on low-SNR UV metal-line measurements or upper limits, with several sources also overlapping AGN or conventional star-forming model predictions. The inferred metallicities are substantially higher than expected for genuinely pristine Pop~III systems.

\item \textbf{\textit{The Blueberry} (EPOCHS-DR2 ID~13176, $z=5.94$) is the strongest Pop~III candidate in our sample.} It is the only candidate whose classification is supported entirely by robust detections of all relevant UV emission lines, without reliance on upper limits. Its spatially resolved stellar populations are consistently very young ($\lesssim5.4$~Myr), blue, weakly attenuated, and characterised by elevated star formation rate surface densities, indicating an intense and recent burst of star formation. Its integrated metallicity, $\log(Z/Z_{\odot})=-1.18\pm0.05$, rules out a globally pristine origin. These properties are consistent with the physical conditions expected during the self-polluted or early hybrid stages of Pop~III evolution, although they are not unique to metal-free stars and can also occur in chemically enriched starbursts.

\item \textbf{Resolved SED fitting reveals that extreme star formation can be spatially localised within otherwise enriched galaxies.} In particular, \textit{The Carrot} (EPOCHS-DR2 ID~13577, $z=5.57$) contains a compact central region that is substantially younger, bluer, and less attenuated than the galaxy as a whole, despite its globally enriched properties. This demonstrates that galaxy-integrated measurements can dilute the signatures of extreme star-forming regions and highlights the value of spatially resolved analysis in future Pop~III searches.

\item \textbf{AGN contamination is limited but remains an important source of uncertainty.} Morphological and SED-based screening identifies only a small number of sources with strong evidence for AGN activity, with ID~7384 providing the clearest case. However, weak or obscured AGN cannot be completely excluded, reinforcing the need for multiple independent diagnostics.

\end{itemize}

Overall, our results provide no definitive evidence for a globally pristine Pop~III galaxy among strong He emitters, but identify \textit{The Blueberry} as a candidate for a surviving Pop~III population within an enriched host. Distinguishing genuine primordial populations from chemically enriched, high-ionisation systems will require combining robust UV emission-line diagnostics with spatially resolved stellar-population analyses and high-resolution spectroscopy. Extending this framework to larger samples, incorporating Pop~III-specific templates into resolved SED fitting, and pursuing integral field spectroscopy will be key to building a more complete census of surviving Pop~III signatures and clarifying their contribution to early galaxy formation.

\section*{Acknowledgements}
We acknowledge support from the ERC Advanced Investigator Grant EPOCHS (788113), as well as two studentships from the STFC.    We thank James Trussler for enlightening discussions on the topics in this paper and Leonardo Ferreira and Fabricio Ferrari for assistance and help with the Morfometryka code. This work is based on observations made with the NASA/ESA \textit{Hubble Space Telescope} (HST) and NASA/ESA/CSA \textit{James Webb Space Telescope} (JWST) obtained from the \texttt{Mikulski Archive for Space Telescopes} (\texttt{MAST}) at the \textit{Space Telescope Science Institute} (STScI), which is operated by the Association of Universities for Research in Astronomy, Inc., under NASA contract NAS 5-03127 for JWST, and NAS 5–26555 for HST. The authors thank all involved with the construction and operation of JWST, without whom this work would not be possible.

Some of the data products presented herein were retrieved from the Dawn JWST Archive (DJA). The DJA is an initiative of the Cosmic Dawn Center, which is funded by the Danish National Research Foundation under Grant No. 140. The authors thank all those involved in the construction and operation of the telescope, as well as those who designed and executed the observations used in this work.
This work makes use of \texttt{Astropy} \citep{Astropy_2013, Astropy_2018, Astropy_2022}, \texttt{matplotlib} \citep{Hunter_2007_matplotlib}, \texttt{SciPy} \citep{SciPy_2020}, \texttt{Photutils} \citep{Bradley_2026_photutils}, \texttt{specfitmsa} \citep{Rusakov2026}, \texttt{JAX} \citep{jax2018github}, \texttt{PROSPECT} \citep{Robotham_2020_Prospect}, \texttt{Morfometryka} \citep{Ferrari_2015}, \texttt{SExtractor} \citep{SExtractor_1,SExtractor_2}, \texttt{piXedfit} \citep{Abdurro_pixedfit_2021, Abdurro_2023} and \texttt{EXPANSE} \citep{EXPANSE}.

%%%%%%%%%%%%%%%%%%%%%%%%%%%%%%%%%%%%%%%%%%%%%%%%%%
\section*{Data Availability}\label{sec:Data Availability}

The resolved SED fitting overview plots generated using the \texttt{EXPANSE}
pipeline for all 62 resolved He~II-emitting galaxies analysed in this work
are publicly available in the accompanying GitHub repository
(\url{https://github.com/rooosaaa/popiii-jades-heii}) and are archived
on Zenodo (\url{https://doi.org/10.5281/zenodo.22287104}). The repository
contains the resolved SED fitting overview plots for both the Pop~III
candidate galaxies and the full sample, including the resolved property maps and star formation histories for each galaxy.



\bibliographystyle{mnras}
\bibliography{main,austind} % if your bibtex file is called example.bib

%\begin{thebibliography}{99}
%\end{thebibliography}

%%%%%%%%%%%%%%%%%%%%%%%%%%%%%%%%%%%%%%%%%%%%%%%%%%

%%%%%%%%%%%%%%%%% APPENDICES %%%%%%%%%%%%%%%%%%%%%

\appendix
\label{appendix}

\section{Pop III Candidate Selection and Emission-Line Diagnostics}
\label{sec:Pop III Candidate Selection and Emission-Line Diagnostics}

\begin{figure*}
\centering
\includegraphics[width=\textwidth]{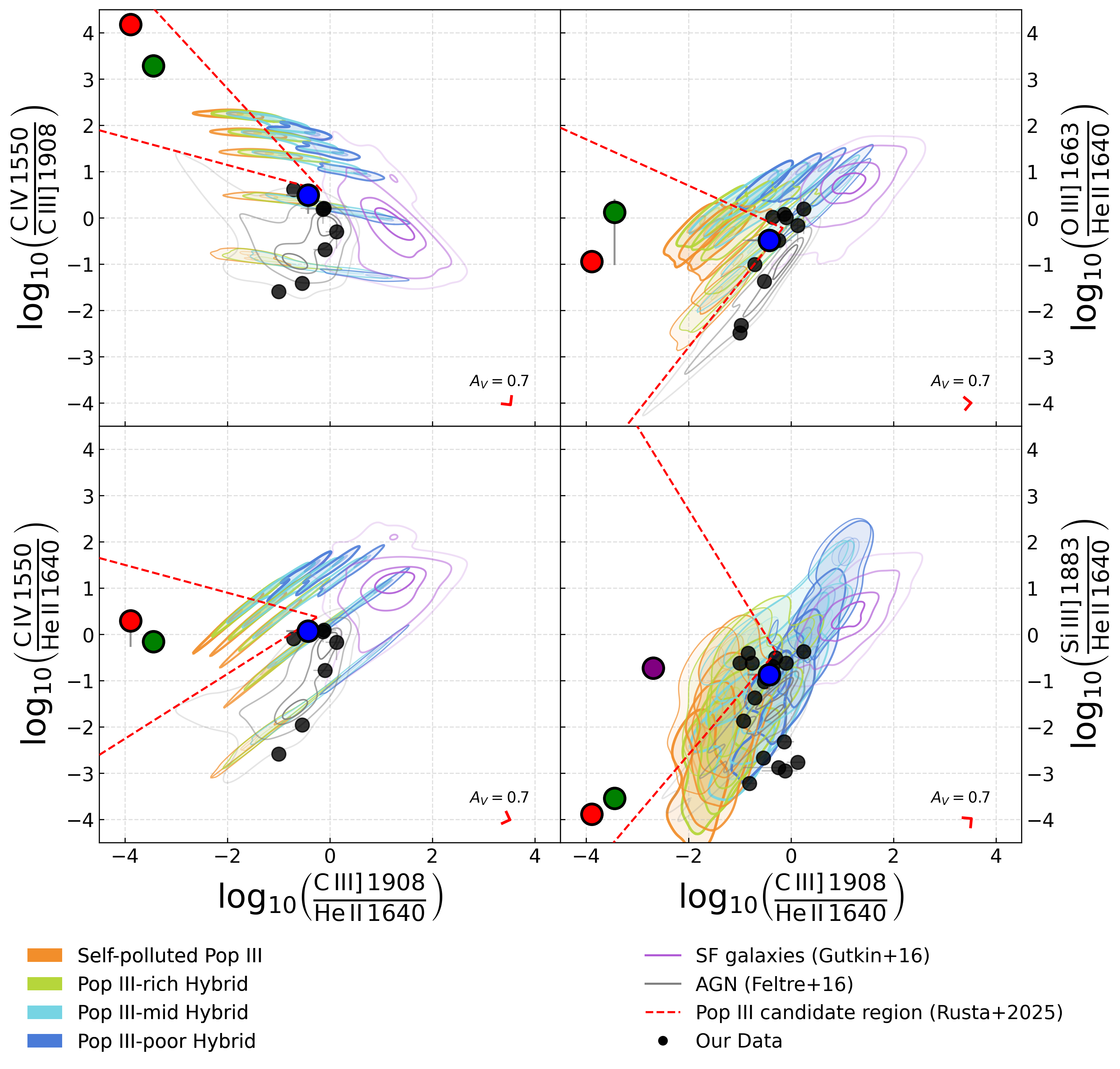}
\caption{UV emission-line diagnostic diagrams using He~II $\lambda1640$ showing our sample in comparison with Pop~III model predictions from \citet{Rusta_2025}. Model density distributions are illustrated with coloured contours, whose thickness increases with ionisation parameter ($\log U = -3, -2, -1, -0.5, 0$). The empty contours represent AGN (grey) and star-forming galaxy (purple) models from \citet{Feltre_2016} and \citet{Gutkin_2016}, respectively. Red dashed lines delineate the region where Pop~III stars are predicted to contribute more than 25\% of the total stellar mass. Black points indicate our observed emission-line ratios, with grey lines showing $1\sigma$ uncertainties. Galaxies flagged as potential Pop~III candidates are highlighted with larger red, green, blue and purple circles. While error bars are plotted for all points in grey, some are too small to be visible. Red arrows in the corner of each panel indicate the effect of dust attenuation assuming $A_V = 0.7$.}
\label{fig:RawContourPlot1}
\end{figure*}

\begin{figure*}
\centering
\includegraphics[width=\textwidth]{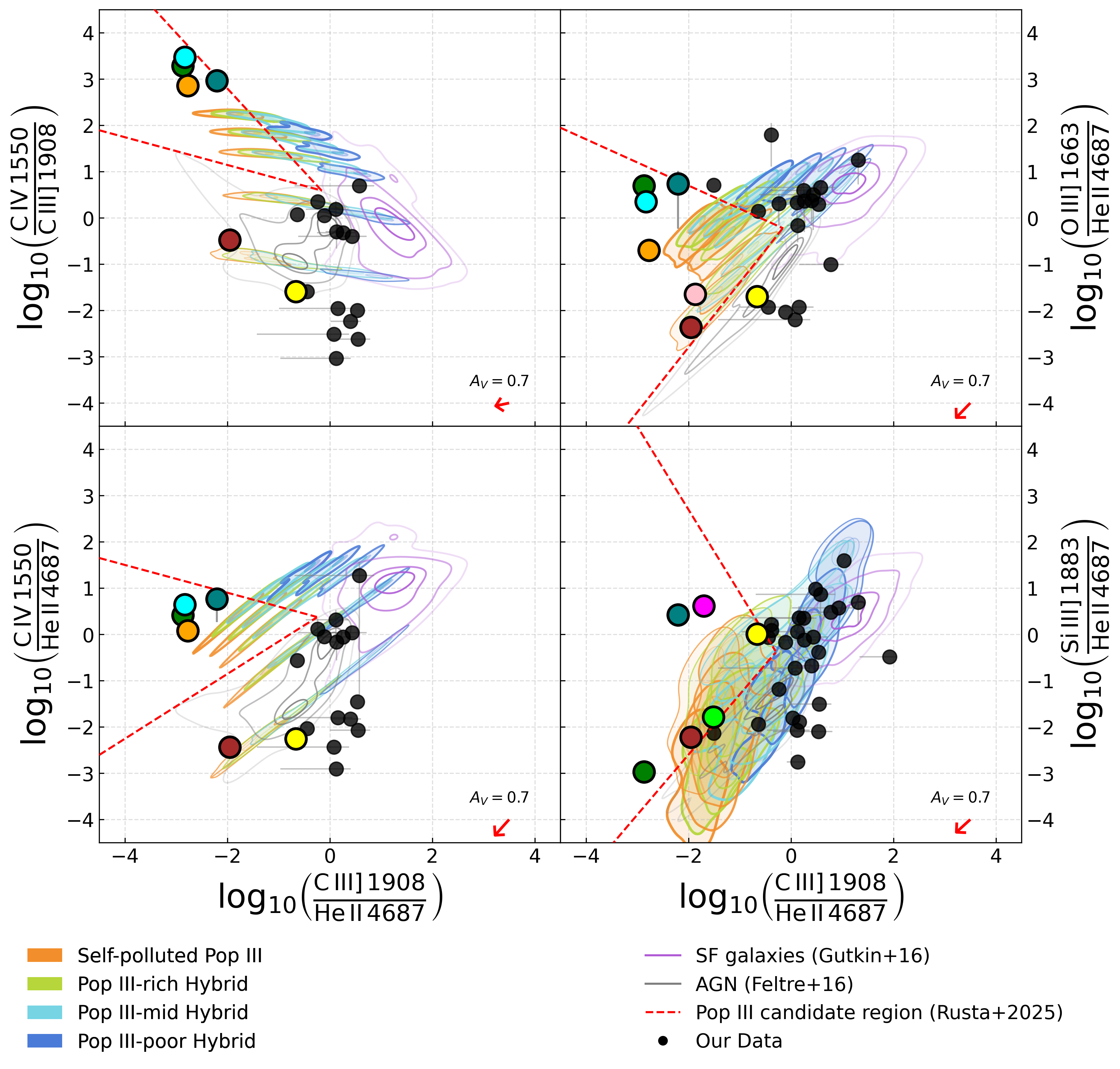}
\caption{UV emission-line diagnostic diagrams using He~II $\lambda4687$ showing our sample in comparison with Pop~III model predictions from \citet{Rusta_2025}. Model density distributions are illustrated with coloured contours, whose thickness increases with ionisation parameter ($\log U = -3, -2, -1, -0.5, 0$). The empty contours represent AGN (grey) and star-forming galaxy (purple) models from \citet{Feltre_2016} and \citet{Gutkin_2016}, respectively. Red dashed lines delineate the region where Pop~III stars are predicted to contribute more than 25\% of the total stellar mass. Black points indicate our observed emission-line ratios, with grey lines showing $1\sigma$ uncertainties. Galaxies flagged as potential Pop~III candidates are shown with larger green, teal, cyan, and orange circles. Brown and yellow circles indicate objects consistently lying within the AGN and polluted/hybrid contours and are selected for further analysis. Pink, magenta, and lime circles denote sources with only 2 ratios measured, but which fall within the candidate region and are therefore also included for further analysis. While error bars are plotted for all points in grey, some are too small to be visible. Red arrows in the corner of each panel indicate the effect of dust attenuation assuming $A_V = 0.7$.}
\label{fig:RawContourPlot2}
\end{figure*}

This appendix presents the UV emission-line diagnostic diagrams underlying the candidate selection in Section~\ref{sec:Pop III Candidates}, showing the nominal line-ratio positions of the selected candidates without the $3\sigma$ upper-limit treatment applied to low-SNR metal lines in the main text. Comparing the nominal diagrams here (Figures~\ref{fig:RawContourPlot1} and \ref{fig:RawContourPlot2}) with their limit-corrected counterparts (Figures~\ref{fig:LimitContourPlot1} and \ref{fig:LimitContourPlot2}) illustrates how sensitive the resulting classifications are to the treatment of non-detections.

We retain twelve galaxies for further investigation based on their location within the Pop~III model region across the UV diagnostic diagrams. The strength of this selection varies between sources. Four galaxies (IDs~9598, 126865, 210262, and 9564) satisfy the candidate criterion in only a single diagnostic plane, but are retained because their line ratios nevertheless place them within the predicted Pop~III region and warrant further independent assessment.

IDs~61712 and 28746 are similarly selected by a single diagnostic, specifically $\log_{10}(\mathrm{Si~III]} \lambda1883/\mathrm{He~II} \lambda4687)$. However, both also overlap the AGN model contours, making their interpretation ambiguous. ID~28746 additionally lies within the self-polluted Pop~III model region. These sources are therefore retained as candidates, but require further constraints to distinguish between stellar and AGN-powered ionisation.

The remaining candidates show more consistent locations within the Pop~III model region across the available diagnostic planes and are therefore prioritised for further investigation. In all cases, the candidate classification is treated as an indication for follow-up rather than definitive evidence for Pop~III stars.

\section{Spectroscopic and Imaging Properties of Selected Galaxies}\label{sec:Spectroscopic and Imaging Properties of Selected Galaxies}

Table~\ref{tab:line_fluxes_snrs} lists the \texttt{specFitMSA} fluxes and
SNRs (Section~\ref{sec:Emission Line Measurements}) for He~II~$\lambda1640$, He~II~$\lambda4687$, and H$\alpha$~$\lambda6565$ for the Pop~III candidates and the AGN candidate ID~7384 (Section~\ref{sec:Morphological AGN Candidates}), included here for comparison, together with the corresponding Balmer decrements and inferred $A_V$. These complement the coordinates, redshifts, and derived UV/metallicity properties listed in Table~\ref{tab:selected_galaxies}.

\begin{figure*}
\centering
\includegraphics[
width=\textwidth,
height=\textheight,
keepaspectratio
]{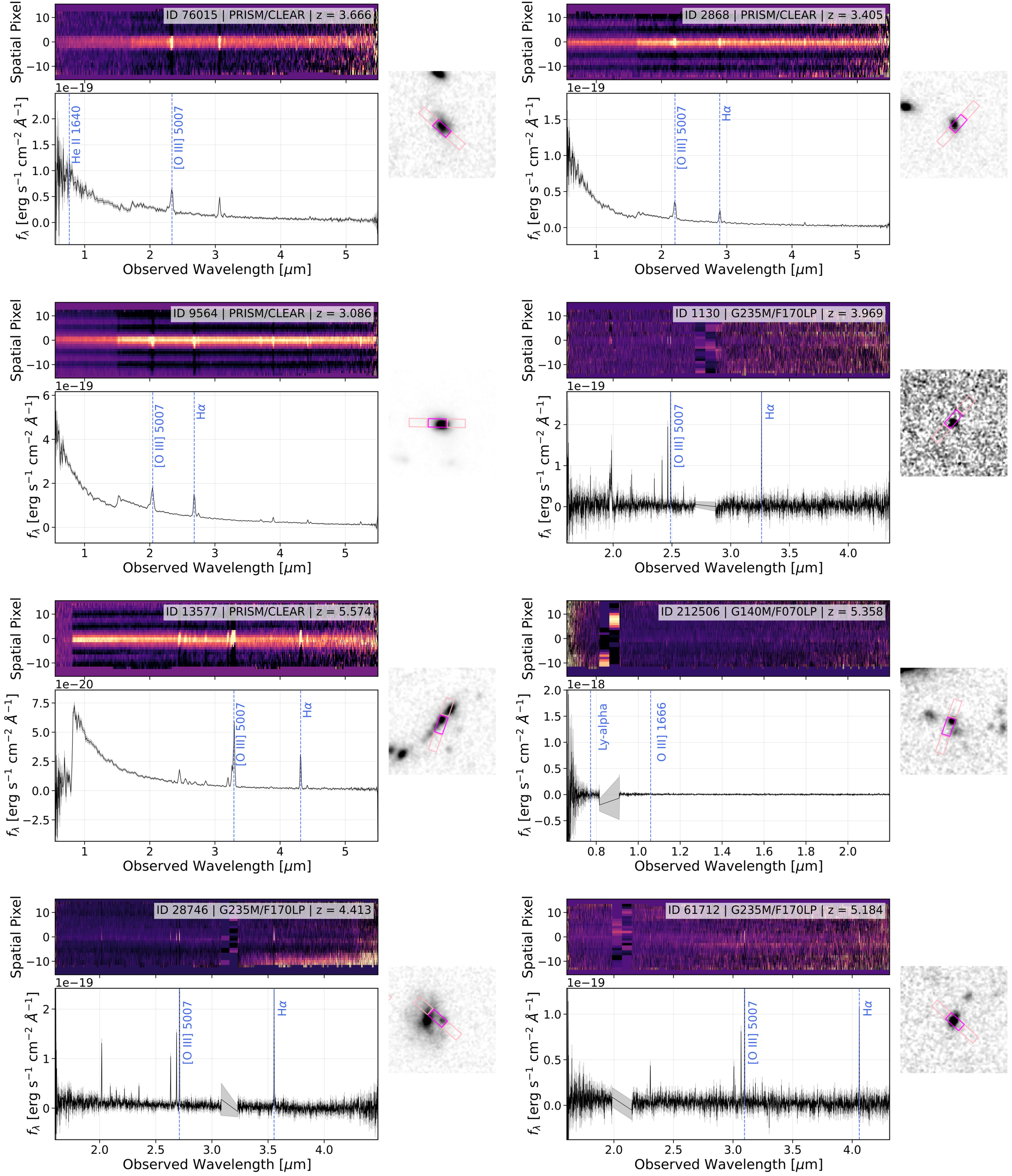}
\caption{Combined spectroscopic and imaging diagnostics for eight selected galaxies from the He~II-emitting sample. For each galaxy, the \textit{top left} panel shows the two-dimensional \textit{JWST}/NIRSpec spectrum and the \textit{bottom left} panel shows the corresponding extracted one-dimensional spectrum in $\mathrm{ergs^{-1}cm^{-2}\mathring{A}^{-1}}$, both obtained with the grating/filter combination used to measure the He~II line. The \textit{right} panel shows the corresponding \textit{JWST}/NIRCam F444W imaging cutout with the NIRSpec MSA slitlet configuration overlaid. The spectroscopic source ID, grating/filter configuration, and redshift are indicated in each panel.}
\label{fig:full_spectra_1}
\end{figure*}

\begin{figure*}
\centering
\includegraphics[
width=\textwidth,
height=\textheight,
keepaspectratio
]{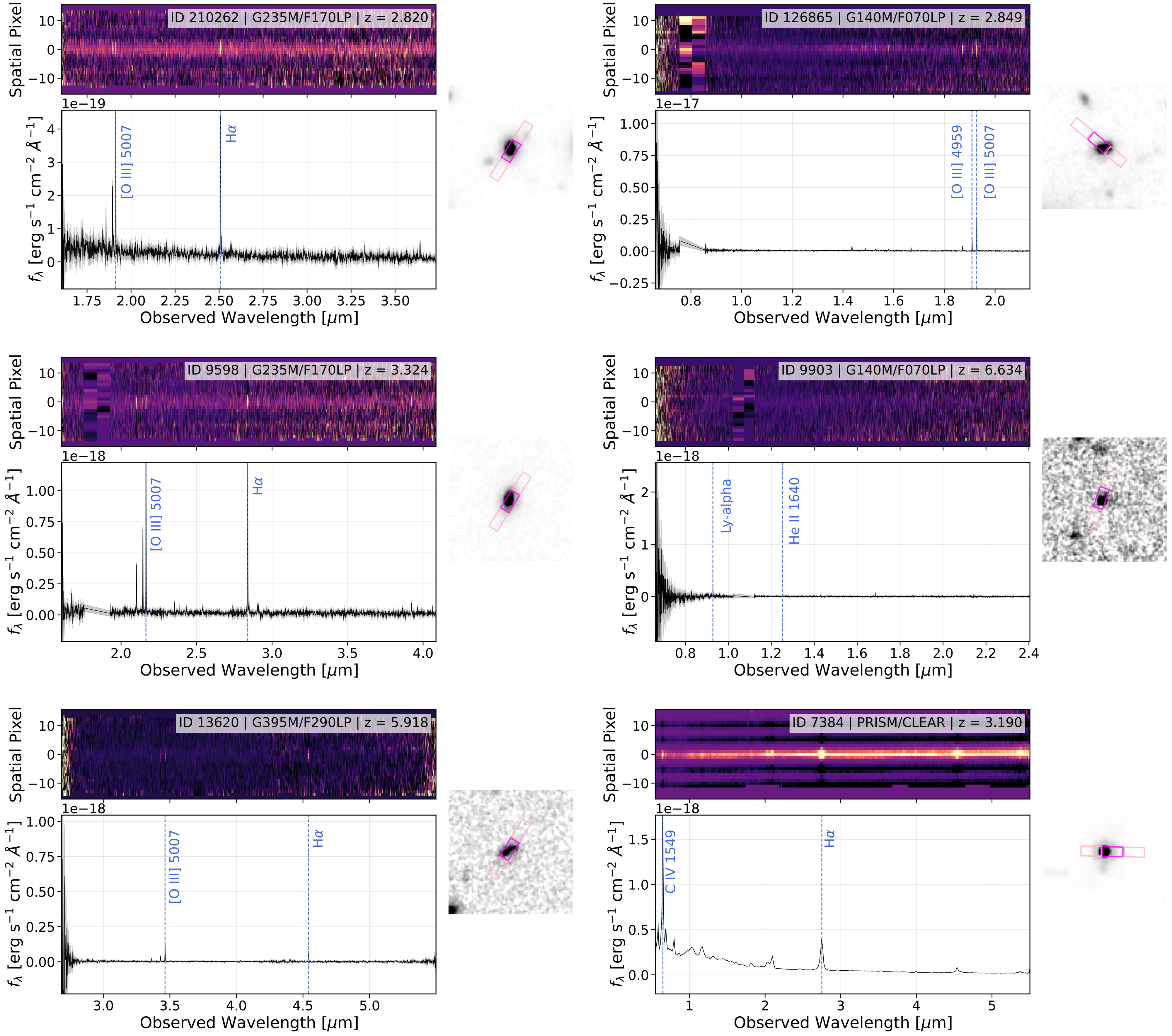}
\caption{Combined spectroscopic and imaging diagnostics for six additional selected galaxies from the He~II-emitting sample. For each galaxy, the \textit{top left} panel shows the two-dimensional \textit{JWST}/NIRSpec spectrum and the \textit{bottom left} panel shows the corresponding extracted one-dimensional spectrum in $\mathrm{ergs^{-1}cm^{-2}\mathring{A}^{-1}}$, both obtained with the grating/filter combination used to measure the He~II line. The \textit{right} panel shows the corresponding \textit{JWST}/NIRCam F444W imaging cutout with the NIRSpec MSA slitlet configuration overlaid. The spectroscopic source ID, grating/filter configuration, and redshift are indicated in each panel.}
\label{fig:full_spectra_2}
\end{figure*}

\begin{table*}
\centering
\caption{Emission-line fluxes, signal-to-noise ratios, Balmer decrements, and inferred dust attenuation for the Pop~III candidates and additional selected galaxies discussed in this work. The final column lists previous literature references reporting or discussing each source, keyed to the numbering scheme below; sources with no prior reference are marked with a dash.}
\label{tab:line_fluxes_snrs}
\renewcommand{\arraystretch}{1.8}
\setlength{\tabcolsep}{5pt}
\resizebox{\textwidth}{!}{%
\begin{tabular}{cccccccccc}
\toprule
\texttt{EPOCHS-DR2} ID &
\begin{tabular}{c}
$F_{\rm 1640}$\\
$(10^{-20}\mathrm{erg\,s^{-1}\,cm^{-2}})$
\end{tabular} &
SNR$_{\rm 1640}$ &
\begin{tabular}{c}
$F_{\rm 4687}$\\
$(10^{-20}\mathrm{erg\,s^{-1}\,cm^{-2}})$
\end{tabular} &
SNR$_{\rm 4687}$ &
\begin{tabular}{c}
$F_{\rm 6565}$\\
$(10^{-20}\mathrm{erg\,s^{-1}\,cm^{-2}})$
\end{tabular} &
SNR$_{\rm 6565}$ &
\begin{tabular}{c}
H$\alpha$/H$\beta$
\end{tabular} &
\begin{tabular}{c}
$A_{V}$\\
(mag)
\end{tabular} &
Reference \\
\midrule
13176 (\textit{The Blueberry}) & $162.75^{+26.04}_{-25.50}$ & 6.89 & -- & -- & $604.41^{+3.16}_{-3.16}$ & 198.04 & $2.99 \pm 0.07$ & $0.15 \pm 0.08$ & (1), (2), (3), (4) \\
13577 (\textit{The Carrot}) & -- & -- & $50.20^{+26.88}_{-25.88}$ & 2.36 & $499.52^{+12.55}_{-12.55}$ & 42.08 & $2.64 \pm 0.27$ & $0.00^{*}$ & (2), (4) \\
76015 & $901.51^{+278.27}_{-278.27}$ & 4.16 & -- & -- & $761.70^{+26.42}_{-26.42}$ & 30.84 & $4.53 \pm 0.70$ & $1.60 \pm 0.53$ & -- \\
2868 & $344.59^{+106.83}_{-106.83}$ & 4.35 & $89.65^{+65.51}_{-65.51}$ & 2.55 & $337.75^{+14.04}_{-14.04}$ & 24.68 & $2.98 \pm 0.37$ & $0.14 \pm 0.44$ & -- \\
9564 & $853.30^{+248.67}_{-248.67}$ & 4.07 & -- & -- & $2757.12^{+48.03}_{-48.03}$ & 62.56 & $4.74 \pm 0.30$ & $1.75 \pm 0.22$ & -- \\
212506 & -- & -- & $3.68^{+1.76}_{-1.76}$ & 2.25 & $156.87^{+1.37}_{-1.37}$ & 116.99 & $2.56 \pm 0.07$ & $0.00^{*}$ & (2) \\
28746 & -- & -- & $29.68^{+16.66}_{-16.66}$ & 2.15 & $789.23^{+11.96}_{-11.97}$ & 67.62 & $2.54 \pm 0.14$ & $0.00^{*}$ & -- \\
61712 & -- & -- & $18.88^{+41.48}_{-41.48}$ & 2.05 & $236.84^{+10.96}_{-10.96}$ & 25.16 & $2.80 \pm 0.27$ & $0.00^{*}$ & -- \\
210262 & -- & -- & $78.08^{+41.95}_{-41.95}$ & 2.19 & $908.47^{+26.18}_{-26.18}$ & 35.38 & $5.04 \pm 0.85$ & $1.96 \pm 0.58$ & -- \\
126865 & -- & -- & $87.48^{+46.07}_{-46.07}$ & 2.26 & $1734.63^{+24.58}_{-24.58}$ & 71.39 & $2.96 \pm 0.16$ & $0.11 \pm 0.19$ & -- \\
9598 & -- & -- & $81.06^{+28.62}_{-28.62}$ & 2.95 & $3895.45^{+62.19}_{-62.19}$ & 64.31 & $4.83 \pm 0.24$ & $1.82 \pm 0.17$ & -- \\
1130 & -- & -- & $57.43^{+27.87}_{-27.87}$ & 2.37 & $566.07^{+21.99}_{-21.99}$ & 26.58 & $2.70 \pm 0.27$ & $0.00^{*}$ & -- \\
7384 & $2558.68^{+164.85}_{-164.85}$ & 25.47 & -- & -- & $3479.55^{+36.92}_{-36.92}$ & 106.79 & $5.57 \pm 0.45$ & $2.31 \pm 0.28$ & -- \\
\bottomrule
\end{tabular}%
}

\medskip
\noindent {\footnotesize \textbf{Note.} $^{*}$Negative $A_V$ floored to zero; the correspondingly low Balmer decrement may reflect non-standard Case~B conditions rather than zero dust attenuation \citep{McClymont_2025}. We reference works that have previously reported these galaxies following the numbering system: (1)~\citet{Witstok_2024}; (2)~\citet{Simmonds_2024}; (3)~\citet{Saxena_2024}; (4)~\citet{Bunker_2024}.}
\end{table*}

\begin{table*}
\centering
\caption{Coordinates, redshifts, gas-phase metallicity, UV continuum slope, and UV absolute magnitude for the selected galaxies discussed in this work. R.A./Decl. are given to six decimal places, $z_{\rm spec}$ and $z_{\rm phot}$ to four, and $\log(Z/Z_\odot)$, $\beta$, and $M_{\rm UV}$ to two; uncertainties are $1\sigma$.}
\label{tab:selected_galaxies}
\renewcommand{\arraystretch}{1.8}
\setlength{\tabcolsep}{16pt}
\resizebox{\textwidth}{!}{%
\begin{tabular}{cccccccc}
\toprule
\texttt{EPOCHS-DR2} ID & R.A. & Decl. & $z_{\rm spec}$ & $z_{\rm phot}$ &
$\log(Z/Z_{\odot})$ & $\beta$ & $M_{\rm UV}$ \\
\midrule
13176 (\textit{The Blueberry}) & 53.121759 & -27.797632 & 5.9355 & 5.9449 & 
$-1.18^{+0.05}_{-0.05}$ & $-2.15 \pm 0.10$ & $-19.85 \pm 0.01$ \\
13577 (\textit{The Carrot}) & 53.130010 & -27.778402 & 5.5738 & 5.7973 & 
$-0.51^{+0.10}_{-0.10}$ & $-2.08 \pm 0.12$ & $-20.52 \pm 0.02$ \\
76015 & 189.281436 & 62.190315 & 3.6661 & 3.7363 & $-0.43^{+0.08}_{-0.09}$ & $-1.37 \pm 0.34$ & $-19.82 \pm 0.08$ \\
2868 & 189.073987 & 62.254812 & 3.4052 & 3.4251 & $-1.14^{+0.12}_{-0.10}$ & $-2.05 \pm 0.16$ & $-19.48 \pm 0.04$ \\
9564 & 53.152592 & -27.793914 & 3.0858 & 3.0922 & $-1.21^{+0.25}_{-0.18}$ & $-1.64 \pm 0.10$ & $-20.85 \pm 0.02$ \\
212506 & 53.155842 & -27.766717 & 5.3540 & 5.3588 & $-1.25^{+0.21}_{-0.16}$ & $-1.91 \pm 0.23$ & $-18.09 \pm 0.04$ \\
28746 & 189.176078 & 62.256325 & 4.4149 &  4.5595 & $-0.83^{+0.66}_{-0.08}$ & $-1.87 \pm 0.20$ & $-19.71 \pm 0.05$ \\
61712 & 189.221208 & 62.216524 & 5.1858 & 5.1400 & $-0.48^{+0.13}_{-0.17}$ & $-2.14 \pm 0.33$ & $-19.29 \pm 0.07$ \\
210262 & 53.195942 & -27.772757 & 2.8189 & 2.8643 &  $-0.42^{+0.09}_{-0.10}$ & $-1.33 \pm 0.31$ & $-18.98 \pm 0.10$ \\
126865 & 53.195391 & -27.779350 & 2.8516 & 2.7579 & $-0.74^{+0.13}_{-0.09}$ & $-2.14 \pm 0.23$ & $-19.70 \pm 0.06$ \\
9598 & 53.161808 & -27.770717 & 3.3263 & 3.3289 & $-0.44^{+0.10}_{-0.14}$ & $-1.36 \pm 0.43$ & $-19.32 \pm 0.13$ \\
1130 & 189.096287 & 62.282485 & 3.9729 & 3.9272 & $-1.31^{+0.21}_{-0.16}$ & $-2.11 \pm 0.34$ & $-19.10 \pm 0.07$ \\
7384 & 53.178503 & -27.784107 & 3.1905 & 0.5539 & $-0.79^{+0.09}_{-0.07}$ & $-0.37 \pm 0.04$ & $-21.61 \pm 0.01$ \\
\bottomrule
\end{tabular}%
}
\end{table*}

\section{Emission-Line Fits for \textit{The Blueberry}}\label{sec:Emission-Line Fits for The Blueberry}

\begin{figure*}
    \centering

    \includegraphics[width=0.48\textwidth]{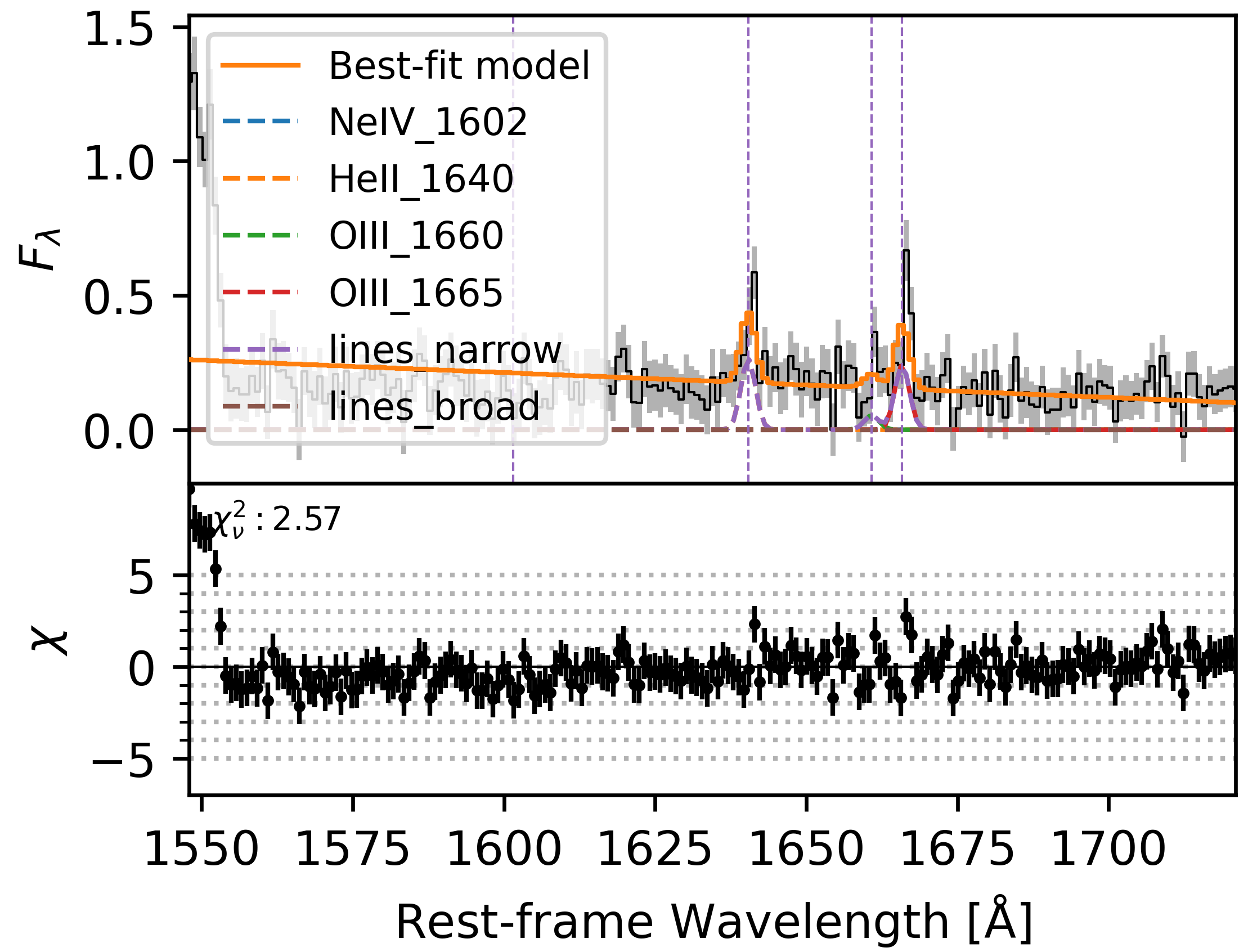}
    \includegraphics[width=0.48\textwidth]{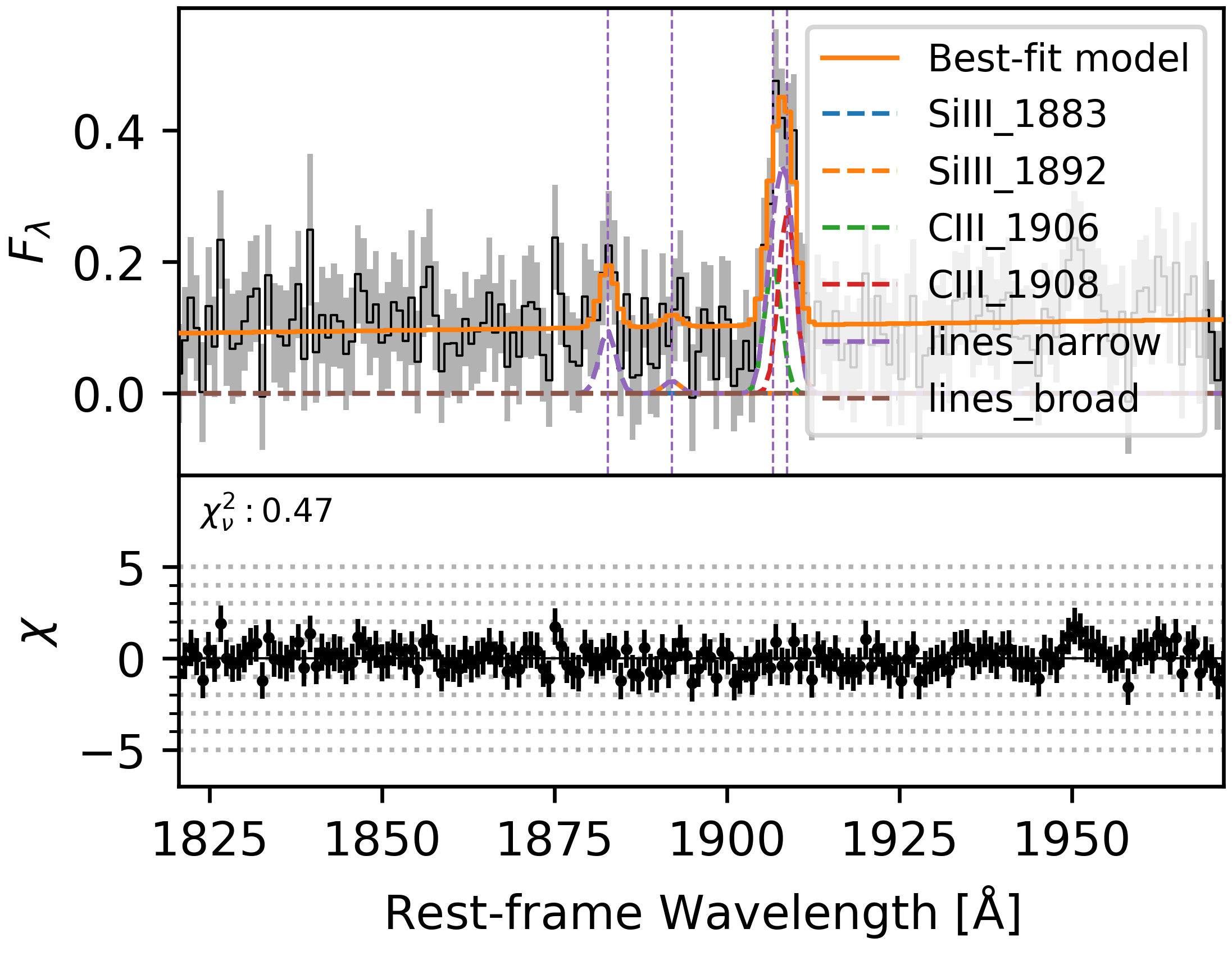}

    \vspace{0.5em}

    \includegraphics[width=0.48\textwidth]{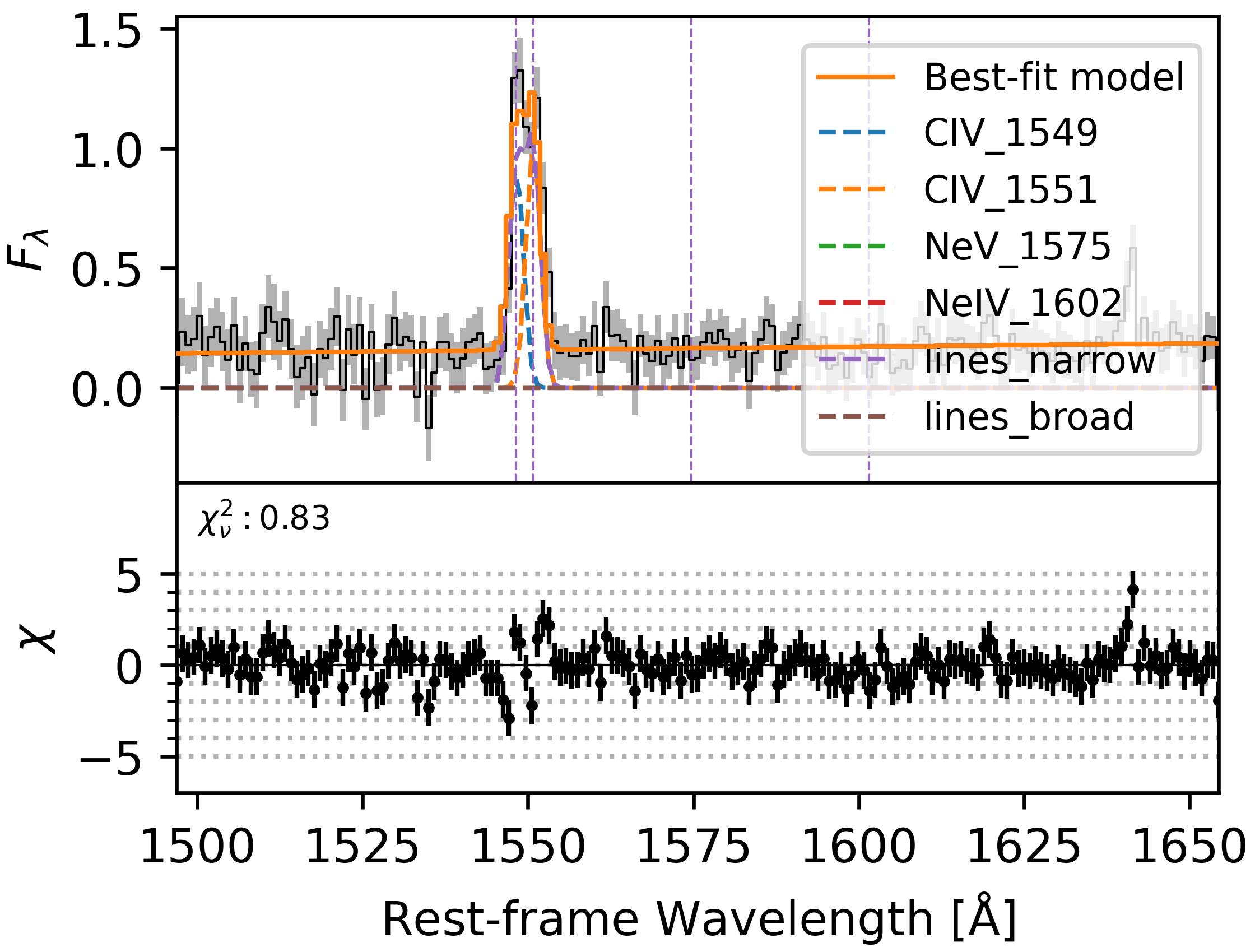}

    \caption{Our \texttt{specFitMSA} emission-line fits for \textit{The Blueberry} obtained from the G140M/F070LP spectrum. Observed fluxes are shown in black with $1\sigma$ uncertainties in grey, while the best-fitting model is shown in orange with its 68\% highest-density interval. Individual Gaussian components are indicated by coloured dashed curves, and vertical lines mark the corresponding rest-frame transitions. Narrow and broad components are distinguished by their respective colours. The lower panels show the residuals normalised by the flux uncertainties. The fits illustrate the strong C~III]~$\lambda1908$, C~IV~$\lambda1551$, Si~III]~$\lambda1883$, O~III]~$\lambda1663$, and He~II~$\lambda1640$ emission used in the UV diagnostic analysis.}
    \label{fig:13176_fits}
\end{figure*}

\textit{The Blueberry} provides the clearest spectroscopic evidence for a potential Pop~III contribution among the candidates identified by our emission-line diagnostics. It is the only source whose observed line ratios place it within the Pop~III candidate region of Figure~\ref{fig:LimitContourPlot1} using exclusively measured line fluxes, without reliance on upper limits. Figure~\ref{fig:13176_fits} presents the \texttt{specFitMSA} fits to the UV emission lines used in the Pop~III diagnostics. The fitted profiles allow the quality of these measurements and, where required, the decomposition into individual Gaussian components to be inspected directly. Together, these lines provide the observational basis for \textit{The Blueberry}'s location within the Pop~III candidate region.

\section{Sources Lacking Strong-Line Metallicity Measurements}\label{sec:Oxygen Spectra}

\begin{figure*}
\centering
\includegraphics[
    width=\textwidth,
    height=\textheight,
    keepaspectratio
]{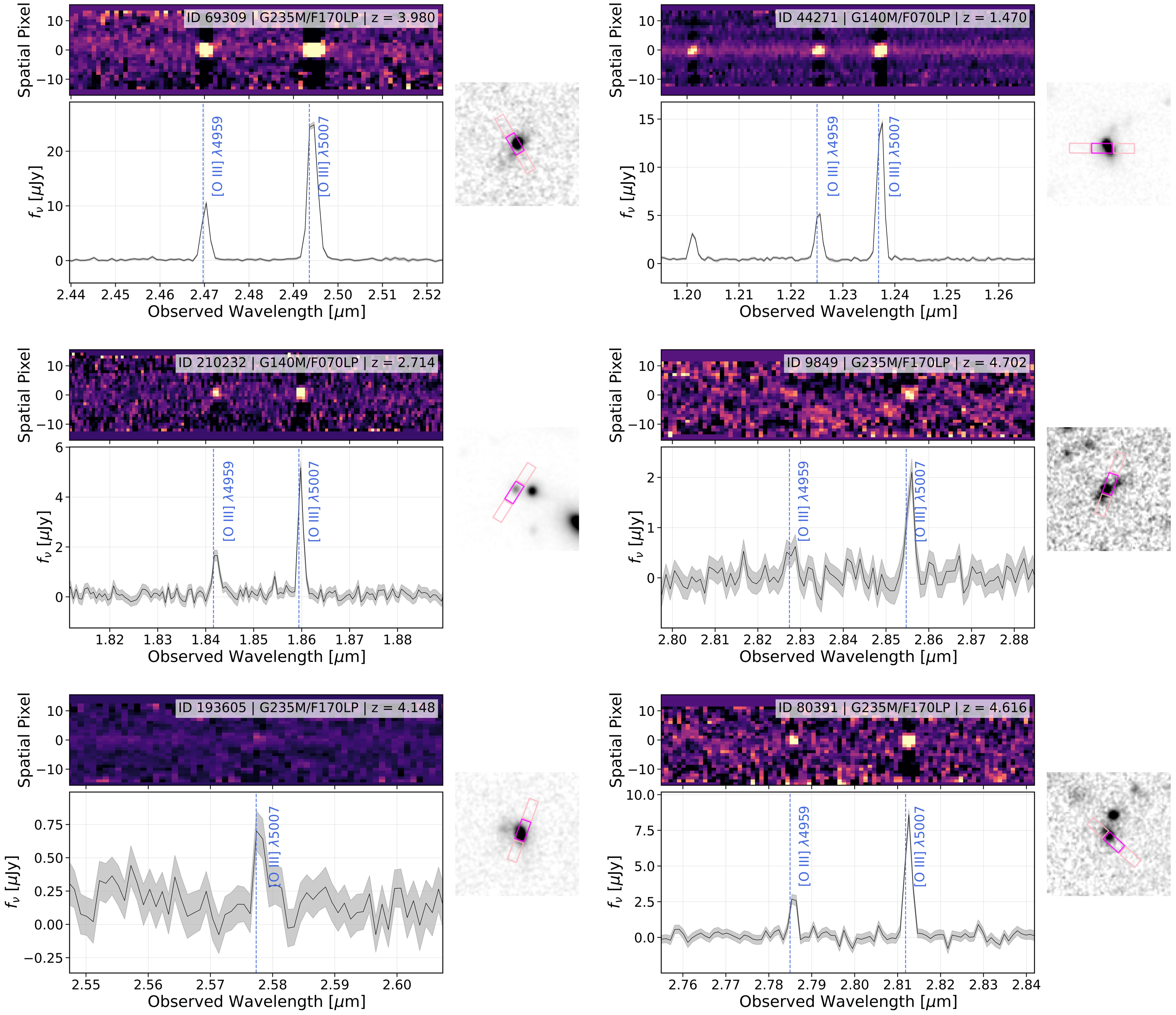}
\caption{Combined spectroscopic and imaging diagnostics for 6 galaxies without reliable gas-phase metallicity measurements. These sources lack sufficient emission-line detections required for the strong-line metallicity diagnostics; however, all show clear evidence of oxygen emission through detections of [O~III] $\lambda4959$ and $\lambda5007$, demonstrating that they are chemically enriched and inconsistent with a pristine primordial gas composition. Each galaxy is shown with three complementary views; \textit{top left}: the 2D spectrum, displaying the spatially resolved emission features and prominent oxygen line traces. The spectroscopic ID, grating/filter combination, and redshift of each galaxy are indicated in a white box; \textit{bottom left}: the extracted 1D spectrum ($\mu$Jy), highlighting the detected emission features used to assess the presence of oxygen; \textit{right}: the corresponding \textit{JWST}/NIRCam F444W imaging cutout with the NIRSpec MSA slitlet configuration overlaid in pink and magenta, indicating the region from which the spectrum was extracted.}
\label{fig:oxygen_spectra}
\end{figure*}

Gas-phase metallicities for the majority of the He~II-emitting sample were derived using the Bayesian strong-line fitting procedure described in Section~\ref{sec:Gas-Phase Metallicity Method}. However, 6 galaxies did not possess a sufficient number of emission-line detections to satisfy the requirements of the fitting procedure, preventing robust metallicity estimates from being obtained. The absence of a strong-line metallicity measurement reflects insufficient spectroscopic information to constrain the abundance reliably, rather than evidence that these galaxies are chemically pristine. At very low metallicity, oxygen emission lines may become intrinsically weak owing to the low oxygen abundance, and therefore their non-detection alone would not necessarily exclude a near-pristine system. In these cases, however, the detection of oxygen emission demonstrates that heavy elements are already present.

As shown in Figure~\ref{fig:oxygen_spectra}, each of these galaxies exhibits clear detections of the [O~III] $\lambda4959$ and $\lambda5007$ emission lines. Their presence demonstrates that these systems have undergone chemical enrichment through previous generations of star formation and therefore cannot represent genuinely primordial, metal-free galaxies. Their exclusion from the metallicity analysis therefore reflects insufficient detections of the full set of emission lines required to constrain the strong-line diagnostics ($R2$, $R3$, $O32$, $R23$, and $\hat{R}$), rather than an absence of heavy elements. Figure~\ref{fig:oxygen_spectra} presents the available spectroscopic and imaging data for each source.

%%%%%%%%%%%%%%%%%%%%%%%%%%%%%%%%%%%%%%%%%%%%%%%%%%

% Don't change these lines
\bsp	% typesetting comment
\label{lastpage}
\end{document}